\documentclass[10pt,aps,prb,amsmath,superscriptaddress,twocolumn,longbibliography,nofootinbib,eqsecnum]{revtex4-2}

\usepackage[toc,page]{appendix}

\usepackage{cases} % Required package
\usepackage{graphicx}
\usepackage{amsmath,amsfonts,amssymb,amsthm,mathtools,mathrsfs}
\usepackage{braket,bm,multirow}
\usepackage[normalem]{ulem}
\usepackage{cancel}
\usepackage{float,enumitem,wasysym,dsfont}

\usepackage[dvipsnames]{xcolor} % Load xcolor once

\makeatletter

\newcommand{\Rmnum}[1]{\expandafter\@slowromancap\romannumeral #1@}
\makeatother

\usepackage{tikz}
\usetikzlibrary{arrows.meta, calc, decorations.markings,angles,quotes,calc,positioning}

\usepackage[colorlinks=true,breaklinks=true]{hyperref}

\usepackage[nameinlink,capitalize]{cleveref}

\hypersetup{
     pdfauthor   = {<Your Name>},
     pdftitle    = {<Paper Title>},
     pdfsubject  = {<Subject>},
     pdfkeywords = {<keywords>},
     colorlinks=true,
     linkcolor=blue,
     citecolor=magenta,
     urlcolor=teal,
     filecolor=teal,
     breaklinks=true}

\begin{document}

\title{Graph-based emulation of $d$-dimensional curved spaces
with superconducting arrays}

\author{Mehmet Dede}
\thanks{These authors contributed equally to this work.}

\author{Guilherme Delfino}
\thanks{These authors contributed equally to this work.}

\affiliation{Department of Physics and Astronomy, Purdue University, West Lafayette, Indiana, 47907, USA}

\author{Andr\'e L.G.\ Mudry}
\affiliation{
ETH Z\"urich, 8092 Z\"urich, Switzerland
            }

\author{Junseok Oh}
\affiliation{The James Franck Institute, University of Chicago, Chicago, Illinois 60637, USA}

\author{Andrew P. Higginbotham}
\affiliation{The James Franck Institute, University of Chicago, Chicago, Illinois 60637, USA}
\affiliation{The Department of Physics, University of Chicago, Chicago, Illinois 60637, USA}

\author{Christopher Mudry}
\affiliation{
Condensed Matter Theory Group,
PSI Center for Scientific Computing,
Theory and Data,
5232 Villigen PSI, Switzerland
            }
\affiliation{
Institut de Physique, EPF Lausanne, CH-1015 Lausanne, Switzerland
            }  

\author{Claudio Chamon} % \orcidlink{0000-0002-8275-2024}}
\affiliation{Department of Physics and Astronomy, Purdue University, West Lafayette, Indiana, 47907, USA}
\affiliation{Purdue Quantum Science and Engineering Institute, Purdue University, West Lafayette, Indiana, 47907, USA}

% \date{\today}

\begin{abstract}
We introduce a framework for emulating graphs and, through them,
curved spaces of arbitrary dimension, using arrays of superconducting
wires. The array consists of two stacked layers of wires, horizontal
and vertical, such that wires are parallel within each layer and
perpendicular between layers.
By discretizing a space into a graph, assigning
a superconducting wire with a rigid phase to each vertex, and coupling
pairs of wires through Josephson junctions according to the graph edges,
arbitrary geometries and topologies can be engineered in a controlled
setting. The superconducting phases then realize scalar field theories
on the emergent geometry. We establish experimentally realistic
conditions for implementing these architectures and develop a
dictionary relating measurable circuit observables to quantities in
the emulated field theory. As an application, we develop the
implementation of hyperbolic (Anti-de Sitter) spaces of constant
negative curvature and use them as an experimentally accessible
platform to explore holographic duality in arbitrary dimensions. We
investigate the effects of disorder in the Josephson couplings, which
translate into metric variations in the bulk-boundary correspondence,
and analyze their impact on boundary scaling exponents both
analytically and numerically, finding that holographic duality remains
robust even in the presence of strong disorder. Beyond holography, the
framework opens a broad range of architectural possibilities,
including the exploration of physics on highly nontrivial graphs and
toy models of dynamical spacetimes.
	
\end{abstract}

 \maketitle

% \tableofcontents

\newpage

%%%%%%%%%%%%%%%%%%%%%%%%%%%%%%%%%%%%%%%%%%%%%%%%%%%%%%%%%%%%%%%%%%%%%%%%%%%%%%%
\section{Introduction}

Physical reality is experienced in three spatial dimensions, with time
serving as a distinct fourth. Yet throughout history, scientists have
found ways to transcend this apparent {(}3+1{)}-dimensional
constraint. Theoretically, there is no limit. Quantum and classical
systems can be conceived in any number of spatial dimensions, from
zero to infinity. In certain theories, extra spatial dimensions are
part of our physical universe, but they are compactified and remain
hidden to us \cite{Kaluza1921,Klein1926}.
Experimentally, numerous systems allow confinement to
reduced dimensions such as, for example, two-dimensional electron
gases in semiconductor heterojunctions \cite{Hiyamizu1980gaas,Klitzing1980},
atomically thin materials like
graphene \cite{Novoselov2004electric,Zhang2005},
one-dimensional quantum wires \cite{Bockrath1999,Moreno2016}
or carbon nanotubes \cite{Radushkevich1952strukture,Iijima1991helical},
and zero-dimensional structures such as quantum dots
\cite{Ekimov1981quantum,Brus1984electron} or buckyballs \cite{Kroto1985}.
On the other hand, synthetic or artificial dimensions can be engineered by
dynamically driving systems, such as with multiple laser frequencies
(see \cite{yu2025comprehensive} for a review),
offering a route to higher-dimensional physics, albeit typically less
robust than static realizations. Beyond dimensionality, geometry can
also be engineered. In particular, curved spaces in two-dimensions
have been realized in a variety of platforms,
including metamaterials, networks,
superconducting and electric circuits
\cite{Kollar2019,Lenggenhager22,Zhang2022,Chen2023hyperbolic,Zhang2023,Patino2024hyperbolic,Yuan2024,Chen2024anomalous,Huang2024hyperbolic,lai2026observation}.

\begin{figure}[t!]
\centering
\includegraphics[width=1.0\linewidth]{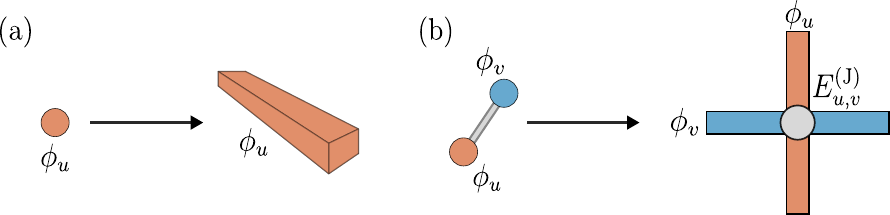}
\caption{
(a) Mapping between  a degree of freedom $\phi_u$
at a vertice {$u$} of a graph and the phase of an extended superconducting
wire {labeled by $u$}.
(b) An edge connecting two vertices {from the graph}
is mapped onto a crossing between {two} wires {$u$ and $v$}
coupled via a Josephson junction $E^{\mathrm{(J)}}_{u,v}$,
indicated as a light circle.
        }
\label{fig:dot_to_wire}
\end{figure}

Here, we introduce a simple scheme to access
\textit{arbitrary geometries in any dimension}
using a matrix array of superconducting wires.
\footnote{
{For that matter,
we could have used any matrix array of wires such that each wire
supports the same long-range order. For example, we could have chosen
wires supporting colinear long-range magnetic order.}
         }
The array consists of two layers of wires,
vertical and horizontal,
with wires on one layer parallel to those in the same layer, but
perpendicular to those on the other layer. The geometry is encoded by
placing Josephson couplings at selected intersections. This
architecture enables programming of general graphs, which in turn
encode discrete geometries. As a result, it provides access to a broad
class of systems, including Euclidean and curved spaces in any
dimension, hybrid structures combining flat and curved regions,
nontrivial topologies such as tori and Klein bottles, and even Cayley
graphs associated with abstract groups. The superconducting matrix
array thus functions as a versatile ``protoboard'' for experimenting
and exploring an entire universe of physical systems, within
geometries and topologies limited only by imagination.

The superconducting matrix array enables the emulation of arbitrary
spaces through three key elements. First, all curved spaces can be
discretized via a mesh or a triangulation \cite{Cairns1935, Whitehead1939},
where the mesh induces a graph structure approximating the space.
Second, each site or vertex
in a graph is represented by a superconducting wire. Any bipartite
graph can be constructed by assigning one subgraph to vertical wires
and the other to horizontal wires. Since every vertical wire
intersects every horizontal one, edges between vertices in opposite
subgraphs can be selectively introduced by placing Josephson couplings
at specific intersections. Third, long-range order (LRO) in the superconductor
over the full length of each wire ensures that it can be described by
a single phase degree of freedom, effectively representing a single vertex.
By reducing the phase stiffness, e.g., by thinning the wires or increasing their
aspect ratio, or by increasing phase fluctuations by raising
temperature, the system would instead transition to a two-dimensional
Euclidean regime when the phase correlation length approaches the wire
length.

The connectivity (or adjacency) matrix of the graph is the mathematical handle on how
to design the physical space. In the case when the vertices correspond
to point-like physical quantities, such as sites in a lattice, one
cannot build geometries in space dimensions larger than the ones in
which the system exists. But when the vertices correspond to line-like
physical quantities, such as the superconducting wires, we can obtain
connectivity matrices for arbitrary geometries. This idea is codified
in Fig.~\ref{fig:dot_to_wire}.

%fig here

Moreover, superconducting wire networks not only enable the
realization of arbitrary geometries, but also provide a physical
setting for probing correlation functions in such spaces.  We show
that measuring current and voltage relations on chosen {pairs of wires} 
provides direct experimental access to two-point correlators on
arbitrary graphs. We also present a detailed and realistic discussion
on how these networks can be built, including design considerations
and measurement protocols.

As a concrete application, we demonstrate that Anti-de Sitter
(AdS){${}_{d+1}$}
spaces with constant negative curvature can be implemented in any
dimension {$(d+1)$ of spacetime}
within this framework. The isometry group of AdS${}_{d+1}$
coincides with the conformal group of
$d$-dimensional spacetime.
It is this match that is central to the duality between AdS bulk
space and conformal field theory (CFT) at the boundary~\cite{Maldacena99}.
Our proposed network of superconducting wire{s}
provides an experimentally
accessible platform to explore holographic duality in a controlled
setting for any dimension $d$ of space.
We note that a flip side of the
bulk/boundary correspondence is that one can use the network of
superconducting wires to design statistical models
with robust criticality at the boundary of a hyperbolic space.

We detail the implementation of a superconducting network realizing a
massive scalar field theory on hyperbolic space, for which the
correspondence between the bulk mass and the boundary scaling exponent
was explicitly derived by Witten \cite{Witten98}. In addition to providing a
laboratory platform to test established predictions of holographic
duality, this realization raises questions that are natural in a
condensed matter setting, but less commonly considered in high-energy
contexts. In particular, we examine the role of disorder in the
Josephson couplings, which translates into variations of the metric in
the bulk-boundary correspondence. 
We show numerically that the relation between the bulk mass and
the boundary scaling dimension in the free theory is continuously deformed
in the presence of disorder. In the weak-disorder regime,
this behavior is consistent with
a random-phase approximation
that we obtain via the replica method \cite{Mezard1988spin}.
As the disorder strength increases, the scaling dimension corrections
depart from those obtained perturbatively. More broadly,
disorder provides a controlled route to accessing interacting effective
theories on arbitrary graph geometries,
opening a range of theoretical questions
in the context of holography and its experimental realization
in superconducting arrays.

The paper is organized as follows.
In Sec.\ \ref{sec: graphs/geometries and JJ arrays},
we explain the general framework and show how
superconducting wire arrays can be used to emulate arbitrary graphs,
and as a consequence, curved
spaces in any dimension. We also discuss explicitly how a
massive scalar field theory on curved spaces is programmed on the
superconducting arrays. In Sec.\
\ref{sec:Crossover between different geometries of space},
we discuss the crossover from the programmed
geometry to {that of two dimensional Euclidean space}
as the phase gradients along the wires
increase due to, for example, thermal fluctuations.
In Sec.\
\ref{sec:The AdS/CFT correspondence for Euclidean hyperbolic spaces},
we present an application of our framework,
namely how to experimentally access the holographic AdS/CFT
correspondence. We discuss how superconducting wire arrays can realize
Anti-de Sitter geometries, and establish that the presence of a critical
boundary persists against disorder. Furthermore, we characterize
the deformation of the relationship between bulk mass and
the boundary scaling exponent across the weak- and strong-disorder regimes.
Finally, in Sec.~\ref{sec:Experimental outlook},
we discuss a pathway to experimental realization using aluminium wires
and give an estimate of the maximum system
size that can be attained for reasonable temperatures.

%%%%%%%%%%%%%%%%%%%%%%%%%%%%%%%%%%%%%%%%%%%%%%%%%%%%%%%%%%%%%%%%%%%%%%%%%%%%%%%

\section{Physics on arbitrary graphs
through Josephson-coupled wires}
\label{sec: graphs/geometries and JJ arrays}

Here, we develop a general framework to implement scalar field
theories on arbitrary discrete graphs using
{networks of superconducting wires}.
The realization of curved spaces in any dimension is a
particular case of the implementation on graphs. For this, 
we encode degrees of freedom as phases of
superconducting wires. When the phase rigidity is sufficiently large,
the phase takes a single value along the wire, up to a certain
temperature-dependent length scale (that will be computed in
Sec. \ref{sec:Crossover between different geometries of space}).  In
this regime, each wire is assigned a ``site'' variable $\phi_u\in
[0,2\pi)$, as illustrated in Fig.~\ref{fig:dot_to_wire}(a). Couplings
between degrees of freedom $\phi_u$ and $\phi_v$ are implemented via
Josephson junctions between the wires, with energy
$E^{(\mathrm{J})}_{u,v}$, as exemplified in
Fig.~\ref{fig:dot_to_wire}(b).

\subsection{Emulating arbitrary simple graphs}

{
W}e consider a
general configuration of compact variables
$\phi_u{\in\mathbb{R}/2\pi\mathbb{Z}}$
defined on a discrete space described by a simple graph 
\footnote{A \emph{simple graph} $G=(V,E)$ consists of a countable set $V$
of vertices and a set
$E\subseteq\bigl\{\{u,v\}\mid u,v\in V,\; u\neq v \bigr\}$, i.e.,
an undirected graph with no loops and no multiple edges, and whose edges are not necessarily assigned weights.}
$G=(V,E)$ with $|V|$
vertices and $|E|$ edges. Here,
each vertex $u\in V$ is assigned the
local degree of freedom {$\phi_u$},
while {each} edge ${\{u,v\}}\in E$ capture{s the interaction
between $\phi_u$ and $\phi_v$}. 
This graph formulation provides a unified description for the
statistical physics with target space given by $\mathbb{R}/2\pi\mathbb{Z}$
and whose base space is a discrete structure, such as a triangulation
of a smooth manifold, a fully connected graph, a Cayley tree,
or related geometries.

The connectivity of the graph is encoded in the
{$|V|\times|V|$}
symmetric adjacency matrix $A=(a_{u,v})$, whose nonzero entries indicate
coupled pairs of degrees of freedom. The matrix element $a_{u,v}$
equals 1 if {the pair of distinct vertices} $u$ and $v$
are connected by an edge, that is
${\{u,v\}}\in E$, and 0 otherwise. Diagonal entries vanish,
reflecting the absence of self-connections.

It is convenient to focus on bipartite graphs,
where the vertex set $V$ can be partitioned into two disjoint
subsets $V_1$ and $V_2$ such that edges connect only vertices across
the partition. In this case, the adjacency matrix takes an
off-diagonal block form
\begin{equation}
A =
\begin{pmatrix}
0 & B \\
B^{\mathsf{T}} & 0
\end{pmatrix},
\end{equation}
where $B $ is the $|V_1|\times|V_2|$
biadjacency matrix encoding connections between nodes in $V_1$ and $V_2$.

A bipartite graph
admits a natural physical realization in terms of
superconducting wire arrays. We assign a superconducting wire to each
vertex, with phases $\varphi_i$ ($i\in V_{1}$) and
$\phi_j$ ($j\in V_{{2}}$). The array is arranged as
$|V_{{1}}|$
horizontal and
$|V_{{2}}|$
vertical wires.
Josephson couplings are added to
selected intersections of wires according to the biadjacency matrix
$B=(b_{i,j})${,}
\begin{subequations}
\label{eq:def array sc wires Josephson coupled f bipartite}
\begin{numcases}{E_{i,j}^{(\mathrm{J})}=}
  |J_{i,j}|>0, & if $b_{i,j}^{\,}= 1$, \label{eq:def array sc wires Josephson coupled f bipartite a} \\
  0,   & if $b_{i,j}^{\,}=0$. \label{eq:def array sc wires Josephson coupled f bipartite b}
\end{numcases}
In the above, $J_{i,j}$ is the Josephson coupling between nodes $i\in V_1$ and $j\in V_2$. 
Condition
(\ref{eq:def array sc wires Josephson coupled f bipartite a})
ensures that a pair of superconducting wires $(i,j)$ is coupled
via the Josephson energy
\begin{equation} {
|J_{i,j}|\,
\left[
1
-
\cos
\left(
\varphi_{i}^{\,}
-
\phi_{j}^{\,}
+
\delta^{\,}_{-1,\mathrm{sgn}(J_{i,j})}\,\pi
\right)
\right]
\geq0  }
\label{eq:def array sc wires Josephson coupled f bipartite c}
\end{equation}
if and only if the biadjacency matrix element $b_{i,j}^{\,}=1$.
Conversely, condition
(\ref{eq:def array sc wires Josephson coupled f bipartite b})
ensures that pairs with $b_{i,j}=0$ are not Josephson coupled. 
{When
\begin{equation}
b_{i,j}=1 \ \Longrightarrow\
J_{i,j}=J>0
\label{eq:def array sc wires Josephson coupled f bipartite d}
\end{equation}
holds,}
\footnote{
When this strong  form of translation invariance does not hold,
we replace the right-hand side of Eq.\
(\ref{eq:def array sc wires Josephson coupled f bipartite e})
with
$
\sum\limits_{i\in V_1}
\sum\limits_{j\in V_2}
b_{i,j}^{\,}\,
|J_{i,j}|\,
\left[
1
-
\cos
\left(
\varphi_{i}^{\,}
-
\phi_{j}^{\,}
+
\delta^{\,}_{-1,\mathrm{sgn}(J_{i,j})}\,\pi
\right)
\right]
$.  %
         }
the total {classical}
Josephson energy of the array of $|V_1|+|V_2|$
superconducting wires {is defined to be}
\footnote{{
An array of quantum Josephson junctions also assigns a charge
operator to any vertex of the graph such that its commutator with the
superconducting phase assigned to the same vertex
is proportional to the imaginary unit
together with a charging energy, as presented in
Appendix~\ref{subsec:Definition of the quantum Josephson array}.
The classical regime presumes a vanishing charging energy for all
the quantum wires, i.e., we work in the classical limit of large
capacitances. We will use finite capacitances when deriving
the current-voltage response functions that can be measured
in Sec.~\ref{sec:Experimental outlook}.   }
         }
\begin{align}
H_{G}^{\,}(J):=&\,
\sum\limits_{{\{i,j\}}\in E} 
E_{i,j}^{(\mathrm{J})}\,
\left[
1
-
\cos
\left(
\varphi_{i}^{\,}
-
\phi_{j}^{\,}
\right)
\right]
\nonumber\\
=&\,
J
\sum\limits_{i\in V_{{1}}}
\sum\limits_{j\in V_{{2}}}
b_{i,j}^{\,}\,
\left[
1
-
\cos
\left(
\varphi_{i}^{\,}
-
\phi_{j}^{\,}
\right)
\right],
\label{eq:def array sc wires Josephson coupled f bipartite e}
\end{align}
\end{subequations}
where $\varphi$ ($\phi$) refers to the rigid superconducting phase
of the $|V_1|$ ($|V_2|$)
horizontal (vertical) superconducting wires.

\begin{figure}[t!]
\centering
\includegraphics[width=0.8\linewidth]{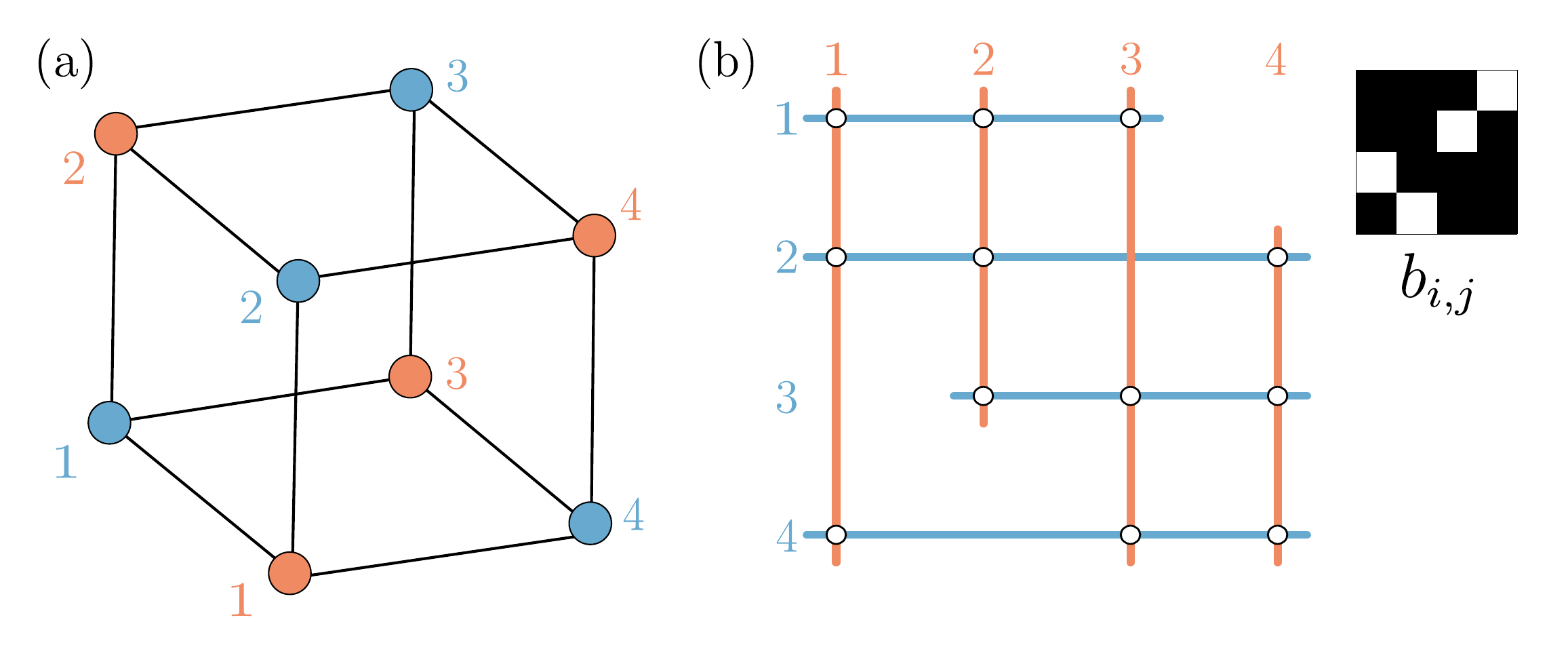}
\caption{ 
(a) Bipartite graph structure {associated to a} single cube{.}
(b) Josephson array using $8$ superconducting wires,
$4$ of which are
either vertical or horizontal.
The pattern of crossings (white circles) reproduces the biadjacency matrix
(shown in the upper right corner),
with black and white squares indicating entries
$b_{i,j}=1$ and $b_{i,j}=0$, respectively.
Selected wires are shortened where no additional crossings are required.
        }
\label{fig:cube_wire}
\end{figure}

As a concrete example, let us consider the
graph corresponding to a single three-dimensional cube.
As a bipartite graph, indicated by
the blue and orange coloring of the vertices in
Fig. \ref{fig:cube_wire}(a), it can be {
realized by} the wire array
through horizontal and vertical wires in
Fig.\ \ref{fig:cube_wire}(b). The pattern of Josephson
junctions reproduces the pattern of nonvanishing elements of the
biadjacency matrix.  In Fig.\ \ref{fig:cube_wire}(b),
some wires are shortened as no additional couplings
{are} required {to realize the biadjacency matrix of the
three-dimensional cube}.

     The construction {(\ref{eq:def array sc wires Josephson coupled f bipartite e})}
 can be generalized to the
        non-bipartite case.  Let $G = \left(V,\, E\right)$ be a simple
        non-bipartite graph with adjacency matrix $A = (a_{u,v})$.  We
        begin by ``duplicating" $G$ to construct a bipartite auxiliary
        graph $G_{\text{aux}} = \left(V_{\text{aux}},\, E_{\text{aux}}
        \right)$.  The set of auxiliary vertices $V_{\text{aux}} := V
        \times \{0,1\} $ is defined by duplicating each original
        vertex.  For every edge $\{u,v\} \in E$ of the non-bipartite
        graph we add the two edges $\left\{ (u,0), \, (v,1) \right\}$
        and $\{ (u,1) ,\, (v,0)\}$ to the auxiliary one.
        The auxiliary graph
        $G_\text{aux}$ is by construction bipartite (with disjoint
        subsets indexed by 0 or 1).
        Next, we associate to each auxiliary vertex of the form $(u,0)$
        \big($(u,1)$\big) a horizontal (vertical) wire with
        superconducting phase $\varphi_{u}$ ($\phi_u$).  This results
        in a square array made of
        $2|V|$ superconducting wires, on
        which we fix the Josephson junctions with
        their coupling energies
     \begin{subequations}
     \begin{align}
       E_{(u,0),(v,0)}^{(\mathrm{J})} =&\,
          E_{(u,1),(v,1)}^{(\mathrm{J})} = 0,
       \label{eq:def array sc wires Josephson coupled nonbipartite a} \\
       E_{(u,0),(u,1)}^{(\mathrm{J})}
          \rightarrow&\,
          \infty,
          \label{eq:def array sc wires Josephson coupled nonbipartite b}  \\
       E_{(u,0),(v,1)}^{(\mathrm{J})} =&\,
          E_{(u,1),(v,0)}^{(\mathrm{J})} = \frac{J}{2}a_{u,v} > 0
     \label{eq:def array sc wires Josephson coupled nonbipartite c} 
     \end{align}
     \end{subequations}
     defined for all vertices $u,v \in V$ of the original graph $G$.

\begin{figure}[t!]
\includegraphics[width=0.8\linewidth]{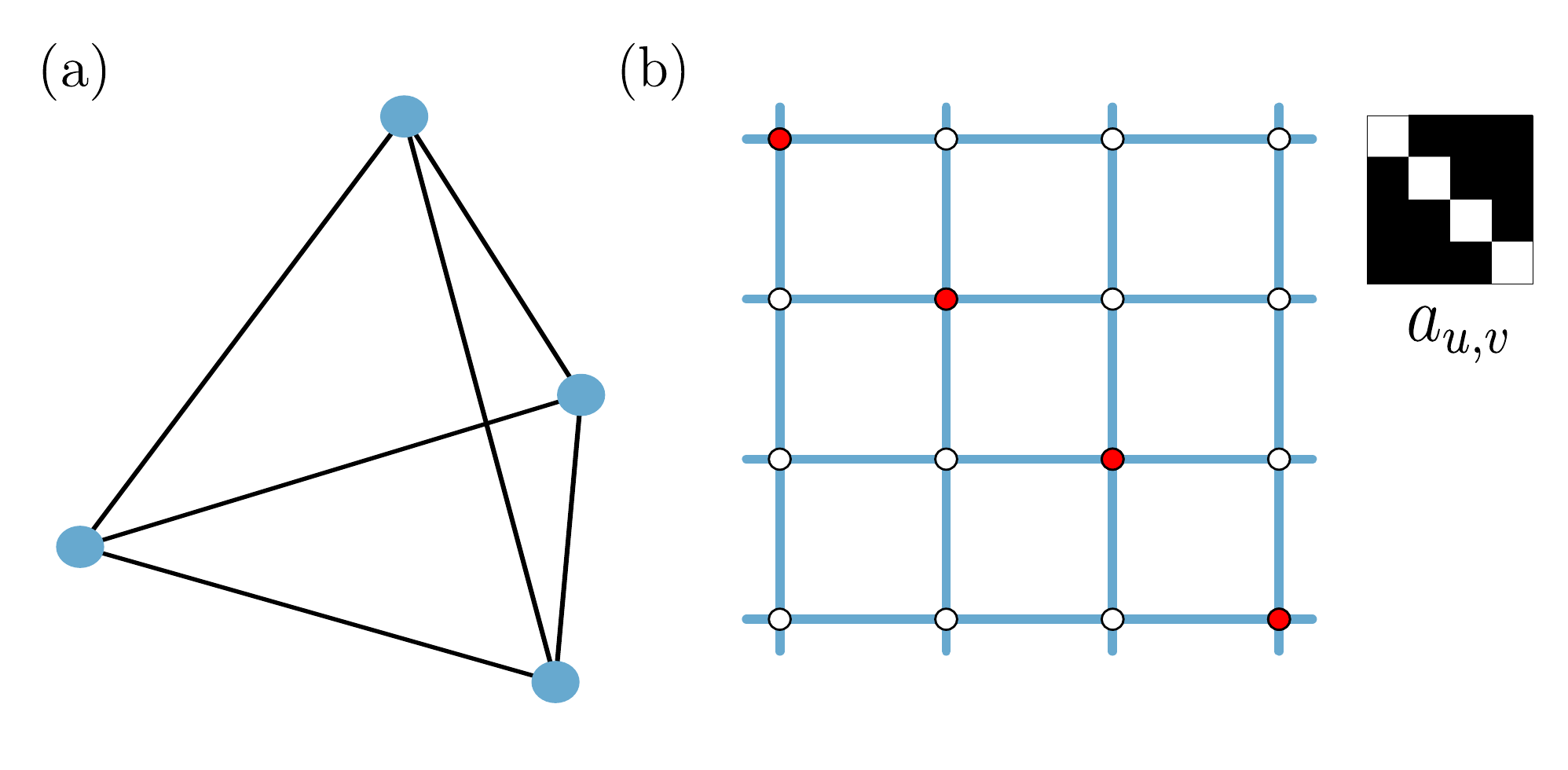}
\caption{
(a) Triangulation of a three-dimensional tetrahedron. 
(b) Josephson array using $8$ superconducting wires.
The red circles denote ``short circuits'' or infinite Josephson couplings.
Phases of the vertical and horizontal wires corresponding to identical vertices are
locked together.
        }
\label{Fig:tetrahedron_wire}
\end{figure}

     Condition~\eqref{eq:def array sc wires Josephson coupled nonbipartite a} preserves the bipartite structure of
     the auxiliary graph by strictly forbidding intra-set couplings
     within $V \times \{0\}$ or $V \times \{1\}$.
     Condition~\eqref{eq:def array sc wires Josephson coupled nonbipartite b} energetically enforces a diagonal pairing between phases
     $\varphi_{u}$ and $\phi_{u}$, effectively merging those
     horizontal and vertical wire pairs into a single entity via an
     ``infinite'' Josephson coupling, essentially a ``short circuit'' of
     those two wires.
     This condition removes the initial duplication of
     degrees of freedom associated to the auxiliary construction.
     Condition~\eqref{eq:def array sc wires Josephson coupled nonbipartite c}  is analogous to the
     bipartite case, except that the coupling $J$ is halved to account
     for the symmetry of $A$.
     The total Josephson energy of the
     non-bipartite array of $|V|$ superconducting wires is thus
     \begin{equation}
          \begin{split}
               H_{G}(J)
               =&\, \sum_{\{u,v\}\in E}
               E_{u,v}^{(\mathrm{J})}
               \bigl[1-\cos(\phi_{u}-\phi_{v})\bigr] \\
               =&\, \frac{J}{2}
               \sum_{u,v\in V}
               a_{u,v}
               \bigl[1-\cos(\phi_{u}-\phi_{v})\bigr],\label{eq:def array sc wires Josephson coupled f}
          \end{split}
     \end{equation}
     where $E_{u,v}^{(\mathrm{J})} := E_{(u,0),(v,1)}^{(\mathrm{J})}$ and
     $\varphi_{u} \equiv \phi_{u}$
     is energetically enforced for all
     $u \in V$. Figure~\ref{Fig:tetrahedron_wire}(a) illustrates an explicit example
     of a non-bipartite graph, namely the tetrahedron. Red circles
     along the diagonal crossings in the wire realization in
     Fig.~\ref{Fig:tetrahedron_wire}(b) represent the ``short circuits'' or infinite
     Josephson couplings.

\begin{figure}[t!]
\centering
\includegraphics[width=1.0\linewidth]{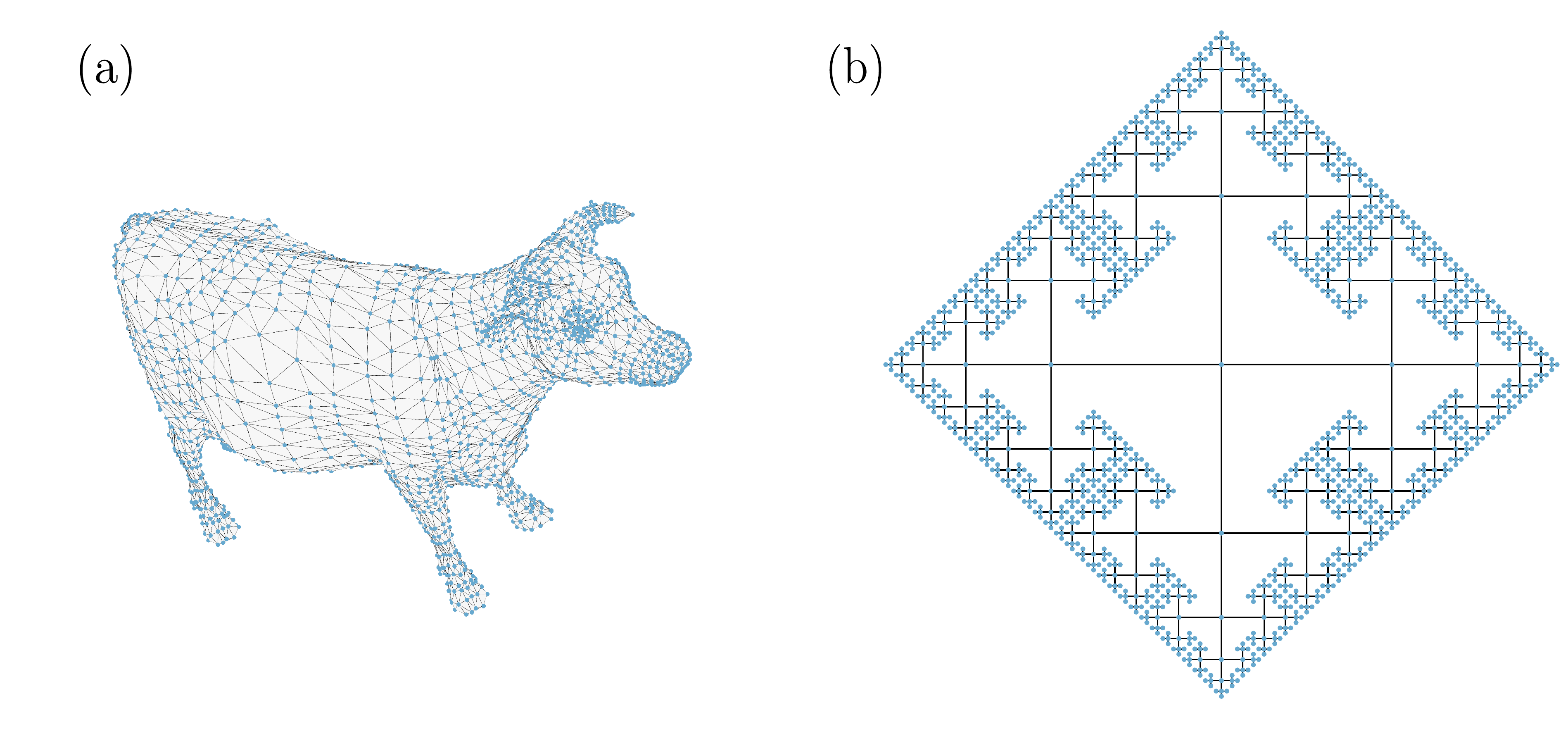}
\caption{
(a) Mesh associated with a triangulation of a smooth surface (cow).
The graph metric is inherited from the underlying continuous metric manifold.
(b) Cayley graph of a free group. The metric is typically defined to reflect
an appropriate property of the graph.
        }
\label{fig:cow_caley}
\end{figure}

\subsection{Discrete geometries}

To relate these discrete constructions to continuum geometries, it is
useful to equip the graph with a distance function $d(u,v)$ that
assigns a non-negative real number to each pair of vertices $u,v \in
V$. This distance function must satisfy the usual metric properties{,
namely}
(i) $d(u,v)=0$ if and only if $u=v$, (ii) symmetry $d(u,v)=d(v,u)$,
and (iii) the triangle inequality $d(u,w)\le d(u,v)+d(v,w)$ for all
vertices $u,v,w\in V$.
Depending on the context, $d(u,v)$ can arise in different ways.
For triangulations of {a Riemannian manifold}, it is
inherited from the manifold's Riemannian metric{.
F}or a Cayley graph, it reflects
algebraic or combinatorial distances \cite{Serre2002trees}.
In the absence of any underlying structure,
it  is the shortest-path graph distance.
Representative examples are shown in Fig.~\ref{fig:cow_caley}.

{Conversely,
starting from a continuum description,
a natural and powerful way to connect smooth manifolds to discrete
combinatorial structures is through triangulation.}
In this approach, a
smooth $n$-dimensional manifold $M$ is decomposed into $n$-simplices
such that (i) the union of all simplices reconstructs $M$, and (ii)
any two simplices intersect in a shared face or not at all. Classic
results by Cairns~\cite{Cairns1935} and Whitehead~\cite{Whitehead1939}
guarantee that any smooth manifold admits a triangulation that is
essentially unique up to piecewise-linear equivalence, meaning that
the manifold’s topological and compatible geometric structures are
faithfully preserved.

{
Each triangulation of the $n$-dimensional manifold $M$ can then be encoded by
an $n$-dimensional simplicial complex $K$, whose top-dimensional
simplices correspond to the $n$-simplices of the triangulation and whose
faces encode their intersections. A useful way to organize the
combinatorial structure of $K$ is through its skeleta. For each
$k=0,1,\dots,n$,
the $k$-skeleton
\begin{equation}
K^{(k)} = \{\text{all simplices of dimension } \le k\}
\end{equation}
captures the information of $K$ up to dimension $k$.
In particular, the
1-skeleton $K^{(1)}$ forms a graph consisting of vertices
(0-simplices) connected by edges (1-simplices). 
Because a simplicial complex does not allow 1-simplices with identical
endpoints, nor more than one 1-simplex connecting the same pair of
distinct vertices, this graph has no loops and no multiple edges. The
1-skeleton is therefore a simple graph. } 
Each smooth manifold therefore
defines a set of simple graphs{,}
one for each admissible triangulation, providing the one-to-many map
\begin{equation}
M \ \rightsquigarrow\
\{\,\text{1-skeleton graphs of its triangulations}\}.
\end{equation}

Triangulation preserves the manifold's topological structure while
remaining compatible with its smooth structure through
piecewise-linear geometry. Extracting the 1-skeleton reduces the
manifold to a combinatorial object, a graph, that encodes adjacency and
connectivity. Equipping this graph with a metric $d(u,v)$ inherited
from a {Riemannian manifold, say,}
further allows the discrete structure to capture
geometric features such as lengths and curvature. Statistical
{physics}
defined on such graphs, when refined in the thermodynamic
limit, can give rise to effective field theories describing low-energy
and long-wavelength physics on an emergent smooth manifold. From this
perspective, microscopic physics on a discrete graph may be regarded
as more fundamental, with spacetime symmetries, such as Galilean,
Poincar\'e, or conformal invariance{,} emerging at long distances from
the underlying discrete structure.

\begin{figure*}[t]
\centering
\includegraphics[width=0.43\linewidth]{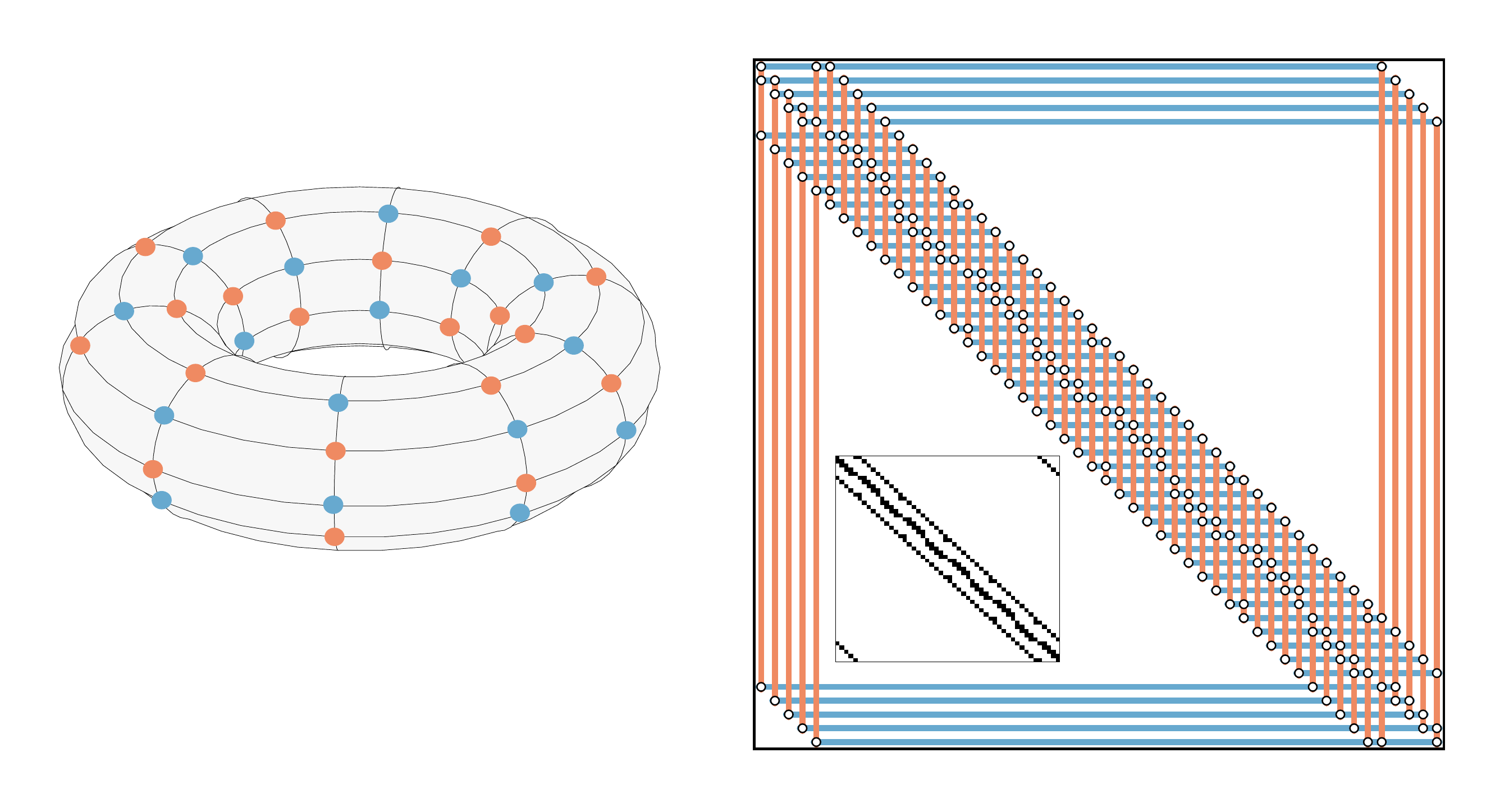}
\includegraphics[width=0.43\linewidth]{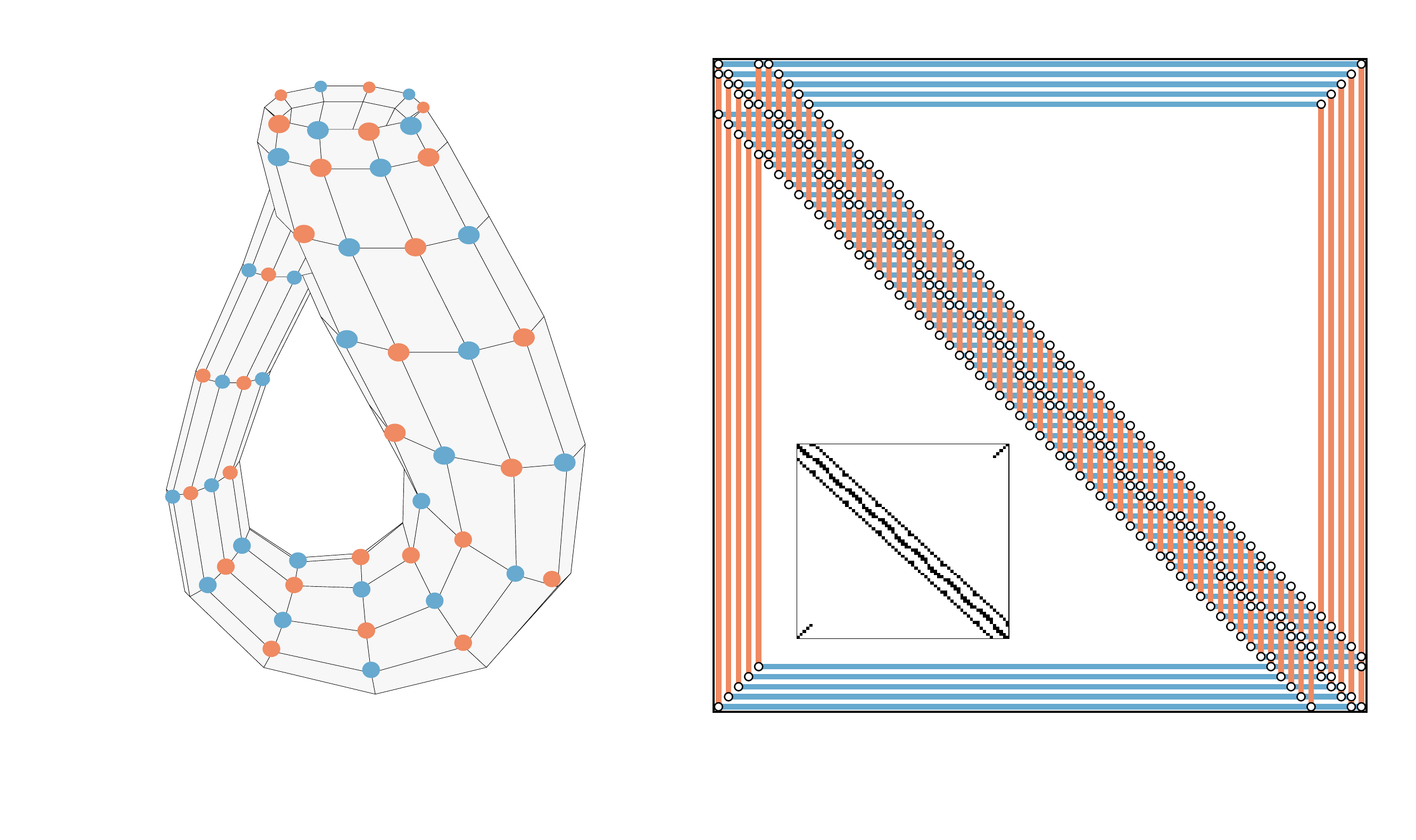}
\caption{
Superconducting wire implementations of bipartite graphs
corresponding to nontrivial topologies: a torus and a Klein bottle.
The inset shows the biadjacency matrix $b_{ij}$
with black and white corresponding to 1 and 0 entries, respectively.}
\label{fig:torus_klein}
\end{figure*}

More elaborate examples of such realizations involving larger graphs
are shown in Fig.~\ref{fig:torus_klein}, where nontrivial topologies
such as the torus and Klein bottle are implemented through
{networks of wires}.
Smooth manifolds can be approximated by discrete structures
such as triangulations and their associated graphs.  

So far, we discussed programming biadjacency/adjacency matrices for
simple graphs using Josephson junctions, which encode only the
connectivity information of the vertices {owing to assumption
(\ref{eq:def array sc wires Josephson coupled f bipartite d})}.
However, for a given
triangulation on a particular geometry, the edge lengths of the
simplices differ. To integrate geometric data{,
we relax assumption
(\ref{eq:def array sc wires Josephson coupled f bipartite d})
by using the weighted Josephson couplings}
\begin{equation}
E^{(\mathrm{J})}_{u,v}=
\begin{cases}
J_{u,v}, & a_{u,v}=1,
\\
0, & a_{u,v}=0,
\end{cases}
\label{eq: JJ arrays and edge length}
\end{equation}
where $J_{u,v}\in\mathbb{R}_{>0}$ is a positive real number that
stores the edge length.  Therefore, instead of using identical
junctions, we use varying coupling strength $J_{u,v}$ to emulate the
distance function on the graph $G$.

To illustrate how the weighted energies encode the
desired geometry, we
consider, without loss of generality, a non-bipartite discretization
of the action of a massive scalar field on a curved space
\cite{Christ1982}
\begin{equation}
S=
\frac{1}{2}
\sum_{\{ u,v\}\in E}
V_{uv}^{\,}\,
\frac{\left(\phi_{u}^{\,}-\phi_{v}^{\,}\right)^{2}}{l_{uv}^{2}}
+
\frac{1}{2}
\sum_{u\in V}
V_{u}^{\,}\,
m_{\,}^{2}\,\phi_{u}^{2},
\label{eq:discritized massive scalar}
\end{equation}
where
$\phi_{u}^{\,}$
is the value of the scalar field at vertex $u$ of the triangulation,
${\{u,v\}}\in E$ runs over edges connecting the {pair of distinct}
vert{ices} $u$ and $v$,
$l_{uv}^{\,}\equiv l_{vu}^{\,}$ is the length of the edge ${\{u,v\}}$
computed from the distance function $l_{uv}\equiv d(u,v)$,
$V_{uv}^{\,}\equiv V_{vu}^{\,}$
is the dual volume associated with edge ${\{u,v\}}$
(typically, the volume of the region in the Voronoi dual cell
intersected by the edge),
$V_{u}^{\,}$ is the dual volume associated with vertex $u$,
and
$m$ is the mass of the scalar field.
These dual volumes are encoded in higher-dimensional skeleton structures
$K^{(k)}$, or can be derived directly from the metric $d(u,v)$.

We observe here that the first summation on the right-hand side of Eq.\
(\ref{eq:discritized massive scalar})
is identical to the expansion of the Josephson
energy
(\ref{eq:def array sc wires Josephson coupled f bipartite c})
to second order in the phase difference,
provided we make the identification
$E_{u,v}^{\mathrm{(J)}}\to V_{uv}^{\,}/l_{uv}^{2}$.
The second summation on the right-hand side of Eq.\
(\ref{eq:discritized massive scalar})
follows {(i)} by adding a Josephson coupling
$E_{u\infty}^{\mathrm{(J)}}[1-\cos(\phi_{u}^{\,}-\phi_{\infty}^{\,})]$
from any one of the superconducting wires
from the square array to a superconducting reservoir
with the superconducting phase $\phi_{\infty}^{\,}\equiv0$
(i.e., all other phases are measured relative
to the superconducting reservoir){, (ii)}
by expanding this Josephson coupling to second order
in its phase difference{,
and (iii)} by identifying
$E_{u\infty}^{\mathrm{(J)}}\to V_{u}^{\,}\,m^{2}$.

In summary, a square array of Josephson-coupled superconducting wires
with couplings satisfying conditions
\eqref{eq:def array sc wires Josephson coupled f bipartite a}%
--\eqref{eq:def array sc wires Josephson coupled f bipartite c}
realizes the biadjacency matrix of any simple bipartite graph
$G$ with $N_1+N_2$ vertices, provided each wire carries a rigid
superconducting phase. By additionally imposing
condition
\eqref{eq:def array sc wires Josephson coupled nonbipartite b},
the adjacency matrix of any simple (possibly non-bipartite) graph
$G$ with $N$ vertices can also be realized. Each such {simple}
graph can be viewed as the $1$-skeleton of a triangulation of a smooth
manifold. This observation establishes a direct mapping between
the discrete Josephson array and arbitrary spatial geometries
(\ref{eq: JJ arrays and edge length}).
Examples include Euclidean or curved spaces in any dimension,
and topologies such as the torus or the Klein bottle
(depicted in Fig.~\ref{fig:torus_klein}).

Thermal fluctuations limit the length scale over which the
superconducting phase can be assumed to be uniform along a wire. This length
scale, in turn, constrains how many wires (i.e., graph vertices) can
be incorporated into the array, given the wire widths and spacings, so
as to still realize the intended geometries. As a result, curved-space
phenomena are accessible only in a crossover regime, with the number
of allowed nodes increasing as the temperature decreases. In
Sec.\ \ref{sec:Crossover between different geometries of space}, we
derive, for a fixed temperature, an upper bound on the number of wires
$N_{\star}$ below which the rigid-phase approximation remains valid. We also
discuss the geometric crossover from the physics of curved space in
arbitrary dimension to two-dimensional Euclidean physics as a function
of temperature and Josephson couplings.

\begin{figure*}
\centering
\includegraphics[width=1.0\linewidth]{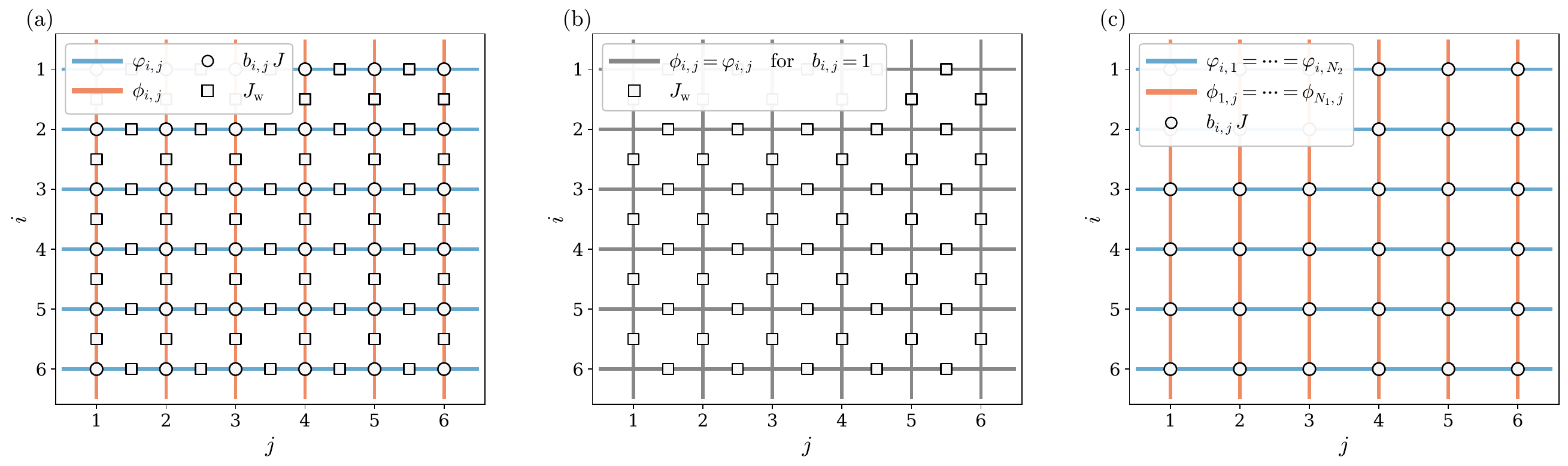}
\caption{
(a)
The square lattice $\Lambda$ is defined by the set of white circles.
The medial square lattice ${\Lambda_{\mathrm{medial}}}$
associated to $\Lambda$
is defined by the set of
white squares. The number $N_1$ ($N_2$)
of coupled superconducting wires of length
$L=N_2\,\mathfrak{a}$
($N_1\,\mathfrak{a}$)
and cross sectional area $(\mathfrak{a}/2)^{2} $
are laid horizontally (vertically) on every 
row (column) of the square lattice $\Lambda$.
{Horizontal (vertical) superconducting wires are colored in blue
(orange). Horizontal and vertical superconducting wires can only interact
through the Josephson coupling $b_{i,j}\, J\geq0$ at their intersections,
the white circles from the set $\Lambda$. }
Each superconducting wire is itself
modeled as a chain of Josephson-coupled superconducting segments of
length $\mathfrak{a}$, whereby each white square
from $\Lambda_{\mathrm{medial}}$
represents the Josephson coupling $J_{\mathrm{w}}^{\,}>0$ between two consecutive
segments of the superconducting wire. 
(b) On the one hand, if $J\gg J_{\mathrm{w}}$, then
the pair of phases $\phi_{i,j}$ and $\varphi_{i,j}$ are energetically enforced
to match, i.e.,
$\phi_{i,j}=\varphi_{i,j}$ (now depicted as gray crosses)
for all $(i,j)\in\Lambda$ such that $b_{i,j}=1$.
The system is effectively mapped to an $XY$-like model
with neighboring coupling constant $J_{\mathrm{w}}>0$.
(c) On the other hand, if $J_{\mathrm{w}}^{\,}\gg J$, then
$\phi_{i,j}=\phi_{j}$ for each column $j=1,\ldots, N_2$
($\varphi_{i,j}=\varphi_{i}$ for each row
$i=1,\ldots, N_1$). This limit delivers $N_1+N_2$
wires that are coupled through the biadjacency matrix
$B$ of a simple bipartite graph $G$. 
         }
\label{Fig:toy model for coupled wires}
\end{figure*}

%%%%%%%%%%%%%%%%%%%%%%%%%%%%%%%%%%%%%%%%%%%%%%%%%%%%%%%%%%%%%%%%%%%%%%%%%%%%%%%
\section{Geometric stability: Superconducting phase rigidity and
dimensional crossover}
\label{sec:Crossover between different geometries of space}

The premise of this work is that any vertex $u$ 
of the simple bipartite graph $G=(V,E)$
can be assigned an independent rigid superconducting phase,
associated to a superconducting wire of
width $r=\mathfrak{a}/2$,
and
cross-sectional area $(\mathfrak{a}/2)^{2}$,
with parallel nearest-neighbor wires spaced by $\mathfrak{a}$.
In this section,
we are going to assume that there are $N_1$ horizontal wires
and $N_2$ vertical wires such that the longest horizontal (vertical) wire
has length $N_2\,\mathfrak{a}$ ($N_1\,\mathfrak{a}$).
The cardinality $|V|$ of the set $V$ is thus $N_1+N_2$,
while the area of the square array of superconducting wires
scales like $N_1\,N_2\,\mathfrak{a}^{2}$.
The assumption according to which each superconducting wire
can be assigned a single rigid phase must 
necessarily break down in the
thermodynamic limit defined by taking $N_1\propto N_2\uparrow\infty$
holding $\mathfrak{a}$ fixed and at any nonvanishing temperature $T>0$.
As $|V|\uparrow\infty$ at fixed $\mathfrak{a}$,
the wire length scales as
\begin{equation}
L:=\frac{|V|}{2}\,\mathfrak{a},
\end{equation}
eventually exceeding the length
$L_{\star}$
over which the phase fluctuates by an angle of
order $\pi$. The cost of this phase variation is determined by the
superfluid stiffness of the wire.

The first goal of this section, in Sec.\
\ref{subsec:Definitions and an upper bound on superconducting phase ...},
is to estimate the value
\begin{equation}
L_{\star}^{\,}(T)\equiv
N_{\star}^{\,}(T)\,\mathfrak{a},
\label{eq:def Lstar and Nstar}
\end{equation}
which is defined to be an upper
bound on the wire length to safely assume a rigid superconducting
phase along the wire at temperature $T$.
The second goal, in Sec.\
\ref{subsec:Geometric crossover},
is to investigate how increasing the number of wires
beyond $N_{\star}$ leads to a loss of phase rigidity, resulting in a
crossover from the effective theory on a curved geometry to a
two-dimensional regime defined by the planar layout of the circuit and
its local Josephson couplings. For disordered aluminum wires at
temperature $0.1\,\mathrm K$, we obtain a conservative estimate
$N_{\star}^{\,}=3.25\times10^{4}$, which will be justified 
in Eq.\ (\ref{eq:Numerical-value-Nstar}).

\subsection{Upper bound on superconducting phase
fluctuations and $N_{\star}$}
\label{subsec:Definitions and an upper bound on superconducting phase ...}

Without loss of generality, we treat a
 bipartite simple graph
{$G=(V,E)$} (with $V$ the disjoint union of $V_1$ and $V_2$)
{that we realize with a square array of
superconducting wires that intersect on the square lattice
\begin{subequations}
\label{eq:def Josephson array on square lattice}
\begin{equation}
\Lambda:=
\Big\{
(i,j)\ \Big|\
i=1,\ldots, N_1,
\qquad
j=1,\ldots, N_2
\Big\}
\label{eq:def Josephson array on square lattice a}
\end{equation}
with lattice spacing $\mathfrak{a}$ (see Fig.\ \ref{Fig:toy model for coupled wires}).  \footnote{We adopt a matrix-like convention when drawing the lattice, in which $i$ labels rows (from top to bottom) and $j$ labels columns (from left to right).}
The sites from the lattice $\Lambda$ have a dual purpose.
On the one hand,
to each site $(i,j)\in\Lambda$ we assign
the pair of superconducting phases
$\phi_{i,j}\in[0,2\pi)$
and
$\varphi_{i,j}\in[0,2\pi)$.
On the other hand, each site $(i,j)\in\Lambda$ hosts the
Josephson coupling $J\geq0$ together with the local Josephson energy
\begin{equation}
b_{i,j}^{\,}\,
J\,
\left[
1
-
\cos
\left(
\phi_{i,j}^{\,}
-
\varphi_{i,j}^{\,}
\right)
\right],
\label{eq:def Josephson array on square lattice b}
\end{equation}
where the number $b_{i,j}=0,1$ originates from the
biadjacency matrix elements of $G = (V,E)$ (with $N_1 = |V_1|$ and $N_2 = |V_2|$) and dictates which ones of the sites
of $\Lambda$ act as Josephson coupling between
the pair of superconducting phases
$\phi_{i,j}$
and
$\varphi_{i,j}$.
Each site from the medial
lattice ${\Lambda_{\mathrm{medial}}}$ associated to $\Lambda$
hosts the Josephson coupling $J_{\mathrm{w}}\geq0$,
where the sites of
${\Lambda_{\mathrm{medial}}}$
are  the midpoints of a nearest-neighbor bonds
from the lattice $\Lambda$
(see Fig.\ \ref{Fig:toy model for coupled wires}(a)).
For the horizontal nearest-bond
with the end points $(i,j),(i,j+1)\in\Lambda$,
we postulate the local Josephson energy
\begin{equation}
J_{\mathrm{w}}
\left[
1
-
\cos
\left(
\varphi_{i,j}^{\,}
-
\varphi_{i,j+1}^{\,}
\right)
\right].
\label{eq:def Josephson array on square lattice d}
\end{equation}
For the vertical nearest-bond
with the end points $(i,j),(i+1,j)\in\Lambda$,
we postulate the local Josephson energy
\begin{equation}
J_{\mathrm{w}}
\left[
1
-
\cos
\left(
\phi_{i,j}^{\,}
-
\phi_{i+1,j}^{\,}
\right)
\right].
\label{eq:def Josephson array on square lattice c}
\end{equation}
In the limit $J_{\mathrm{w}}/J\uparrow\infty$,
each horizontal (vertical) wire is assigned the rigid superconducting phase
$\varphi_{i}$ ($\phi_{j}$) with $i=1,\ldots,N_1$ ($j=1,\ldots,N_2$).
For finite values of $J_{\mathrm{w}}/J$, the superconducting
phases of the horizontal (vertical) wires are allowed to fluctuate
on a length scale of the order of $\mathfrak{a}$.
We may then assign to the square lattice $\Lambda$,
whose topology we choose to be that of the torus, the energy
\begin{equation}
\begin{split}
&
E_{\Lambda,G}(J,J_{\mathrm{w}^{\,}}^{\,}):=
\\
&
\quad
J
\sum_{i=1}^{N_1}
\sum_{j=1}^{N_2}
b_{i,j}^{\,}
\left[
1
-
\cos
\left(
\phi_{i,j}^{\,}
-
\varphi_{i,j}^{\,}
\right)
\right]
\\
&
\quad
+
J_{\mathrm{w}}^{\,}
\sum_{i=1}^{N_1}
\sum_{j=1}^{N_2}
\left[
1
-
\cos
\left(
\phi_{i,j}^{\,}
-
\phi_{i+1,j}^{\,}
\right)
\right]
\\
&
\quad
+
J_{\mathrm{w}}^{\,}
\sum_{i=1}^{N_1}
\sum_{j=1}^{N_2}
\left[
1
-
\cos
\left(
\varphi_{i,j}^{\,}
-
\varphi_{i,j+1}^{\,}
\right)
\right].
\end{split}
\label{eq:def Josephson array on square lattice e}
\end{equation}
\end{subequations}  
}

{
The energy (\ref{eq:def Josephson array on square lattice e})
admits an alternative interpretation if we assign to each site
from $\Lambda$ a pair of two-components unit vectors,
one unit vector colored in blue
and the other unit vector colored in orange.
The coupling $J$ is then to be interpreted
as a ferromagnetic coupling between blue and orange unit vectors on
the same site of $\Lambda$, while $J_{\mathrm{w}}$ is then 
to be interpreted as a nearest-neighbor ferromagnetic coupling
between unit vectors of the same color.
}

Setting a large $J\gg J_{\mathrm{w}}$ corresponds to a number no larger
than $N_1\,N_2$ (this number is fixed by the biadjacency matrix $B$ of
the graph $G$) of
strongly coupled superconducting islands shaped like a cross
[colored in gray in Fig.\ \ref{Fig:toy model for coupled wires}(b)].
The effect of large $J\gg J_{\mathrm{w}}$ is to
energetically enforce the condition
$\phi_{i,j}=\varphi_{i,j}$ for every $(i,j)\in\Lambda$
such that $b_{i,j}=1$, effectively merging the blue and
orange degrees of freedom into a single gray degree of freedom.
Equivalently, each cross in
Fig.\ \ref{Fig:toy model for coupled wires}(b) can be thought of as a
single classical $XY$ spin that couples with neighboring sites through
$J_{\mathrm{w}}\geq0$. In the strict limit $J_{\mathrm{w}}/J=0$,
the ferromagnetic alignment for each decoupled cross is arbitrary from
cross to cross.  Thus, there is no ferromagnetic long-range order at
zero temperature, $T=0$, in the thermodynamic limit
${N_1,N_2}\uparrow\infty$ for $J_{\mathrm{w}}^{\,}/J=0$.  Moreover,
thermal fluctuations destroy the alignment of the two spins making up
a cross for every cross.  The point
$J_{\mathrm{w}}^{\,}/J=k_{\mathrm{B}}^{\,}\,T/J=0$ in the
two-dimensional coupling space defined by the positive values of the
dimensionless coupling $J_{\mathrm{w}}^{\,}/J$ and dimensionless
temperature $k_{\mathrm{B}}^{\,}\,T/J$ is a singular one, as it is
assumed throughout this paper that the effect of the biadjacency
matrix $B$ on Fig.\ \ref{Fig:toy model for coupled wires}(b) is to
dilute the square lattice in such a way that there exists at least one
path of nearest-neighbor bonds that passes through all the diluted
sites [this is a weaker assumption than demanding that two percolating
paths of nearest-neighbor bonds exist along the two nontrivial
cycles of the toroidal topology of Fig.\
\ref{Fig:toy model for coupled wires}(b)].
If so, the ground state is ferromagnetic
(superconducting) for any nonvanishing but finite value of
$J_{\mathrm{w}}^{\,}/J$.

Conversely, setting a $J_{\mathrm{w}}\gg J$ yields $N_1+N_2$
weakly coupled linear chains. In this regime, each chain models
a superconducting wire composed of segments of length
$\mathfrak{a}$. These chains are coupled at every vertex $(i,j)\in\Lambda$
where $b_{i,j}=1$ via the Josephson coupling $J\geq 0$, as is
illustrated in Fig.~\ref{Fig:toy model for coupled wires}(c).
This is the limit of interest in this paper for which the
stiffness is sufficiently large so as to ensure that each wire is
characterized by a single phase.
In the strict limit of decoupled wires $J/J_{\mathrm{w}}=0$ and at
zero temperature $T=0$, each chain is ferromagnetically long-range
ordered, but the spontaneous breaking of the $XY$ global symmetry is
arbitrary from chain to chain.  Thus, there is no long-range order at
zero temperature in the thermodynamic limit ${N_1,N_2}\uparrow\infty$
for $J=0$.  Moreover, thermal fluctuations destroy the ferromagnetism
of each chain in the thermodynamic limit ${N_1,N_2}\uparrow\infty$.
The point
$J/J_{\mathrm{w}}^{\,}=k_{\mathrm{B}}^{\,}\,T/J_{\mathrm{w}}^{\,}=0$
in the two-dimensional coupling space defined by the positive values
of the dimensionless coupling $J/J_{\mathrm{w}}^{\,}$ and
dimensionless temperature $k_{\mathrm{B}}^{\,}\,T/J_{\mathrm{w}}^{\,}$
is a singular one, as it is assumed throughout this paper that the
biadjacency matrix $B$ specifies a connected network of couplings
among the one-dimensional chains.  If so, the ground state is
superconducting (or, equivalently, ferromagnetic) for any
nonvanishing but finite value of $J/J_{\mathrm{w}}^{\,}$.

 By design, we have the identity
\begin{equation}
\lim_{\frac{J_{\mathrm{w}}^{\,}}{J}\uparrow\infty}
E_{\Lambda,G}(J,J_{\mathrm{w}^{\,}}^{\,})=
E_{G}(J)
\label{eq:def limit phase rigidity along wires}
\end{equation}
with $E_{G}(J)$ defined in Eq.\
(\ref{eq:def array sc wires Josephson coupled f}),
for the limit $J_{\mathrm{w}}^{\,}\uparrow\infty$
holding $J>0$ fixed imposes the conditions
for one-dimensional (wire) phase rigidity
\begin{subequations}
\label{eq:conditions for phase rigidity along wires}
\begin{equation}
\phi_{1,j}^{\,}=
\cdots=
\phi_{N_1,j}^{\,},
\qquad
j=1,\ldots,N_2,
\label{eq:conditions for phase rigidity along wires a}
\end{equation}
and
\begin{equation}
\varphi_{i,1}^{\,}=
\cdots=  
\varphi_{i,N_2}^{\,},
\qquad
i=1,\ldots,N_1.
\label{eq:conditions for phase rigidity along wires b}
\end{equation}
\end{subequations}

We define the partition function
\begin{subequations}
\label{eq:def partition function square lattice}
\begin{equation}
\begin{split}
&
Z_{\Lambda,G}^{\,}(K,K_{\mathrm{w}}^{\,})\equiv
\\
&\quad
\mathrm{Tr}\, e^{-\beta\,H_{\Lambda,G}^{\,}(J,J_{\mathrm{w}}^{\,})}:=
\\
&\quad\quad
\left[
\prod\limits_{i,j}
\int\limits_{0}^{2\pi}\!\!
\mathrm{d}\phi_{i,j}^{\,}
\right]
\left[
\prod\limits_{i,j}
\int\limits_{0}^{2\pi}
\mathrm{d}\varphi_{i,j}^{\,}
\right]\,
e^{-\beta H_{\Lambda,G}^{\,}(J,J_{\mathrm{w}}^{\,})}
\end{split}
\label{eq:def partition function square lattice a}
\end{equation}
at temperature $T$, where
\begin{equation}
K:=
\beta\,J,
\qquad
K_{\mathrm{w}}^{\,}:=
\beta\,J_{\mathrm{w}}^{\,},
\qquad
\beta:=\frac{1}{k_{\mathrm{B}}^{\,}\,T},
\end{equation}
together with the statistical average
\begin{equation}
\langle
O
\rangle:=
\frac{
\mathrm{Tr}\, e^{-\beta\,H_{\Lambda,G}^{\,}(J,J_{\mathrm{w}}^{\,})}\, O
     }
     {
\mathrm{Tr}\, e^{-\beta\,H_{\Lambda,G}^{\,}(J,J_{\mathrm{w}}^{\,})}\,\hphantom{O}
     }.
\end{equation}
\end{subequations}

Focusing on the limit $J/J_{\mathrm{w}}^{\,}\downarrow0$,
we employ the upper bound (valid for any value of $J/J_{\mathrm{w}}^{\,}$)
\begin{subequations}
\label{eq:upper bound}
\begin{equation}
\hbox{\small
$
\left\langle
\left(
\phi_{i,j}^{\,}
-
\phi_{i',j}^{\,}
\right)^{2}
\right\rangle\leq
\lim\limits_{J/J_{\mathrm{w}}\downarrow0}
\left\langle
\left(
\phi_{i,j}^{\,}
-
\phi_{i',j}^{\,}
\right)^{2}
\right\rangle
$
     }
\label{eq:upper bound a}
\end{equation}
for the fluctuations of the phase
between the sites %
\begin{equation}
i,i'=
1,\ldots,N_1
\label{eq:upper bound b}
\end{equation}
of the row $j$ with
\begin{equation}
j=1,\ldots,N_2.
\label{eq:upper bound c}
\end{equation}
\end{subequations}
The upper bound (\ref{eq:upper bound}) follows from the fact that the
energy (\ref{eq:def Josephson array on square lattice e}) with $J>0$
is greater than or equal to the energy with $J=0$ for all phase
values. We show in Appendix \ref{appssubsec:Proof of the upper bound A}
that this energy hierarchy is a sufficient condition for the upper bound
(\ref{eq:upper bound}).

In order to assume phase rigidity, we seek a condition on the
dimensionless stiffness of a wire $K_{\mathrm{w}}^{\,}$ and on $N_1$
such that
\begin{equation}
\lim_{J/J_{\mathrm{w}}\downarrow0}
\left\langle
\left(
\phi_{i,j}^{\,}
-
\phi_{i',j}^{\,}
\right)^{2}
\right\rangle\ll\pi_{\,}^{2}
\label{eq:condition for graph vertex is rigid phase of sc wire}
\end{equation}
for any pair $i,i'=1,\ldots,N_1$.
Thus, the assumption that the
superconducting wires coherently behave as single degrees of freedom
(Fig.\ \ref{fig:dot_to_wire}) 
is met if condition
(\ref{eq:condition for graph vertex is rigid phase of sc wire})
holds. 
We show in Appendix \ref{appssubsec:Proof of the upper bound B}
that this condition reads
\begin{equation}
\frac{2N_1}{K_{\mathrm{w}}^{\,}}\leq1
\ \Longleftrightarrow\
N_1\leq
\frac{K_{\mathrm{w}}^{\,}}{2}.
\label{eq:condition on N for small phase fluctuations}
\end{equation}

In contrast,
focusing on the limit $J_{\mathrm{w}}^{\,}/J\downarrow0$,
we show in Appendix
\ref{appssubsec:Proof of the upper bound C}
that the condition 
\begin{equation}
\lim_{J_{\mathrm{w}}^{\,}/J\downarrow0}
\left\langle
\left(
\phi_{i,j}^{\,}
-
\phi_{i',j}^{\,}
\right)^{2}
\right\rangle\ll\pi_{\,}^{2}
\label{eq:condition for graph vertex is rigid phase of sc wire 2}
\end{equation}
for any pair {$i,i'=1,\ldots,N_1$
along the row $j=1,\ldots,N_2$}
yields the bound \begin{equation}
\frac{\mathrm{const}}{K_{\mathrm{w}^{\,}}}\,\ln N_1\leq 1
\ \Longleftrightarrow\
N_1\leq e^{\frac{K_{\mathrm{w}^{\,}}}{\mathrm{const}}}.
\label{eq:condition for rigid phases of crosses}
\end{equation}
The number $\mathrm{const}$ is here of order one.
The bound
(\ref{eq:condition on N for small phase fluctuations})
is much stronger than the bound
(\ref{eq:condition for rigid phases of crosses})
when
$K_{\mathrm{w}^{\,}}\gg1$,
as Gaussian fluctuations increase with decreasing
dimensionality of Euclidean space. 

The same conditions hold for
$\varphi_i$ {with $i=1,\ldots,N_1$},
which implies that the phase condition
(\ref{eq:condition for graph vertex is rigid phase of sc wire})
is valid whenever  
\begin{subequations}
\begin{equation}
N_{\mathrm{max}}\equiv
\max(N_1,N_2) \leq \frac{K_\mathrm{w}}{2}.
\end{equation}
Hence, we identify
\begin{equation}
N_{\star}(T)\equiv
\frac{J_{\mathrm{w}}^{\,}}{2\,k_{\mathrm{B}}^{\,}\,T}
\label{eq:theory answer for Nstar}  
\end{equation}
as the dimensionless number
that enters on the right-hand side of
Eq.\ (\ref{eq:def Lstar and Nstar}).
Increasing
\begin{equation}
L_{\star}(T):=
N_{\star}(T)\,\mathfrak{a}
\label{eq:theory answer for Lstar}  
\end{equation}
\end{subequations}
is achieved by either increasing $J_{\mathrm{w}}$ or decreasing $T$,
whereby $L_{\star}\uparrow\infty$ if
$\frac{J_{\mathrm{w}}^{\,}}{k_{\mathrm{B}}^{\,}\,T}\uparrow\infty$.

At zero temperature, there are two mechanisms that prevent
assigning a single rigid phase to each wire
from the square array of superconducting wires.

First,
as the length $L$ and cross sectional area of a superconducting wire $u$
is decreased, its charging energy $E^{(\mathrm{C})}$ increases.
Increasing $E^{(\mathrm{C})}$ increases the quantum fluctuations of the
superconducting phases owing to the Heisenberg uncertainty relations
\begin{equation}
\Delta n_{u}\,\Delta\phi_{u}\sim1
\end{equation}
between the charge uncertainty $\Delta n_{u}$ 
and the superconducting phase uncertainty $\Delta\phi_{u}$
of wire $u$. As a quantum superconducting wire at zero temperature
is equivalent to a two-dimensional classical $XY$ model
at the effective temperature $\hbar\,\omega_{\mathrm{jp}}/k_{\mathrm{B}}$
(see Appendix \ref{appssubsec:Proof of the upper bound C}),
we can use the bound (\ref{eq:condition for rigid phases of crosses})
with $k_{\mathrm{B}}\,T$ substituted by the geometric mean
of the Josephson coupling $J_{\mathrm{w}}$ and charging energies $E^{(\mathrm{C})}$,
\begin{subequations}
\begin{align}
&
k_{\mathrm{B}}\,T\to\hbar\,\omega_{\mathrm{jp}}\equiv
\sqrt{J_{\mathrm{w}}\,E^{(\mathrm{C})}},
\\
&
K_{\mathrm{w}}\to
\frac{J_{\mathrm{w}}}{\hbar\,\omega_{\mathrm{jp}}}=
\sqrt{\frac{J_{\mathrm{w}}}{E^{(\mathrm{C})}}}.
\end{align}
This gives the estimate
\begin{equation}
L_{\mathrm{qu}\,\star}=
e^{
\frac{1}{\mathrm{const}}
\sqrt{\frac{J_{\mathrm{w}}}{E^{(\mathrm{C})}}}
  }\,
\mathfrak{a}
\end{equation}
\end{subequations}
for the length of a wire above which
quantum fluctuations destroy the phase stiffness of the
superconducting quantum wire.

Second, at zero temperature but in the presence of a uniform magnetic field
of magnitude $B$ that is perpendicular to the array of wires,
the superconducting phase along a wire of length $L$ must twist.
This twist can be neglected if $L$ is smaller than
the characteristic length
\begin{equation}
L_{B\,\star}:=
\sqrt{\frac{h\,c}{2\,e\,B}},
\label{eq:def-LofB}
\end{equation}
where $L_{B\,\star}$ is the length of the side of a square
threaded by the superconducting flux quantum
\begin{equation}
\Phi_0:=\frac{h\,c}{2\,e}.
\label{eq:def-sc-quantum-flux}
\end{equation}

We stress that the inequality
(\ref{eq:condition on N for small phase fluctuations})
is valid for the massless case, i.e., the compact
fields $\phi_{i,j}$
and
$\varphi_{i,j}$ do not have any mass. In
practice, we can add a large mass term as discussed in
(\ref{eq:discritized massive scalar}) for both fields and increase the
maximum number of wires $N_{\star}$ arbitrarily.

\begin{figure*}[t]
\centering
\includegraphics[scale=0.18]{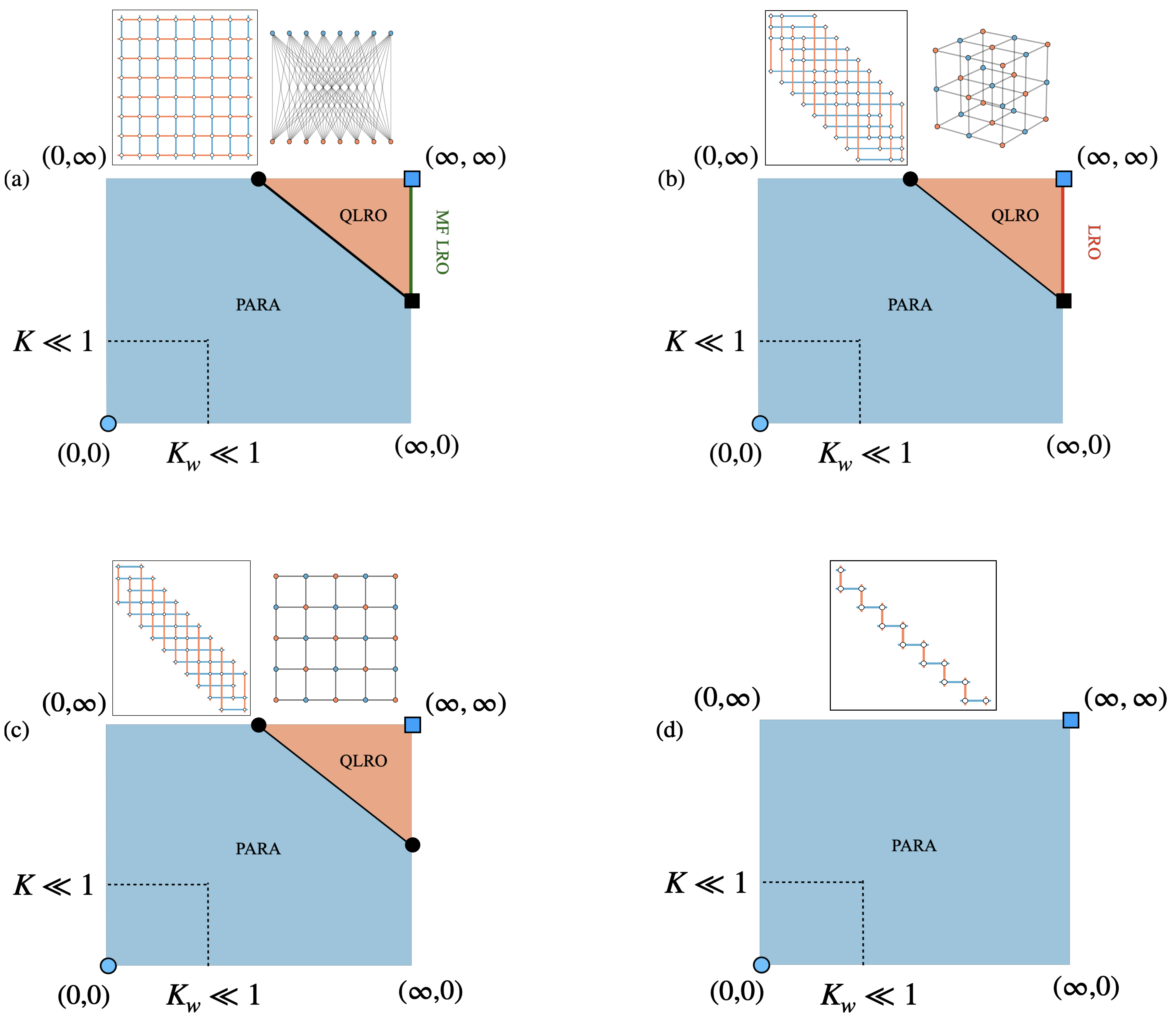}
\caption{
Phase diagram at nonvanishing temperature $T>0$
for the partition function
(\ref{eq:def partition function square lattice a})
(a)
when the adjacency matrix is given by Eq.\
{
(\ref{eq:mean field adjacency matrix a})},
i.e., it corresponds to the tessellation of infinite-dimensional
Euclidean space;
(b)
when the adjacency matrix corresponds to the tessellation of
three-dimensional Euclidean space;
(c)
when the adjacency matrix corresponds to the tessellation of
two-dimensional Euclidean space;
(d)
when the adjacency matrix corresponds to the tessellation of
one-dimensional Euclidean space.
The blue color is reserved for a disordered (paramagnetic) phase.
The color red is reserved for a long-range-ordered
 phase.
The orange color is reserved for a quasi-long-range-ordered phase.
A black disk symbolizes
a Berezinskii-Kosterlitz-Thouless (BKT) critical point.
A black square symbolizes a critical point.
The infinite (zero) temperature fixed point is symbolized by a
blue disc with a black perimeter (blue square with a black perimeter).}
\label{Fig:phase diagrams}
\end{figure*}

\subsection{Geometric crossover}
\label{subsec:Geometric crossover}

We shall assume that $J>0$
{and that $N_1=N_2\equiv N$}
for the theory defined
by the partition function (\ref{eq:def partition function square lattice a}).
Hamiltonian
(\ref{eq:def Josephson array on square lattice e})
then depends on the biadjacency matrix $B$ and on the dimensionless coupling
$J_{\mathrm{w}}^{\,}/J\geq0$
if all energies are measured in units of
$J$.

To provide a more concrete discussion, we will focus on the complete
bipartite graph $G_{\mathrm{MF}}$, extensively studied in
\cite{Vinokur87,feigel1987random,Sohn93a,Sohn93b,chandra1995possible,chandra1996glass, Shea97}, that is defined
by the biadjacency matrix
\begin{equation}
b_{i,j}^{\,}=1,
\qquad
i,j=1,\ldots,N.
\label{eq:mean field adjacency matrix a}
\end{equation}
Given
the Hamiltonian
(\ref{eq:def Josephson array on square lattice e})
with the partition  function
(\ref{eq:def partition function square lattice a}), we are going to show that
(i) the limit
$J_{\mathrm{w}}^{\,}/J\uparrow\infty$
is the mean-field theory, i.e., the ferromagnetic $XY$ model
when the dimension of space is effectively infinite, on the one hand,
(ii) while the limit
$J_{\mathrm{w}}^{\,}/J\downarrow0$
is the nearest-neighbor ferromagnetic $XY$ model
on the square lattice,
on the other hand.

In the limit (\ref{eq:def limit phase rigidity along wires}),
{the} simple bipartite graph $G_{\mathrm{MF}}^{\,}$
whose biadjacency matrix $B = (b_{i,j})$ in 
(\ref{eq:mean field adjacency matrix a})
delivers the energy term
\begin{subequations}
\label{eq:mean field adjacency matrix}
\begin{align}
E_{G_{\mathrm{MF}}^{\,}}(J)\equiv&\,
\lim_{\frac{J_{\mathrm{w}}^{\,}}{J}\uparrow\infty}
E_{\Lambda,G_{\mathrm{MF}}^{\,}}(J,J_{\mathrm{w}^{\,}}^{\,})
\nonumber\\
=&\,
J
\sum_{i=1}^{N}
\sum_{j=1}^{N}
\left[
1
-
\cos\left(\bar{\phi}_{j}-\bar{\varphi}_{i}\right)
\right],
\label{eq:mean field adjacency matrix b}
\end{align}
where
\begin{align}
\bar{\phi}_{j}:=
\frac{1}{N}
\sum_{i=1}^{N}
\phi_{i,j},
\label{eq:mean field adjacency matrix c}\\
\bar{\varphi}_{i}:=
\frac{1}{N}
\sum_{j=1}^{N}
\varphi_{i,j}.
\label{eq:mean field adjacency matrix d}
\end{align}
\end{subequations}
This is the same energy as that of a square array
with the lattice spacing  $\mathfrak{a}$
of superconducting dots
with the infinite-range Josephson coupling $J>0$ or, equivalently,
a mean-field ferromagnetic $XY$ model on the square lattice
with the lattice spacing $\mathfrak{a}$ with the exchange coupling $J>0$. 

In contrast, for any nonvanishing but finite value of $J_{\mathrm{w}}/J$,
Hamiltonian (\ref{eq:def Josephson array on square lattice e})
with the mean-field choice
{
(\ref{eq:mean field adjacency matrix a})}
for the biadjacency matrix
behaves like a two-dimensional $XY$ model with
short-range ferromagnetic exchange couplings
in the thermodynamic limit $N\uparrow\infty$. This claim
is best understood from perturbing the singular point
$J_{\mathrm{w}}/J=0$
for the graph $G_{\mathrm{MF}}^{\,}$, 
encoded as it is by Fig.\ \ref{Fig:toy model for coupled wires}(b),
with $J_{\mathrm{w}}/J\ll1$.
When $J_{\mathrm{w}}/J=0$ and $k_{\mathrm{B}}^{\,}T\ll J$,
we can safely impose the conditions
\begin{equation}
\phi_{i,j}^{\,}=
\varphi_{i,j}^{\,},
\qquad
i,j=1,\ldots,N.
\end{equation}
The effect of the perturbation
$J_{\mathrm{w}}/J\ll1$ on the singular point
$J_{\mathrm{w}}/J=0$ for the graph $G_{\mathrm{MF}}^{\,}$
and for temperatures $k_{\mathrm{B}}^{\,}T\ll J$
in Fig.\ \ref{Fig:toy model for coupled wires}(b)
is to induce an effective ferromagnetic nearest-neighbor coupling
between the effective $XY$ spin assigned to any cross
in Fig.\ \ref{Fig:toy model for coupled wires}(b).
In this way, the phase rigidity of an isolated cross
is extended to all crosses at zero temperature
by the weak coupling between the crosses.
At any nonvanishing temperature, thermal fluctuations in the form
of spin waves, if we adopt the spin interpretation, downgrade the
ferromagnetic long-range order to quasi-long-range order.
Vortices destroy the quasi-long-range ordered phase above the
Berezinskii-Kosterlitz-Thouless (BKT) transition temperature.

The phase diagram corresponding to the graph with the mean-field
adjacency matrix
{
(\ref{eq:mean field adjacency matrix a})}
is thus controlled by two fixed points.
First, there is the zero-temperature fixed point
\begin{equation}
K=K_{\mathrm{w}}^{\,}=\infty
\label{eq:fixed point K=Kw=infty}
\end{equation}
that corresponds to the long-range ordered phase that breaks
spontaneously the global $\mathrm{U}(1)$ symmetry
shown in Fig.\ \ref{Fig:phase diagrams}(a)
[superconducting or ferromagnetic depending on the interpretation of
the $\mathrm{U}(1)$ degrees of freedom in the partition function
(\ref{eq:def partition function square lattice a})].
Second, there is the infinite-temperature fixed point
\begin{equation}
K=K_{\mathrm{w}}^{\,}=0
\label{eq:fixed point K=Kw=0}
\end{equation}
that realizes a disordered phase with vanishing correlation length.
A necessary condition for the long-range-ordered phase
or a quasi-long-range downgraded version of it to extend beyond
(\ref{eq:fixed point K=Kw=infty})
is that 
\begin{equation}
K,K_{\mathrm{w}}^{\,}\gtrsim1,
\label{eq:fixed point K=Kw gg1}
\end{equation}
i.e., $k_{\mathrm{B}}^{\,}\,T$ is smaller than both $J$ and $J_{\mathrm{w}}^{\,}$.
Along the right vertical boundary of
Fig.\ \ref{Fig:phase diagrams}(a),
the critical point represented by a black square
is the mean-field transition from the paramagnetic to
the ferromagnetic $XY$ long-range ordered phase upon increasing
$K$.
Along the top horizontal boundary of
Fig.\ \ref{Fig:phase diagrams}(a),
the critical point represented by a black disk
is the BKT transition temperature that separates
the paramagnetic from the ferromagnetic $XY$ quasi-long-range ordered phase
upon increasing $K_{\mathrm{w}}^{\,}$.
These two critical points are the end points of a phase
boundary in Fig.\ \ref{Fig:phase diagrams}(a) that separates
the paramagnetic phase at high temperatures from the quasi-long-range
ordered phases at low temperatures.  

When the number $2N$ of superconducting wires is finite, the sharp
phase transitions of Fig.\ \ref{Fig:phase diagrams}(a) are replaced by
crossovers.  When
\begin{subequations}
\label{eq:def partition function Graph}
\begin{equation}
K\lesssim K_{\mathrm{w}}^{\,}
\label{eq:def partition function Graph a}
\end{equation}
and
\begin{equation}
N\leq K_{\mathrm{w}}^{\,},
\label{eq:def partition function Graph b}
\end{equation}
we may approximate the partition function
(\ref{eq:def partition function square lattice a})
by the partition function
\begin{align}
Z_{G}^{\,}(K):=&\,
\left[
\prod\limits_{j=1}^{N}
\int\limits_{0}^{2\pi}\!\!
\mathrm{d}\bar{\phi}_{j}^{\,}
\right] \left[
\prod\limits_{i=1}^{N}
\int\limits_{0}^{2\pi}\!\!
\mathrm{d}\bar{\varphi}_{i}^{\,}
\right]
e^{-\beta H_{G}^{\,}(J)},
\label{eq:def partition function Graph c}
\\
H_{G}^{\,}(J):=&\,
J
\sum_{i,j=1}^{N}
b_{i,j}^{\,}
\left[
1
-
\cos\left(\bar{\varphi}_{i}-\bar{\phi}_{j}\right)
\right]
\label{eq:def partition function Graph d}
\end{align}
\end{subequations}
with the {mean-field} adjacency matrix
{
(\ref{eq:mean field adjacency matrix a})}.
In effect, when conditions
(\ref{eq:def partition function Graph a})
and
(\ref{eq:def partition function Graph b})
hold, observables behave as those of the
unfrustrated $XY$ model on the space tessellated
by the graph $G_{\mathrm{MF}}^{\,}$.
This is the regime of validity of Sec.\
\ref{sec: graphs/geometries and JJ arrays}.
As a function of increasing $N\gtrsim K_{\mathrm{w}}^{\,}$,
observables undergo a crossover (here merely dimensional)
to those of the unfrustrated $XY$ model
in two-dimensional Euclidean space that rules the regime
$K\gg K_{\mathrm{w}}^{\,}$.

Replacing the mean-field biadjacency matrix
{
(\ref{eq:mean field adjacency matrix a})}
by a sparser matrix changes (i) the geometry and
topology of space by reducing its dimensionality and modifying its
local curvature in the limit $J_{\mathrm{w}}^{\,}/J\uparrow\infty$, on
the one hand, (ii) while it removes (dilutes) sites from the square
lattice of crosses in Fig.\ \ref{Fig:toy model for coupled wires}(b)
on which the effective nearest-neighbor ferromagnetic $XY$ model is
defined in the limit $J_{\mathrm{w}}^{\,}/J\downarrow0$, on the other
hand.  Because we are assuming that the biadjacency matrix is not too
sparse, ferromagnetic long-range order holds for any nonvanishing and
finite values of $J_{\mathrm{w}}^{\,}/J$ at zero temperature. However,
if this long-range order originates from a one-dimensional
connectivity instead of a two-dimensional connectivity, thermal
fluctuations downgrade the ferromagnetic long-range order to
paramagnetism.  Different choices for the biadjacency matrix modify
the phase diagram \ref{Fig:phase diagrams}(a) to the phase diagrams
\ref{Fig:phase diagrams}(b)-\ref{Fig:phase diagrams}(d).

Finally, if we consider a non-bipartite simple graph $G$ and adjacency
matrix $A$, the framework described in Sec.~\ref{sec:
  graphs/geometries and JJ arrays} can be directly applied to obtain a
wire realization of $G$. The analysis of this case carries over with
two modifications: the biadjacency matrix is replaced by the adjacency
matrix, and the two-field description in terms of $\phi$ and $\varphi$
reduces to a single field $\phi$.

%%%%%%%%%%%%%%%%%%%%%%%%%%%%%%%%%%%%%%%%%%%%%%%%%%%%%%%%%%%%%%%%%%%%%%%%%%%%%%%
\section{Application: Probing $\mathrm{AdS/CFT}$ correspondence
in hyperbolic spaces}
\label{sec:The AdS/CFT correspondence for Euclidean hyperbolic spaces}

A route to studying field theories on negatively curved manifolds is
provided by discretization on regular hyperbolic lattices
\cite{Brower21,Brower22,Asaduzzaman20}. In this context, theoretical
efforts have explored implications for band theory
\cite{Maciejko2021, Maciejko2022, Boettcher2022,Urwyler2022,Zhang2023,Shankar2023, Yuan2024,Qin2024boundary,Sun2024, Chen2025,Rajbongshi2025topological,Djordjevic2025symmetry,Leong2025amplified,Leong2025strained,Leong2025global,Bashmakov2025superconductivity, Pavliuk2025superconductivity},
Anderson localization,
\cite{Chen2024Anderson,Li2024anderson,Altland2026}, real time evolution~\cite{Asaduzzaman2024}, 
phase transitions
\cite{Callan90,Series90, Wu1996ising,Rietman1992ising, dAuriac2001spin,Doyon2004ising, Madras05,Shima2006geometric, Shima2006dynamic, Ueda2007corner,Baek2007phase,Fujita2008disordered, Gendiar2008,baek2009phase, Baek2009curvature,Sakaniwa09, Iharagi2010phase, Baek2012,Mnasri15, Lee2016boundary,Paulos2016conformal, Wu2000ising, Breuckmann2020critical, Okunishi2024holographic,Wang2025emergence,Gotz2024hubbard, Sela2025failure,Petermann2026inherent,samlodia2026quantum}
and topological states
\cite{Danivska2016analysis,Breuckmann2017hyperbolic, Yan2019hyperbolic,yan2019hyperbolicII,Yan2020geodesic,Ebisu2022z,Zhang2022,Lenggenhager2025hyperbolic,Shokeeb2025hyperbolic},
alongside proposals for table-top realizations of the AdS/CFT %
correspondence \cite{Dey2024simulating}. Such geometries have
been implemented on circuit boards
\cite{Kollar2019,Lenggenhager22,Zhang2022,Chen2023hyperbolic,Zhang2023,
Yuan2024,lai2026observation,Chen2023ads}, superconducting circuits \cite{xu2025scalable},
networks \cite{Chen2024anomalous},
photonic devices \cite{Huang2024hyperbolic} and mechanical
metamaterials \cite{Patino2024hyperbolic}.

We note that all these are intrinsically planar representations of
hyperbolic spaces on a Poincaré disk. Therefore, as additional radial
generations of the hyperbolic lattice are introduced, the size of the
constituent elements must be progressively reduced, reflecting the
finite planar area available to represent a space of intrinsically
exponential growth that ``does not fit'' in two-dimensional Euclidean
space. This limitation is overcome in the superconducting wire arrays,
which encode the geometry (in any dimension) differently, as discussed
in Sec.~\ref{sec: graphs/geometries and JJ arrays}.

As a concrete application, we construct a superconducting circuit
realization of a scalar field on a hyperbolic lattice that can be
scaled to thousands of sites. Within this setting, we probe the
holographic relation between the bulk mass $m$ and the boundary
scaling dimension $\Delta$ through correlation functions. Numerically,
we show that boundary-boundary correlators computed on the lattice
reproduce the continuum prediction for a massive free scalar field.

Any experimental realization of this scheme will inevitably exhibit
deviations from idealized couplings. Motivated by this, we study the
effects of disorder in the Josephson couplings on the resulting
boundary correlation functions.  In the weak-disorder regime, we find
agreement between
a random-phase approximation  (making use of
replica theory) and numerical simulations, showing that disorder
effectively renormalizes the bare mass and increases the boundary
scaling dimension.  Beyond the perturbative regime, numerical results
indicate that the power-law behavior of boundary-boundary correlators
remains robust against disorder in the couplings, preserving the
boundary conformal structure while continuously deforming the scaling
dimension.

\subsection{Continuum quantum field theory picture}
\label{subsec:Continuum quantum field theory picture}

We consider {A}nti-de Sitter (AdS)${}_{d+1}$ in $(d+1)$-dimensional
spacetime, corresponding to a negative cosmological constant.
AdS{${}_{d+1}$}
is a maximally symmetric spacetime
of constant negative curvature, with isometry group
$\mathrm{SO}(d,2)$. It can be realized geometrically as a
$(d+1)$-dimensional hyperboloid embedded in a
$(d+2)$-dimensional space
with two timelike directions. A distinctive feature of AdS is the
presence of a timelike boundary at spatial infinity, which allows for
well-posed boundary value problems and a precise definition of
conserved quantities. Although AdS does not describe the observed
expansion of the universe, it plays a central role in modern
theoretical physics, particularly in the context of the AdS/CFT
correspondence. The curvature scale of AdS is set by the cosmological
constant, with radius inversely proportional to the square root of its
magnitude.

Henceforth, we take the $(d+1)$-dimensional Riemannian manifold to
be a $(d+1)$-dimensional hyperbolic space $\mathbb{H}^{d+1}$, i.e., a
Riemannian space of constant negative curvature obtained by performing
a Wick rotation to imaginary time on a $(d+1)$-dimensional AdS
spacetime. Correlation functions can be probed in this space using the
framework described in Sec.~\ref{sec:Linear response theory for a
  quantum Josephson array}, where a ``flat'' time direction defines
the time evolution of the system.

Our interest in hyperbolic space $\mathbb{H}^{d+1}$ is twofold.
First, hyperbolic spaces are homogeneous, which makes their
tessellations particularly simple to encode on a Josephson array.
Second, the free massive scalar field theory on a $(d+1)$-dimensional
hyperbolic space is the simplest setting for which the holographic
identity
\begin{equation}
Z_{\mathrm{bulk}}[\phi_{\mathrm{bd}}^{\,}]=
\left\langle
e^{
\int\mathrm{d}^{d}\bm{x}\,
\phi_{\mathrm{bd}}^{\,}(\bm{x})\,
\mathcal{O}(\bm{x})
  }
\right\rangle_{\mathrm{CFT}}^{\,}
\label{eq:Witten holography}
\end{equation}
holds. On the left-hand side, $Z_{\mathrm{bulk}}[\phi_{\mathrm{bd}}]$
is the partition function of the free massive scalar field theory on
$\mathbb{H}^{d+1}$ subject to the boundary condition that the field
takes the value $\phi_{\mathrm{bd}}$ on the boundary. On the
right-hand side, the expression refers to a conformal field theory on
the $d$-dimensional boundary, with $\mathcal{O}$ a boundary
operator. Knowledge of the dependence on $\phi_{\mathrm{bd}}$ in the
bulk partition function allows one to compute correlation functions of
$\mathcal{O}$ on the boundary.

Following Witten \cite{Witten98}, we outline a proof of the
holographic identity (\ref{eq:Witten holography}) for a free massive
scalar field on $(d+1)$-dimensional hyperbolic space
$\mathbb{H}^{d+1}$. No gravitational dynamics are involved {as} the metric
of $\mathbb{H}^{d+1}$ is treated as a fixed background. To proceed, we
employ the Poincar\'e half-space coordinates
\begin{subequations}
\begin{equation}
X\equiv(z,\bm{x})
\end{equation}
on $\mathbb{H}^{d+1}$ defined by the squared length
\begin{equation}\label{Hdmetric}
\mathrm{d}s^{2}=
\frac{\mathrm{d}z^{2}+\mathrm{d}\bm{x}^{2}}{z^{2}},
\qquad
z>0,
\qquad
\bm{x}\in\mathbb{R}^{d},
\end{equation}
\end{subequations}
of an infinitesimal line element. 
Here, the conformal boundary is located at $z\to 0$.  After a
conformal compactification, the boundary is isometric to
$\mathbb{R}^{d}$ (or $S^{d}$).

Consider a massive scalar field $\phi$ with Euclidean action
\begin{equation}
S[\phi]=
\frac{1}{2}
\int\limits_{\mathbb{H}^{d+1}}
\mathrm{d}^{d+1}X\,\sqrt{g}\,
\left(
g_{\,}^{\mu\nu}\partial_{\mu}^{\,}\phi\,\partial_{\nu}^{\,}\phi
+
m^{2}\,\phi_{\,}^{2}
\right),
\label{eq:BulkAction}
\end{equation}
where $g^{\mu\nu}$ is the Poincar\'e metric on $\mathbb{H}^{d+1}$.
The Euler-Lagrange equation is the Klein-Gordon equation
\begin{equation}
(\Box_{\mathbb{H}^{d+1}}^{\,}-m^{2})\phi=0,
\label{eq:KGE}
\end{equation}
where $\Box_{\mathbb{H}^{d+1}}^{\,}$ denotes the Laplace-Beltrami operator
on $\mathbb{H}^{d+1}$. We fix the boundary condition so that, as $z\to 0$,
\begin{equation}
\phi(z,\bm{x})\sim z^{\,d-\Delta}\,\phi_{\mathrm{bd}}(\bm{x}),
\label{eq:boundarybehavior}
\end{equation}
where $\Delta$ will be determined momentarily.

Near the boundary $z\to 0$, we search for separable solutions of the form
\begin{equation}
\phi(z,\bm{x})\sim z^{\lambda}\,f(\bm{x}).
\end{equation}  
Inserting this ansatz into the $z\to0$ limit of the 
Klein-Gordon equation (\ref{eq:KGE})
yields the indicial relation
\begin{subequations}
\begin{equation}
\lambda\,(\lambda-d)=m^{2},
\end{equation}
whose pair of solutions delivers the two characteristic exponents 
\begin{equation}\label{eq: holograph-dimension}
d-\Delta,
\qquad
\Delta=\frac{d}{2}+\sqrt{\frac{d^{2}}{4}+m^{2}}.
\end{equation}
\end{subequations}
The general near-boundary expansion to the resulting
second-order differential equation is thus given by
\begin{equation}
\phi(z,\bm{x})
\sim 
z^{d-\Delta}\,
\phi_{\mathrm{bd}}(\bm{x})
+
z^{\Delta}\,\varphi(\bm{x})
+\cdots.
\label{eq:AsymptoticExpansion}
\end{equation}
The first term on the right-hand side is the leading one as a power expansion
of $z$ and it is not normalizable because
the pole of $\sqrt{g}\,z^{2(d-\Delta)}=z^{2(d-\Delta)-d}$ at the boundary $z=0$
is not integrable at the origin.
This divergence is what forces the mode $\phi_{\mathrm{bd}}(\bm{x})$
to be fixed as external boundary data,
it cannot be part of the physical Hilbert space of bulk excitations.
The second term
$\varphi(\bm{x})$ on the right-hand side is the subleading one as
a power expansion of $z$ and it is normalizable because
the pole of $\sqrt{g}\,z^{2\Delta}=z^{2\Delta-d}$ at the boundary $z=0$
is integrable at the origin.

The solution of Eq.\ (\ref{eq:KGE}) with boundary value
$\phi_{\mathrm{bd}}^{\,}$ is given by
\begin{subequations}
\label{eq:BulkToBoundaryRepresentation}
\begin{equation}
\phi(z,\bm{x})=
\int \mathrm{d}^{d}\bm{y}\,
K_{\Delta}^{\,}(z,\bm{x};\bm{y})\,\phi_{\mathrm{bd}}^{\,}(\bm{y}),
\label{eq:BulkToBoundaryRepresentation a}
\end{equation}
where the bulk-to-boundary propagator is
\begin{equation}
K_{\Delta}^{\,}(z,\bm{x};\bm{y})=
c_{\Delta}^{\,}
\left(
\frac{z}{z^{2}+|\bm{x}-\bm{y}|^{2}}
\right)^{\Delta},
\label{eq:BulkToBoundaryRepresentation b}
\end{equation}
with the normalization constant
\begin{equation}
c_{\Delta}^{\,}=
\frac{\Gamma(\Delta)}
     {\pi^{d/2}\,\Gamma\!\left(\Delta-\frac{d}{2}\right)}.
\label{eq:BulkToBoundaryRepresentation c}
\end{equation}
Indeed, as $z\to 0$, $K_{\Delta}^{\,}$ satisfies
\begin{equation}
K_{\Delta}^{\,}(z,\bm{x};\bm{y})\sim
z^{\,d-\Delta}\,
\delta^{(d)}(\bm{x}-\bm{y}).
\label{eq:BulkToBoundaryRepresentation d}
\end{equation}
\end{subequations}

Using the equation of motion to simplify the bulk action, one obtains
a pure boundary term:
\begin{equation}
S_{\mathrm{on\text{-}shell}}^{\,}=
\frac{1}{2}
\int\limits_{\partial\mathbb{H}^{d+1}}
\mathrm{d}^{d}\bm{x}\,
\sqrt{\gamma}\;\phi\,n^{\mu}\partial_{\mu}\phi.
\label{eq:SonShellBoundary}
\end{equation}
For the Poincar\'e metric,
\begin{equation}
n^{\mu}\partial_{\mu}^{\,}=-z\,\partial_{z}^{\,},
\qquad
\sqrt{\gamma}=z^{-d}.
\end{equation}
Thus, the boundary contribution at $z=\varepsilon$ becomes
\begin{equation}
S_{\mathrm{on\text{-}shell}}=
-
\frac{1}{2}
\int\limits_{z=\varepsilon}
\mathrm{d}^{d}\bm{x}\,
z^{1-d}\,
\phi\,\partial_{z}^{\,}\phi.
\end{equation}
Substituting the expansion~\eqref{eq:AsymptoticExpansion} and extracting
the finite part as $\varepsilon\to 0$ yields
\begin{equation}
S_{\mathrm{on\text{-}shell}}^{\,}=
\frac{2\Delta-d}{2}
\int\mathrm{d}^{d}\bm{x}\,\mathrm{d}^{d}\bm{y}\;
\frac{\phi_{\mathrm{bd}}^{\,}(\bm{x})\,\phi_{\mathrm{bd}}^{\,}(\bm{y})}
{|\bm{x}-\bm{y}|^{2\Delta}}.
\label{eq:SonShellFinal}
\end{equation}
This expression is fixed up to overall normalization by conformal
symmetry. 
The bulk partition function at the classical (saddle-point) level is
therefore
\begin{equation}
Z_{\mathrm{bulk}}^{\,}[\phi_{\mathrm{bd}}^{\,}]=
e^{
-
\frac{2\Delta-d}{2}
\int\mathrm{d}^{d}\bm{x}\,\mathrm{d}^{d}\bm{y}\,
\frac{\phi_{\mathrm{bd}}^{\,}(\bm{x})\phi_{\mathrm{bd}}^{\,}(\bm{y})}
{|\bm{x}-\bm{y}|^{2\Delta}}
  }.
\label{eq:ZbulkExplicit}
\end{equation}

\begin{figure*}[t!]
\centering
\includegraphics[width=0.8\textwidth]{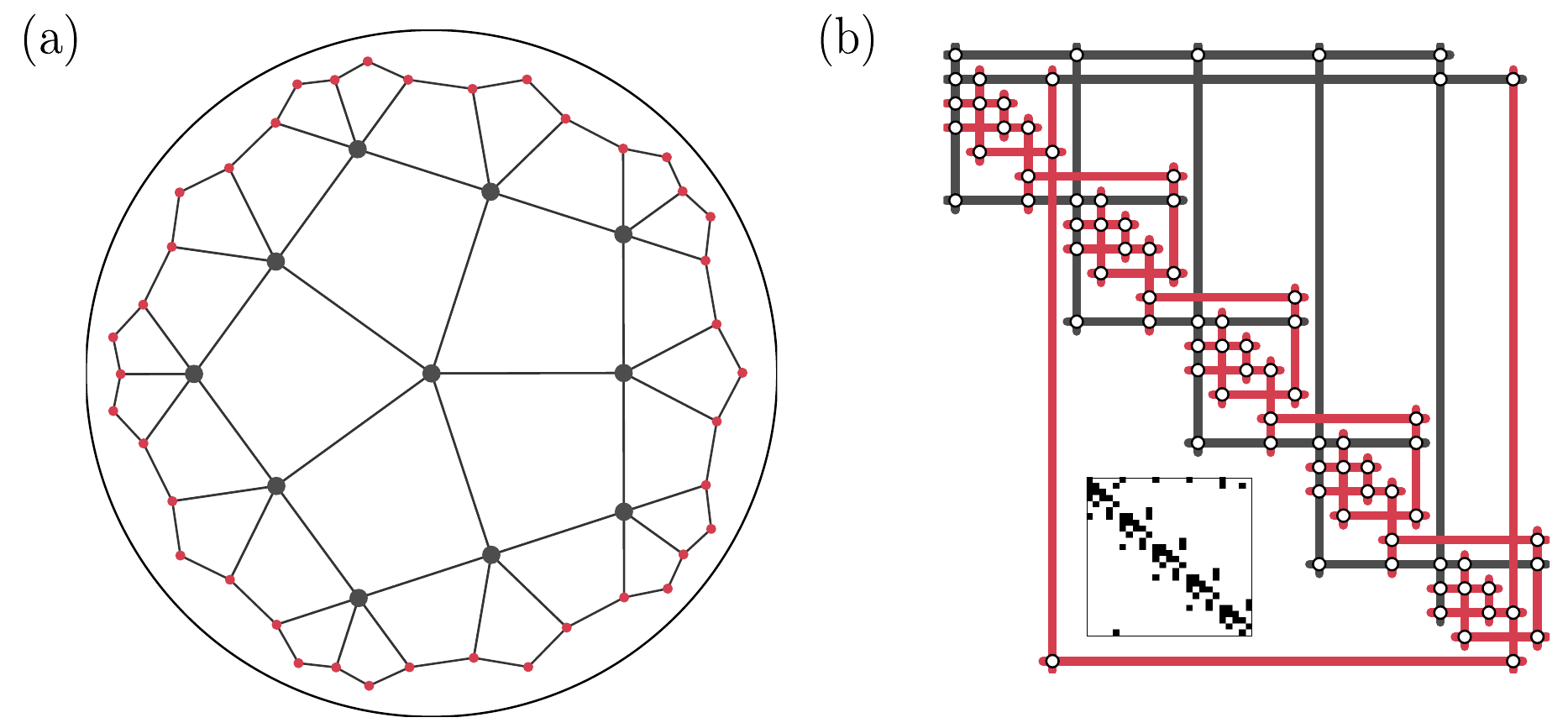}
\caption{
(a) Bipartite graph associated to the $\{4,5\}$ tessellation with three
radial layers. Boundary (bulk) vertices are colored red
(black).
(b) Superconducting wire array construction of the $\{4,5\}$
tessellation. Wires are colored red (black) if they correspond to
boundary (bulk) vertices, and white circles denote Josephson junction
couplings between wires.  The binary biadjacency matrix corresponding to
the tessellation is shown in the bottom left corner, where ones (zeros) values are denoted by black (white) squares.
       }
\label{boundary-tess-boundary-wire}
\end{figure*}

In a conformal field theory deformed by the source $\phi_{\mathrm{bd}}$,
the generating functional is
\begin{equation}
Z_{\mathrm{CFT}}^{\,}[\phi_{\mathrm{bd}}^{\,}]:=
\left\langle
e^{
\int
\mathrm{d}^{d}\bm{x}\,
\phi_{\mathrm{bd}}^{\,}(\bm{x})\,
\mathcal{O}(\bm{x})
  }
\right\rangle_{\mathrm{CFT}}.
\label{eq:Zcft}
\end{equation}
Identifying $Z_{\mathrm{bulk}}^{\,}=Z_{\mathrm{CFT}}^{\,}$
leads to the holographic identity
(\ref{eq:Witten holography}).
Correlation functions of
$\mathcal{O}$ follow by functional differentiation, e.g.,
\begin{equation}\label{eq: scaling-dimension-cft}
\begin{split}
\langle
\mathcal{O}(\bm{x}_{1}^{\,})\,
\mathcal{O}(\bm{x}_{2}^{\,})
\rangle=&\,
\frac{\delta^{2}S_{\mathrm{on\text{-}shell}}}
     {\delta\phi_{\mathrm{bd}}(\bm{x}_{1}^{\,})
       \delta\phi_{\mathrm{bd}}(\bm{x}_{2}^{\,})}
\\
=&\,
\frac{C_{\Delta}^{\,}}{|\bm{x}_{1}^{\,}-\bm{x}_{2}^{\,}|^{2\Delta}},
\end{split}
\end{equation}
as required by conformal invariance.

The holographic
identity (\ref{eq:Witten holography})
is believed to hold in the presence of interactions,
a claim that has been verified whenever interactions
can be treated perturbatively.

\subsection{Discrete formulation and realization with Josephson junction arrays}

To probe holographic properties discussed in
Sec.\
\ref{subsec:Continuum quantum field theory picture},
 we choose to
discretize {$\mathbb{H}^{d+1}$
with $d=1$}
via regular $\{p,q\}$ tessellations,
where {
the integer $p$ denotes the number of sides for the elementary polygon
and the integer $q$ denotes the coordination number of the vertices.}   
These tilings must satisfy
\begin{equation}
\frac{1}{p}+\frac{1}{q}<\frac{1}{2}
\end{equation}
to ensure {a} constant negative curvature,
leading to hyperbolic lattices. Such
structures naturally define a graph $G=(V,E)$ with inherited distance
$d_{u,v}$ that serves as the blueprint for the Josephson junction
arrays discussed in Sec.~\ref{sec: graphs/geometries and JJ
  arrays}. In practice, we generate the adjacency matrix for these
tilings using Coxeter reflection groups
(see Appendix~\ref{Appendix subsec: Tessellations of homogeneous spaces}
for a thorough discussion of tessellations of homogeneous spaces).

As an example, we consider the $\{4,5\}$ hyperbolic
tessellation [Fig.~\ref{boundary-tess-boundary-wire}(a)]. This lattice
is bipartite, enabling a direct implementation using superconducting
arrays composed of intersecting horizontal and vertical wires
[Fig.~\ref{boundary-tess-boundary-wire}(b)]. We partition the
set $V$ {of vertices} into 
the disjoint union of the sets
$V_{\mathrm{bulk}}$
and
$V_{\mathrm{bd}}$
for the vertices from the bulk
[colored in black in Fig.~\ref{boundary-tess-boundary-wire}(a)]
and for the vertices from the boundary  
[colored in red in Fig.~\ref{boundary-tess-boundary-wire}(a)],
respectively.  
Experimentally, the boundary wires
correspond to the outer radial layer and define the set of
{vertices}
used to probe boundary-boundary correlation functions.

We model the classical Josephson junction array
on the bipartite {simple} graph
$G=(V,E)$
{corresponding to the $\{4,5\}$ hyperbolic tessellation}
with the Hamiltonian
\begin{align}
H_{G}=&\,
\frac{1}{2}
\sum_{u,v\in V}
a_{u,v}\,
J\,
\left[
1
-
\cos(\phi_u-\phi_v)
\right]
\nonumber\\
&\,
+
\sum_{u\in V}
m^{2}\,
\left[
1
-
\cos(\phi_u-\phi_{\infty})
\right]
\nonumber\\
&\,
+
\sum_{u\in V_{\mathrm{bd}}}
M^{2}\,
\left[
1
-
\cos(\phi_u-\phi_{\infty})
\right],
\label{eq:def HG for tesselation 4,5}
\end{align}
{where $A=(a_{u,v})$ is the adjacency matrix}
and
$m^{2}$
($M^{2}$)
represent{s the} coupling of each vertex
(boundary vertex)
to a superconducting reservoir $\phi_\infty$,
effectively acting as mass terms for phase fluctuations. We enforce
Dirichlet boundary conditions energetically by taking $M\gg m$. In the
regime of large mass terms
($m^{2},M^{2}\gg J$),
the phase fluctuations remain small, justifying a harmonic expansion of the
cosine potentials. Shifting the phases to eliminate $\phi_\infty$, the
leading-order dynamics are governed by the quadratic action
\begin{subequations}
\label{eq:linear_theory}
\begin{align}
S_0[\phi]=
\frac{1}{2}
\sum_{u,v\in V}&
\phi_u\,
\Big[
-
a_{u,v}\,
J
+
d_u\,
\delta_{u,v}
\nonumber\\
&
+
\left(
m^{2}
+
M^{2}\,
\delta_{u\in V_{\mathrm{bd}}}
\right)\,
\delta_{u,v}
\Big]\,
\phi_v,
\label{eq:linear_theory a}
\end{align}
where
\begin{equation}
d_{u}\equiv
\sum\limits_{w\in V}
a_{u,w}\,
J
\label{eq:linear_theory b}
\end{equation}
denotes the weighted degree of site $u$ and
\begin{equation}
\delta_{u\in V_{\mathrm{bd}}}\equiv
\begin{cases}
1,&\hbox{ if $u\in\mathrm{V}_{\mathrm{bd}}$,}
\\
0,&\hbox{otherwise.}
\end{cases}
\label{eq:linear_theory c}
\end{equation}
\end{subequations}
The kernel in Eq.~(\ref{eq:linear_theory}) thus corresponds
to a graph Laplacian augmented by mass terms, such that the system
realizes a massive scalar field on a hyperbolic lattice.

Consider the two-point correlation function between
{any two pair of vertices}
$u,v\in V$
\begin{equation}
\langle\phi_u\,\phi_v\rangle_0=
\frac{
\int[\mathcal{D}\phi]\,
e^{- S_0[\phi]}\,
\phi_u\,\phi_v 
     }
     {
\int[\mathcal{D}\phi]\,
e^{- S_0[\phi]}\,
\hphantom{\phi_u\,\phi_v} 
     }.
\label{eq:def two-point fct}
\end{equation}
Following \cite{Asaduzzaman20},
we extract the scaling dimension
$\Delta$
{governing the algebraic decay of the two-point function
(\ref{eq:def two-point fct}), if any,}
by analyzing the boundary-boundary correlator
averaged over all equidistant boundary
pairs at {the} distance $r$
{as defined by}
\begin{subequations}
\label{eq:clean_correlator}
\begin{equation}
G^{(0)}_{\mathrm{edge}}(r):=
\frac{
\sum\limits_{u,v\in V_{\mathrm{bd}}}
\delta_{r,d(u,v)}\,
\langle\phi_u\,\phi_v\rangle_0
     }
     {
\sum\limits_{u,v\in V_{\mathrm{bd}}}
\delta_{r, d(u, v)}\,
\hphantom{
\langle\phi_u\,\phi_v\rangle_0
         }
     }.
\label{eq:clean_correlator a}
\end{equation}
This averaging restores approximate translational invariance
along the boundary, facilitating direct comparison with continuum CFT
predictions.  Let $r_{\mathrm{max}}$
be the perimeter of the boundary and $r_{\mathrm{min}}$
be the perimeter of the boundary for the
hyperbolic tesselation with one radial layer,
the former (latter) serving as an infrared (ultraviolet) cutoff of the theory.
\footnote{ 
In Fig.~\ref{boundary-tess-boundary-wire}(a),
$r_{\mathrm{max}}=40\,\mathfrak{a}$
and
$r_{\mathrm{min}}=10\,\mathfrak{a}$
with $\mathfrak{a}$ the lattice spacing.   
         } 
For $r_{\min}\ll r\ll r_{\max}$,
Eq.\ \eqref{eq: scaling-dimension-cft}
predicts the scaling law~%
\footnote{  
The distance
$2\sin^{2}\left(\pi\,r/r_{\max }\right)$
is called the conformal chordal distance.
It is defined so that it vanishes when $r=0$,
while reaching its maximum $2$ for $r=r_{\max}/2$. 
         }
\begin{align}
G^{(0)}_{\mathrm{edge}}(r)\sim&\,
\left[
\frac{1}{2\,\sin^{2}\left(\pi\,r/r_{\max }\right)}
\right]^{\Delta(m^{2})} 
\nonumber\\
\approx&\,
|r|^{-2\Delta(m^{2})},
\label{eq:clean_correlator b}
\end{align}     
\end{subequations}
which we are going to validate numerically.

\begin{figure}[t!]
\centering
\includegraphics[width=1.0\linewidth]{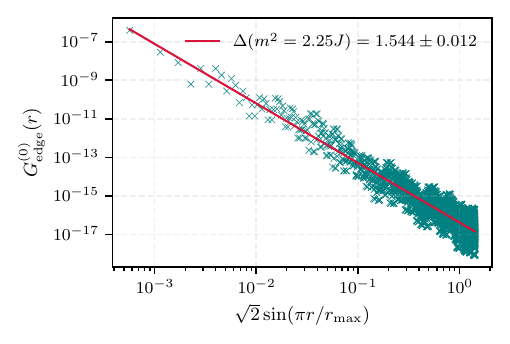}
\caption{
Extraction of the boundary scaling dimension from the boundary-boundary
correlator $G^{(0)}_{\mathrm{edge}}(r)$ at $m^{2}=2.25\,J$.
Points show the numerically computed data averaged over boundary sites,
while the red line is a linear fit in log-log scale.
The slope of the fit determines the scaling dimension $\Delta$.
        }
\label{bdr-bdr-corr-m=0}
\end{figure}

Numerical results for a bulk mass $m^{2}=2.25\,J$ are presented in
Fig.~\ref{bdr-bdr-corr-m=0}. The figure displays the binned values of
the boundary correlator $G_{\mathrm{edge}}^{(0)}(r)$, computed from
Eq.\ \eqref{eq:clean_correlator a} by inverting the graph Laplacian and
expressing distances along the boundary as the conformal chordal distance
$2\,\sin^{2}\left(\pi\,r/r_{\max }\right)$.
The scaling dimension $\Delta$ is then extracted
from the slope of a linear fit in a log-log plot, with the associated
uncertainty obtained from the regression.  Across all numerical
simulations presented here, we use a hyperbolic lattice with seven
radial layers (generations), comprising $|V| = 10651$ sites, and set
{the} boundary mass {to be} $M^{2} = 1600\,J$.

\begin{figure}
\centering
\includegraphics[width=1.0\linewidth]{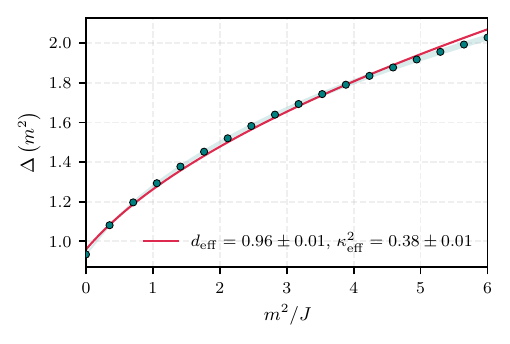}
\caption{
Scaling dimension $\Delta({m^{2}})$ as a function of the bulk mass
$m^{2}$. Numerical data (teal circles) are obtained from the power-law
decay of the boundary-to-boundary correlator at discrete mass
values. The shaded region indicates the error uncertainty derived from
the log-log regression. The solid line (crimson) represents a
non-linear fit to the effective continuum relation in
Eq. \eqref{eq:curve-fit}, with fitting parameters $d_{\mathrm{eff}}$
and $\kappa^{2}_{\mathrm{eff}}$.
        }
\label{fig:mass_sweep}
\end{figure}

As shown in Fig.~\ref{fig:mass_sweep}, the extracted scaling dimension
$\Delta({m^{2}})$ varies with the bulk mass.  To account for lattice
discretization and finite-size effects, we propose 
{the} effective functional form
\begin{equation}
\Delta({m^{2}})=
\frac{d_{\mathrm{eff}}}{2}
+
\sqrt{
\frac{d_{\mathrm{eff}}^{2}}{4}
+
\kappa_{\mathrm{eff}}^{2}\,
m^{2}
     },
\label{eq:curve-fit}
\end{equation}
where $d_{\mathrm{eff}}$ and $\kappa_{\mathrm{eff}}$ are treated as
fitting parameters. This form mirrors the continuum AdS/CFT relation
in Eq.~\eqref{eq: holograph-dimension}, with $\kappa_{\mathrm{eff}}$
playing the role of an effective curvature scale. Our numerical
results yield
\begin{equation}
d_{\mathrm{eff}}=0.96\pm 0.01,
\end{equation}
which closely approximates the continuum boundary dimension $d=1$.
Furthermore,
the extracted curvature
\begin{equation}
\kappa^{2}_{\mathrm{eff}}=0.38\pm 0.01
\end{equation}
is consistent with the geometric radius of the $\{4,5\}$ tiling,
$\kappa_{\{4,5\}}\approx0.7976\,{\mathfrak{a}}$,
{with $\mathfrak{a}$ the lattice spacing}.
\footnote{
For a $\{p,q\}$ regular tessellation, the radius of
curvature $\kappa_{\{p,q\}}$ is defined by
$\cosh({\mathfrak{a}}/2\kappa_{\{p,q\}})=\cos(\pi/p)/\sin(\pi/q)$
\cite{Brower21}.
         }

While the fit is robust for moderate masses, a divergence from the
continuum prediction appears at large $m^{2}$. This deviation reflects
lattice discretization effects rather than a breakdown of the
continuum relation \eqref{eq: holograph-dimension}. As the bulk mass
increases, the correlation length $\xi\sim 1/m$ becomes comparable to
{the short-distance cutoff $r_{\mathrm{min}}$},
rendering the boundary scaling sensitive to
the underlying discrete grid rather than the emergent smooth AdS
geometry. While this UV sensitivity could be mitigated through lattice
refinement with smaller polygons \cite{Brower22}, the current
discretization remains sufficient to capture the relevant holographic
phenomenology investigated in this work.

\subsection{Quenched disorder and replica theory}

In realistic experimental realizations of Josephson junction arrays,
fabrication tolerances inevitably introduce variations in the
Josephson energies. To model these effects,
we replace $J$ in Eq.\
(\ref{eq:def HG for tesselation 4,5})
by
\begin{subequations}
\label{eq:def Josephson array quenched randomness}
\begin{equation}
J_{u,v}=J+\delta J_{u,v},
\qquad
\delta J_{u,v}=\delta J_{v,u},
\label{eq:def Josephson array quenched randomness a}
\end{equation}
where
$J>0$ is the mean coupling and
$\{\delta J_{u,v}\}_{u,v\in V}$
are identically and independently distributed (iid)
Gaussian random variables of vanishing mean and variance $\sigma^{2}$.
The resulting disordered action is
\begin{equation}
S[\phi;\delta J]=S_{0}[\phi]+S_{\delta J}[\phi],
\label{eq:def Josephson array quenched randomness b}
\end{equation}
where the contribution 
\begin{equation}
\begin{split}
S_{\delta J}[\phi]=
\frac{1}{2}
\sum\limits_{u,v\in V}
\phi_u
\bigg[&\,
-
a_{u,v}\,\delta J_{u,v}
\\
&\,
+
\delta_{u,v}
\sum_{w}
a_{u,w}\,
\delta J_{u,v}
\bigg]
\phi_v
\end{split}
\label{eq:def Josephson array quenched randomness c}
\end{equation}
accounts for the random fluctuations of the Josephson couplings.
The two-point correlation functions for each disorder realization is
\begin{equation}
\langle\phi_u\,\phi_v\rangle_{\delta J}:=
\frac{
\int[\mathcal{D}\phi]\,
e^{-S[\phi;\delta J]}\,
\phi_u\,\phi_v 
     }
     {
\int[\mathcal{D}\phi]\,
e^{-S[\phi;\delta J]}\,
\hphantom{\phi_u\,\phi_v}
     }.
\label{eq:def Josephson array quenched randomness d}
\end{equation}
\end{subequations}
The question we want to address is if the conformal invariance of the
boundary survives in the presence of random fluctuations of
the Josephson couplings.

We are interested in the effect of average over quenched disorder on
boundary correlation functions. Let
\begin{subequations}
\label{eq:def Gaussian white noise averaging}
\begin{equation}
\overline{( \cdots )}
\label{eq:def Gaussian white noise averaging a}  
\end{equation}
denote {the Gaussian white-noise disorder averaging.  To} obtain
analytical control over th{is} disorder average,
we employ replica theory
\cite{Mezard1988spin, Dotsenko2005introduction}
and write
\begin{align}
\overline{\langle\phi_u\,\phi_v\rangle}=&\,
\lim_{n\rightarrow 0}
\int\prod^n_{a=1}\left[\mathcal{D}\phi^a\right]\,
\overline{
e^{-\sum\limits^n_{a=1}S[\phi^a;\delta J]}
         }\,
\phi^1_u\,\phi^1_v
\nonumber\\
=&\,
\lim_{n\rightarrow 0}
\int \prod^n_{a=1}\left[\mathcal{D}\phi^a\right]\,
e^{- S_{\mathrm{eff}}[\phi^a]}\,
\phi^1_u\,\phi^1_v,
\label{eq:def Gaussian white noise averaging b}
\end{align}
where we have introduced $n-1$ replicas of the system and relabeled
the original $\phi$ as $\phi^1$. Using the fact that only the second
cumulant of the {Gaussian distribution is nonvanishing}, we have the
resulting effective theory between replicas after averaging over disorder that is given by
\begin{align}
S_{\mathrm{eff}}[\phi^a]=&\,
-\ln
\left(
\overline{
e^{-\sum\limits^n_{a=1}S[\phi^a;\, \delta J]}
         }\,
\right)
\nonumber\\
=&\,
\sum_{a}
S_0[\phi^a]
\nonumber\\
&\,
+
\frac{\sigma^{2}}{2}
\sum_{{u,v\in V}}
{a_{u,v}}\,
\left[
\frac{1}{4}
\sum_a
\left(\phi_u^a - \phi_v^a\right)^{2}
\right]^{2}.
\label{eq:def Gaussian white noise averaging c}
\end{align}
\end{subequations}
The presence of disorder manifests {itself} as a quartic interaction among
replicas, where the variance $\sigma^{2}$
captures the correlations induced by
the random fluctuations of the Josephson couplings.

\begin{figure}[t!]
\centering
\includegraphics[width=1.0\linewidth]{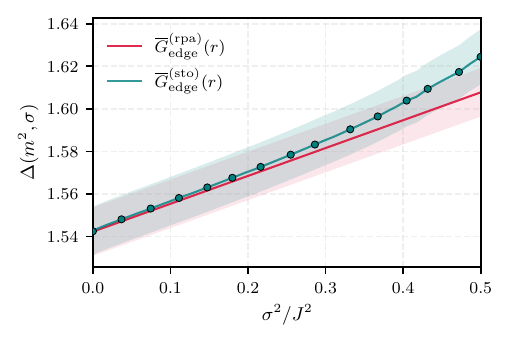}
\caption{
Scaling dimension
{$\Delta(m^{2},\sigma)$}
as a function of the variance $\sigma^{2}/J^{2}$
of the Gaussian white-noise disorder
with fixed mass
$m^{2}=2.25\,J$
for both 
$\overline{G}^{\mathrm{(sto)}}_{\mathrm{edge}}(r)$ (with $N_{\mathrm{dis}}=100$)
and
$\overline{G}^{\mathrm{(rpa)}}_{\mathrm{edge}}(r)$.
The shaded regions represent the standard error of the linear fit.
        }
\label{fig:delta_comparison}
\end{figure}

{
Similarly to what we did in Eq.\ \eqref{eq:clean_correlator a},
we define the disordered boundary-boundary correlator
\begin{subequations}
\label{eq:disordered_correlator}
\begin{equation}
\overline{G}_{\mathrm{edge}}(r):=
\frac{
\sum\limits_{u,v\in V_{\mathrm{bd}}}\,
\delta_{r,d(u,v)}\,  
\overline{\langle\phi_u\,\phi_v\rangle}
     }
     {
\sum\limits_{u,v\in V_{\mathrm{bd}}}\,
\delta_{r,d(u,v)}\,
\hphantom{\overline{\langle\phi_u\,\phi_v\rangle}}
     }.
\label{eq:disordered_correlator a}
\end{equation}
If the conformal invariance of the boundary 
holds in the presence of disorder for
$r_{\mathrm{min}}\ll r\ll r_{\mathrm{max}}$,
we again expect the algebraic decay 
\begin{align}
\overline{G}_{\mathrm{edge}}(r)\sim&\,
\left[
\frac{1}{2\,\sin^{2}\left(\pi\,r/r_{\max }\right)}
\right]^{\Delta(m^{2},\sigma)} 
\nonumber\\
\approx&\,
r^{-2\Delta(m^{2},\sigma)}
\label{eq:disordered_correlator b}
\end{align}
\end{subequations}
with the scaling exponent $\Delta(m^{2},\sigma)$
a function of both $m^{2}$ and $\sigma$.
We are going to provide support for the scaling encoded by
Eq.\ (\ref{eq:disordered_correlator b})
by evaluating the right-hand side of Eq.
(\ref{eq:disordered_correlator a})
with the help of two distinct approximations. 
} 

{
The first approximation that we shall perform to evaluate
the right-hand side of Eq.\
(\ref{eq:disordered_correlator a})
consists in replacing the Gaussian integration over
the iid random variables by an arithmetic average over
$N_{\mathrm{dis}}$ stochastic samplings of the
iid random variables. To this end, we define
\begin{subequations}
\label{eq:stochastic samplings of Gaussian white-noise disordered G(r)}
\begin{align}
\overline{G}^{\mathrm{(sto)}}_{\mathrm{edge}}(r):=
\frac{
\sum\limits_{u,v\in V_{\mathrm{bd}}}\,
\delta_{r,d(u,v)}\,
\left[
N_{\mathrm{dis}}^{-1}\,
\sum\limits_{i=1}^{N_{\mathrm{dis}}}
\langle\phi_u\,\phi_v\rangle_{\delta J_i}
\right]
     }
     {
\sum\limits_{u,v\in V_{\mathrm{bd}}}\,
\delta_{r,d(u,v)}\,
\hphantom{
\left[
N_{\mathrm{dis}}^{-1}\,
\sum\limits_{i=1}^{N_{\mathrm{dis}}}
\langle\phi_u\,\phi_v\rangle_{\delta J_i}
\right]
         }
     },
\label{eq:stochastic samplings of Gaussian white-noise disordered G(r) a}
\end{align}
where the two-point function
$\langle\phi_u\,\phi_v\rangle_{\delta J_i}$
is, according to Eq.\
(\ref{eq:def Josephson array quenched randomness d}),
the matrix element of the inverse of a quadratic kernel.
For a sufficiently large number $N_{\mathrm{dis}}$
of stochastic samplings, we expect that
\begin{equation}
\overline{G}_{\mathrm{edge}}(r)\approx
\overline{G}^{\mathrm{(sto)}}_{\mathrm{edge}}(r).
\label{eq:stochastic samplings of Gaussian white-noise disordered G(r) b}
\end{equation}
\end{subequations}
We report in Fig.\ \ref{fig:delta_comparison}
the dependence on $\sigma^{2}/J^{2}$ of $\Delta(m^{2},\sigma)$
for $m^{2}=2.25\,J$ by fitting the dependence on $r$ of
$\overline{G}^{\mathrm{(sto)}}_{\mathrm{edge}}(r)$ with a power law decay.
}  

{
The second approximation that we shall perform to evaluate
the right-hand side of Eq.\
(\ref{eq:disordered_correlator a})
starts from the Dyson series for the two-point function of
an interacting theory, here defined by the action
(\ref{eq:def Gaussian white noise averaging c}).
Define the operators $G^{\,}_{0}$ and $\overline{G}$
by their matrix elements
\begin{equation}
G^{\,}_{0\,u,v}\equiv
\lim_{\sigma^{2}\downarrow0}\overline{G}_{u,v},
\qquad
\overline{G}_{u,v}\equiv
\overline{\langle\phi_u\,\phi_v\rangle},
\end{equation}
for any $u,v\in V$. Here, the two-point function
$\overline{\langle\phi_u\,\phi_v\rangle}$
is computed by performing the replica limit as defined by
Eq.\ (\ref{eq:def Gaussian white noise averaging b}).
The self-energy is the operator $\overline{\Sigma}$ that relates
$G^{\,}_{0}$ to $\overline{G}$ through the Dyson relation
\begin{subequations}
\label{eq:Dyson relation}
\begin{equation}
\overline{G}=G^{\,}_{0}+G^{\,}_{0}\ \overline{\Sigma}\ \overline{G}.
\label{eq:Dyson relation a}
\end{equation}
Knowledge of $\overline{\Sigma}$ is equivalent to knowledge of
$\overline{G}$ through the geometric series
\begin{equation}
\overline{G}=\left(1-G^{\,}_{0}\ \overline{\Sigma}\right)^{-1}\,G^{\,}_{0}.
\label{eq:Dyson relation b}
\end{equation}
\end{subequations}
We shall call the random-phase approximation (RPA),
the approximation
\begin{subequations}
\label{eq:RPA approximation}
\begin{equation}
\overline{G}\approx
\overline{G}^{\mathrm{(rpa)}},
\label{eq:RPA approximation a}
\end{equation}
whereby
\begin{align}
\overline{G}^{\mathrm{(rpa)}}:=&\,
\left(1-G^{\,}_{0}\ \overline{\Sigma}^{\mathrm{(rpa)}}\right)^{-1}\,
G^{\,}_{0}
\nonumber\\
=&\,
\left(G^{-1}_{0} -\overline{\Sigma}^{\mathrm{(rpa)}}\right)^{-1},
\label{eq:RPA approximation b}
\end{align}
by which the exact self-energy $\overline{\Sigma}$
is substituted by the RPA self-energy $\overline{\Sigma}^{\mathrm{(rpa)}}$.
Here, the matrix elements of $\overline{\Sigma}^{\mathrm{(rpa)}}$ are defined by
\begin{equation}
\overline{\Sigma}^{\mathrm{(rpa)}}_{z,w}:=
\frac{\sigma^{2}}{2}\,
a_{z,w}\,
\langle(\phi_w-\phi_z)^{2}\rangle_0,
\qquad
z,w\in V.
\label{eq:RPA approximation c}
\end{equation}
coming from the perturbative expansion of the interacting effective theory (\ref{eq:def Gaussian white noise averaging c})
\begin{align}
&
\overline{\langle\phi_u\,\phi_v\rangle}
-
\langle\phi_u\,\phi_v\rangle_0=
\nonumber\\
&\quad
+
\frac{\sigma^{2}}{2}
\sum_{{z,w\in V}}
{a_{z,w}}\,
\langle
\phi_u\,\phi_z
\rangle_0\,
\langle
(\phi_w-\phi_z)^{2}
\rangle_0\,
\langle
\phi_w\,\phi_v
\rangle_0
\nonumber\\
&\quad
{+\mathcal{O}(\sigma^{4})}
\label{eq:RPA approximation d}
\end{align}
\end{subequations}
to the order $\sigma^{2}$ for any pair of vertices $u,v\in V$ of
the hyperbolic tesselation. At last, we arrive at our second
approximation of the right-hand side of Eq.\
(\ref{eq:disordered_correlator a})
that is defined by the estimate
\begin{align}
\overline{G}^{\mathrm{(rpa)}}_{\mathrm{edge}}(r):=
\frac{
\sum\limits_{u,v\in V_{\mathrm{bd}}}\,
\delta_{r,d(u,v)}\,
\overline{G}^{\mathrm{(rpa)}}_{u,v}
     }
     {
\sum\limits_{u,v\in V_{\mathrm{bd}}}\,
\delta_{r,d(u,v)}\,
\hphantom{
\overline{G}^{\mathrm{(rpa)}}_{u,v}
         }
     }.
\label{eq:rpa approximation}
\end{align}
Hereto, we report in Fig.\ \ref{fig:delta_comparison}
the dependence of $\Delta(m^{2},\sigma)$ on $\sigma^{2}/J^{2}$
for fixed mass  $m^{2}=2.25\,J$ when fitting 
$\overline{G}^{\mathrm{(rpa)}}_{\mathrm{edge}}(r)$ with a power law decay on $r$.
} 

\begin{figure}[t!]
\centering
\includegraphics[width=1.0\linewidth]{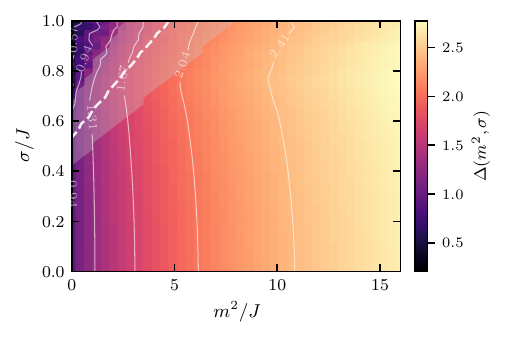}
\caption{
Heatmap of $\Delta({m^{2}},\sigma)$
versus bare mass $m^{2}/J$ and disorder standard deviation $\sigma/J$. 
Constant $\Delta$ levels are indicated by solid white contours. 
The dashed white line denotes the phase boundary
for Laplacian positive definiteness,
while the gray shaded area highlights
the region of realization-dependent stability.
       }
\label{heatmap}
\end{figure}

{
The most important message of Fig.~\ref{fig:delta_comparison}
is that the disorder preserves the scale invariance of the
boundary theory in the clean limit,
although it modifies the critical behavior
by increasing the value of the scaling exponent of the two-point
function relative to its value in the clean limit.
This effect can be captured by a renormalization of the
bulk mass by the disorder strength.
We also infer from Fig.~\ref{fig:delta_comparison}
that the stochastic and RPA estimates for the scaling exponent
exhibit nearly perfect agreement for small disorder strengths
($\sigma^{2}/J^{2}\ll 1$). The RPA approximation underestimates
the increase of the scaling exponent induced by disorder
and this underestimation gets more pronounced with increasing
disorder strength.
}

The global behavior of the scaling dimension $\Delta({m^{2}},\sigma)$
is summarized in the heatmap {from} Fig.~\ref{heatmap}, computed using the
stochastic estimator
\eqref{eq:stochastic samplings of Gaussian white-noise disordered G(r) a}.
We identify two distinct regimes
separated by a stability boundary (dashed white line). To the left of
this boundary, the Laplacian ceases to be positive definite, signaling
a breakdown of the harmonic (cosine) expansion. From an experimental
perspective, a physical Josephson junction array is not expected to
access this unstable regime. However, magnetic flux through the
plaquettes could, in principle, induce frustration in the Josephson
couplings and push the system toward this boundary. Whether such
conditions can modify the bulk theory, and how the boundary CFT would
respond, remains an interesting question for future investigation.

In the physically relevant regime, to the right of the stability
boundary ($m^{2}\gg\sigma$), $\Delta({m^{2}},\sigma)$ increases
monotonically with the bulk mass $m$. For low $\sigma$, the contours
are nearly vertical, indicating that the boundary physics is
relatively insensitive to disorder and remains primarily governed by
the mass.  In the strong-disorder regime, on the other hand, the
scaling dimension becomes sensitive to $\sigma$ and exhibits a
``repulsion'' from the instability region. Crucially, the crossover
behavior is smooth, suggesting that it is not associated with a phase
transition but instead corresponds to a continuous deformation of the
boundary CFT data induced by bulk disorder. These results suggest
that, while disorder modifies the specific holographic scaling of the
free scalar theory, the boundary conformal structure remains robust.

%%%%%%%%%%%%%%%%%%%%%%%%%%%%%%%%%%%%%%%%%%%%%%%%%%%%%%%%%%%%%%%%%%%%%%%%%%%%%%%
\section{Experimental outlook}
\label{sec:Experimental outlook}

The design of an array of superconducting wires can be naturally
realized using the two-angle deposition technique, also known as the
Manhattan process
\cite{potts2001cmos,costache2012lateral,kreikebaum2020improving}.
Both horizontal and vertical wires are
patterned using standard lithography with an MMA/PMMA bilayer resist
to create an undercut.  In the first deposition step, Al wires are
deposited at an azimuthal angle $\theta$, parallel to the horizontal
wires (polar angle $\phi = 0^\circ$).  With a nonzero azimuthal angle,
the vertical wire patterns are masked by the resist wall, and only the
horizontal wires are deposited with Al.  Subsequently, oxidation is
performed in situ, creating an AlOx insulating layer on top of the
horizontal wires.  The second Al layer is deposited at an angle
$\theta$ and $\phi = 90^\circ$ to deposit the vertical Al wires.  The
result is a network of horizontal and vertical wires forming
superconducting-insulating-superconducting (SIS)
Josephson junctions at the crossings.  The junction inductances and
capacitances are most commonly controlled by tuning the area of a
junction, although the thickness of the oxide layer can be varied for
further adjustment.

% (Avoiding unwanted crossings)
Unwanted wire crossings may occur depending on the connectivity of the
graph.  By permuting the rows/columns of the matrix, a matrix
configuration without unnecessary crossings may be found while
preserving the connectivity.  For higher-dimensional graphs, this may
be challenging or impossible, as discussed in
Sec.\ \ref{subsec:Grid-intersection-graphs}. 
In these cases, overpasses, or air bridges
\cite{koster1989investigations,abuwasib2013fabrication,kwon2001low},
allow two orthogonal wires to cross without forming a junction.  An
extra lithography and deposition step is required to create air
bridges.

A fundamental limit on the size of the structure
is set by the phase stiffness of the bulk of a wire
from the array. We are going to estimate a first upper bound
on the length of the wires so that each of them can be described
by a single superconducting phase
assuming the absence of stray magnetic fields.
We will then provide a second such upper bound
that arises from the presence of stray magnetic fields.

Needed is a numerical estimate for the upper bound
$N_{\star}(T)$
defined in Eq.\ (\ref{eq:theory answer for Nstar})
on the number of Josephson-coupled superconducting wires
below which we may assume that their superconducting phases
are rigid for realistic material parameters.
This upper bound is caused by the fluctuations of the
superconducting phase of a condensate that are brought about
by thermal fluctuations at the temperature $T$.
We consider $2N_{\star}(T)$ dirty aluminum wires,
each one of length no longer than
\begin{subequations}
\label{eq:numbers for an al wire}
\begin{equation}
L=N_{\text{max}}\,\mathfrak{a},
\qquad
\mathfrak{a}=1\,\mu\mathrm{m}=1\times10^{-6}\,\mathrm{m},
\end{equation}
with
\begin{equation}
N_{\mathrm{max}}\lesssim N_{\star}(T)
\end{equation}  
and whose cross-section has the area
\begin{equation}
A_{\mathrm{w}}^{\,}\equiv
\,r^{2},
\quad
r=300\,\mathrm{nm}=3\times10^{-7}\,\mathrm{m}.
\end{equation}
At temperature $T=0.1\,\mathrm{K}$,
a dirty wire is superconducting with
the three-dimensional superconducting phase stiffness
\begin{equation}
\rho_{\mathrm{s}}^{\,}\equiv
\frac{J_{\mathrm{w}}\,\mathfrak{a}}{A_{\mathrm{w}}}=
10^{-12}\,\frac{\mathrm{J}}{\mathrm{m}}.
\end{equation}
\end{subequations}
With the relation
($k_{\mathrm{B}}^{\,}=1.380649\times10^{-23}\,\mathrm{J}/\mathrm{K}$)
\begin{align}
J_{\mathrm{w}}^{\,}\equiv&\,
\frac{A_{\mathrm{w}}^{\,}\,\rho_{\mathrm{s}}^{\,}}{\mathfrak{a}}
\nonumber\\
=&\,
r\,\frac{r}{\mathfrak{a}}\,\rho_{\mathrm{s}}^{\,}
\nonumber\\
=&\,
9\times10^{-20}\,\mathrm{J}
\nonumber\\
\approx&\,
k_{\mathrm{B}}^{\,}\, 6.52\times10^{3}\,\mathrm{K}
\end{align}
and with the help of Eq.\
(\ref{eq:theory answer for Nstar}),
we find the conservative condition
\begin{equation}
T=0.1\,\mathrm{K},
\qquad
N_{\star}^{\,}(T)\equiv
\frac{J_{\mathrm{w}}^{\,}}{2\,k_{\mathrm{B}}^{\,}\,T}\approx3.25\times10^{4},
\label{eq:Numerical-value-Nstar}
\end{equation}
i.e.,
$N_{\star}(T)$ should be no larger than $3.25\times10^{4}$
if the superconducting phase of each one of the
$N_{\star}(T)$ [$N_{\star}(T)$]
vertically (horizontally) aligned wires of length
$L_{\mathrm{max}}\equiv
N_{\mathrm{max}}\,\mathfrak{a}\lesssim
N_{\star}(T)\,\mathfrak{a}$
is to be treated as a single rigid phase at the temperature
$T=0.1\,\mathrm{K}$.

%%%%%%%%%%%%%%%%%%%%%%%%%%%%%%%%%%%%
% (Flux/flux screening)
Although the fundamental limit on the size of the structure is
set by the phase stiffness of the bulk of the wire, stray
magnetic fields may impose a lower practical upper bound.  The
connectivity of the wiring inherently creates superconducting loops of
varying areas. As the array dimensions scale, the average loop size
increases.  For experimentally unavoidable stray magnetic fields,
sufficiently large loops will produce substantial magnetic flux
noise. For a general geometry, loop sizes will vary from the scale of
a single unit cell up to that of the entire sample.  As a consequence,
the net effect of stray magnetic fields will be flux offsets whose
magnitude depends on the particular embedding.
Assuming that screening of stray magnetic fields
can be attained at the nano Tesla level
(typical for superconducting qubits \cite{Malevannaya2025engineering}),
we obtain a characteristic length $L_{B\mathrm{max}}$ by demanding that
a square with edges of length $L_{B\mathrm{max}}$
traps one superconducting flux quantum
(\ref{eq:def-sc-quantum-flux})
for a magnetic field of magnitude $B=1\,\mathrm{nT}$,
\begin{align}
L_{B\mathrm{max}}:=
\sqrt{\frac{\Phi_0}{B}}\sim
1.44\,\mathrm{mm}\;.
\end{align}
For a lattice a lattice spacing
$\mathfrak{a}\sim\mu\mathrm{m}$,
we find that nanoTesla screening
translates into an upper bound on the number of
horizontal (vertical) wires of the order
\begin{equation}
N_{\mathrm{c}\,\star}\sim
\frac{L_{B\mathrm{max}}}{\mathfrak{a}}\sim
1.44\times 10^{3}.
\end{equation} 
There are several paths for studying structures beyond this scale.
The first path is to work with graphs that admit a quasi-local
embedding, with loop sizes that do not grow extensively with system
size.  The second path is to seek flux-insensitive hardware designs
incorporating, for instance, gradiometric loops.  Both of these paths
naturally lead to interesting embedding problems for future study.
The third path is to view the presence of unavoidable flux disorder as
a feature, and focus on systems where the role of disorder itself
poses interesting questions.

Aside from the experimental constraints associated with building large
wire networks, it is interesting to discuss experimental measurement
approaches.  Electrical transport measurements have been used
extensively to characterize the physics of Josephson arrays, and we
expect this technique to continue to be appealing for our system, as
discussed in Sec.\
\ref{sec:Linear response theory for a quantum Josephson array}.
Although conceptually the simplest, transport
measurements are not without challenges.  Finite-frequency microwave
response measurements are an alternative.  For large
systems, microwave probes give natural access to collective modes
\cite{kuzmin2019quantum,mukhopadhyay2023superconductivity}.
For example, we expect that the boundary CFTs in Sec.\
\ref{sec:The AdS/CFT correspondence for Euclidean hyperbolic spaces}
are naturally detectable through gapless collective modes.  For smaller systems,
such as the cube in Fig.~\ref{gig-notgig}, it would be natural to
perform spectroscopy using the circuit quantum electrodynamics
framework \cite{blais2021circuit}.  By engineering the capacitance of
the wires, and coupling at least one to a readout resonator, we expect
that it will be possible to measure the excitation spectrum and
perform quantum gates.  The exciting possibility of performing, for
instance, spectroscopy of a high-dimensional artificial atom is
potentially within reach.

\subsection{Grid intersection graphs}
\label{subsec:Grid-intersection-graphs}

{In Sec.\ \ref{sec: graphs/geometries and JJ arrays}, w}e
discussed how to program the biadjacency matrices $B{\equiv(b_{i,j})}$
of bipartite {and simple} graphs $G{=(V,E)}$
using {arrays of} superconducting wire{s}.
Th{is} construction requires extending a superconducting wire with phase
$\varphi_i$ ($\phi_j$) from the first non-zero entry of $B$ to the final
non-zero entry
for its row $i$ (column $j$)
and  coupling the wires through
Josephson junctions according to the non-zero entries of the
biadjacency matrix.
As some horizontal and vertical wires may not be connected through
a Josephson junction, an overpass or an air bridge
is needed at their intersection. While
these overpasses can be realized by introducing a thicker insulating
layer between the superconducting wires, the simplest initial
implementation is to design arrays that avoid them.

\begin{figure}[t!]
\includegraphics[width=\columnwidth]{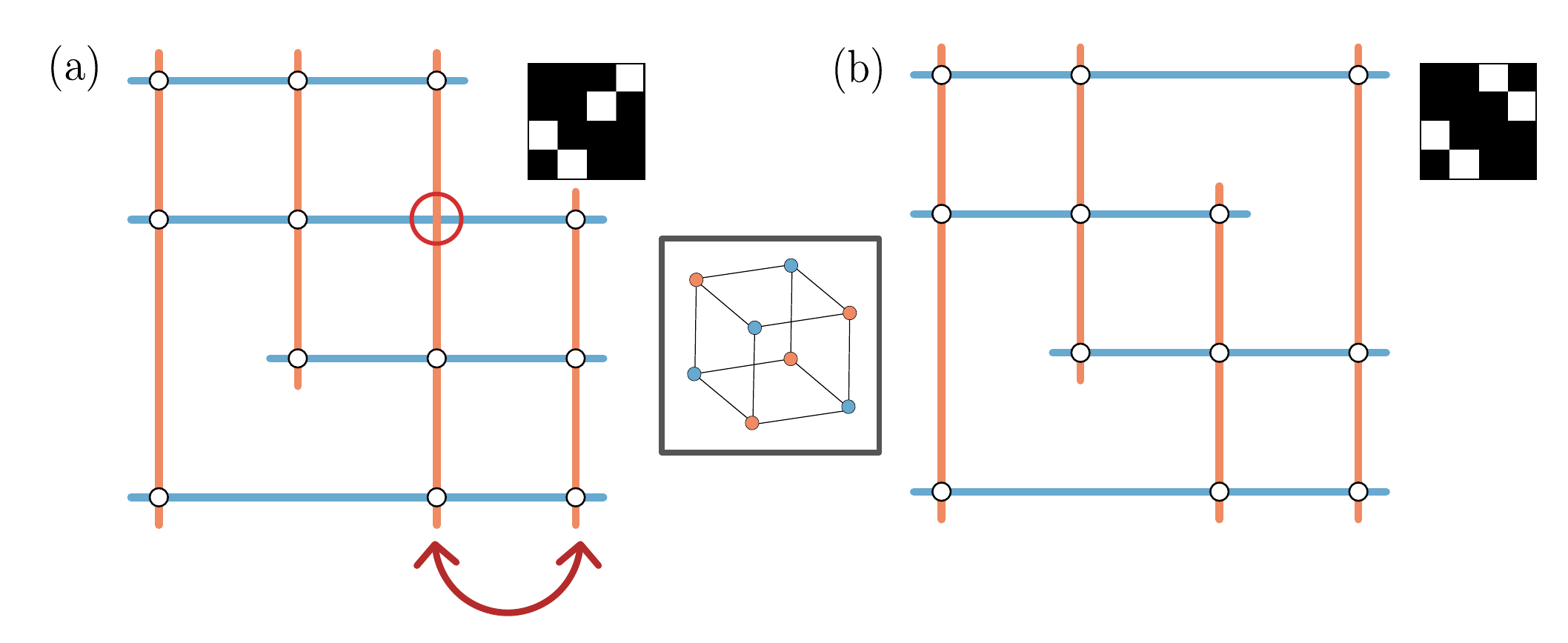}
\caption{
{(a) Wire construction of a cube with a single overpass.
The bipartite graph structure of a cube from
Fig.\ \ref{fig:cube_wire} is a non-GIG representation
with the red circle indicating a single} overpass and white
circles showing the Josephson junctions. The red arrow shows the
correct permutation of wires to eliminate overpasses.
(b) Wire construction of a cube without any overpass.
{For the} permuted biadjacency matrix with GIG representation,
no overpasses occur.
The biadjacency matrices are given as a reference at the
corners. Black (white) denote ones (zeros) entries of the matrix.
        }
\label{gig-notgig}
\end{figure}

We can formalize the notion of an overpass as follows. Given a
biadjacency matrix $B$, we assume that for a fixed column $k$, the
entries {$b_{i,k}= b_{f,k}=1$}, where $i$ and $f$ represent initial and
final non-zero entries in that column, respectively. We call an entry
an overpass if there exists a row $r$, $i<r<f$, with entries
$b_{r,i^\prime}=b_{r,f^\prime }=1$, but $b_{r,k}=0$
such that $i^\prime<k<f^\prime$.
We give an example of overpasses in
Fig.\ \ref{gig-notgig}(a).

Since every permutation of the rows and columns of $B$ yields the same
(i.e., isomorphic) {bipartite and simple} graph,
we can use an{y
permutation of the rows and columns} of $B$
to represent the underlying tessellation. Therefore, we can try to
reduce the number of overpasses by reordering rows and columns of
$B$. In other words, we can try to find a 
{permutation of the rows and columns of $B$}
to eliminate submatrices $\mathrm{Sub}(B)$ of $B$ in the form of
\begin{eqnarray}
\mathrm{Sub}(B)=
\begin{pmatrix}
& & & \vdots & & & \\
&  & & 1 & & &  \\
& & & \vdots & & & \\
\cdots &1 & \cdots & 0 & \cdots & 1  & \cdots\\
& & & \vdots & & & \\
&  & & 1 & & &  \\
& & & \vdots & & &
\end{pmatrix},
\end{eqnarray}
where the center $0$ corresponds to an overpass.

In order to decide if a bipartite and simple graph
$G=(V,E)$ can be realized by a Josephson array of superconducting
wires without overpasses, we use the following terminology.
A graph \(G=(V,E)\) is called a grid intersection graph (GIG)
if there exist two disjoint sets \(\mathcal H\) and \(\mathcal V\)
of line segments in the Euclidean plane such that
\(V=\mathcal H\sqcup\mathcal V\)
and for distinct vertices \(u,v\in V\) one has \(\{u,v\}\in E\) if and
only if one of \(u,v\) corresponds to a segment in \(\mathcal H\), the
other corresponds to a segment in \(\mathcal V\), and the two
corresponding segments intersect in at least one point.
It follows that a GIG is a bipartite and simple graph.
We will also need the definition of a planar graph.
A graph \(G=(V,E)\) is called planar if there exists an embedding of
\(G\) into the Euclidean plane
such that vertices are represented by distinct points,
edges are represented by simple continuous curves connecting
their endpoints, and no two edges intersect except possibly
at a common endpoint.

Given any bipartite and simple graph $G=(V,E)$, 
if a permutation of its biadjacency matrix $B$
without any overpass exists, then $G{=(V,E)}$ is a GIG, as in
Fig.\ \ref{gig-notgig}(b).
The task of finding a configuration of a square array
of superconducting wires with no overpasses
is equivalent to checking whether the
bipartite and simple graph $G$ is a GIG or not.
In general, not all {bipartite and simple} graphs have a GIG
representation. However, it is known that
every finite planar bipartite graph is a GIG
(the converse is not true)
\cite{HARTMAN199141,defraysseix:hal-00005620}. Moreover, there
exists an algorithm to find a vertex ordering that puts the
adjacency matrix of $G$ in GIG representation in linear time
\cite{10.1007/BF02574056}.
Hence, for any planar bipartite graph, such as the graph associated with a
hyperbolic tessellation $\{p,q\}$ with even $p$ (for which all faces have
even number of edges and the graph is therefore bipartite),
one can find a permutation with no overpass in linear time.

We emphasize that planarity is not a necessary condition for a graph
to admit a GIG representation. For example,
the complete bipartite graph $K_{N,N}$, with biadjacency matrix
$b_{i,j}=1$ for all $i,j$, is non-planar. Nevertheless, it is a GIG,
since all vertices in one subgraph can be represented by parallel
horizontal segments intersecting all vertical segments corresponding
to the vertices in the other subgraph. In general, deciding whether a
graph admits a GIG representation is NP-hard
\cite{kratochvil1994planar}. However, the number of overpasses can be
reduced in practice using heuristic optimization methods, such as
simulated annealing.

\begin{figure}[t!]
\includegraphics[width=0.8\columnwidth]{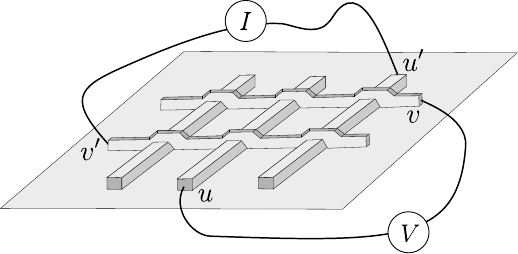}
\caption{
{Square} array of superconducting wires
coupled through Josephson junctions at crossing points.
Voltage drop between wires $u$ and $v$
and electric current between wires $u'$ and $v'$
can be both inserted and/or probed in the system.
        }
\label{Fig:shea-PRL-97}
\end{figure}

\subsection{Linear response theory for a quantum Josephson array}
\label{sec:Linear response theory for a quantum Josephson array}

{
Figure~\ref{Fig:shea-PRL-97}
shows a typical  experimental setup used to measure current-voltage
relations for a square array of superconducting wires {coupled through
Josephson junctions}.
We are going to show that such} measurements provide direct
experimental access to the equal-time two-point correlation function
between the superconducting phases of any two wires in the array. The
central idea is that linear response theory allows one to probe the
underlying connectivity encoded in the adjacency matrix of the array{.
B}y perturbing the system with an applied current (between wires $u'$
and $v'$) and measuring the induced voltage response (between wires
$u$ and $v$), one can reconstruct the two-point correlations between
superconducting phases. The discussion below is carried out with
finite capacitances, i.e., we bring back quantum fluctuations.

The time-independent quantum Hamiltonian
(\ref{eq:def Josephson array f})
for the square array of Josephson-coupled
superconducting wires is defined in 
Appendix~\ref{subsec:Definition of the quantum Josephson array}.
The Kubo formula is built from a pair of operators
\begin{equation}
\widehat{A},
\qquad
\widehat{H}'(t)=f(t)\,\widehat{B},
\end{equation}
where $\widehat{A}$
is the observable in the closed world {described by
Hamiltonian
(\ref{eq:def Josephson array f})}
and 
$\widehat{H}'(t)=f(t)\,\widehat{B}$
is the {time-dependent} probe to the environment.
{It is given in Appendix
\ref{appendix:Linear response theory for a quantum Josephson array}. }

To identify the appropriate observable and probe for
the quantum Josephson array, we consider the experimental setup of Fig.\
\ref{Fig:shea-PRL-97}. 
An electric current $I$ is injected  into wire $u'$
and extracted from  wire $v'\neq u'$, thereby driving a rate of change in $\partial_{t}^{\,}\hat{n}_{u'}^{\,}$ and 
$\partial_{t}^{\,}\hat{n}_{v'}^{\,}$.
The response is measured through the expectation value of the charge imbalance
\begin{equation}
\left(
\hat{n}_{u}^{\,}
-
n_{u}^{\mathrm{(bg)}}
\right)
-
\left(
\hat{n}_{v}^{\,}
-
n_{v}^{\mathrm{(bg)}}
\right)
\end{equation}
for the pair $u,v$ of wires.

Accordingly, we couple the quantum Josephson array defined
in Appendix \ref{subsec:Definition of the quantum Josephson array}
to the external currents by postulating the equations of motion
\begin{subequations}
\label{eq:mofidied coupled equations of motion}
\begin{align}
\frac{\partial}{\partial t}
\hat{n}_{u}^{\,}\equiv&\,
\frac{1}{\mathrm{i}\hbar}\,
\left[
\hat{n}_{u}^{\,},
\widehat{H}_{\mathrm{C,J}}^{\,}
\right]
\nonumber\\
=&\,
+
\frac{1}{\hbar}\,
\sum_{v}
E_{u,v}^{\mathrm{(J)}}
\sin
\left(
\hat{\phi}_{v}^{\,}
-
\hat{\phi}_{u}^{\,}
-
A_{v,u}^{\mathrm{(bg)}}
\right)
+
I_{u}^{\mathrm{ext}}(t),
\label{eq:mofidied coupled equations of motion a}
\\
\frac{\partial}{\partial t}
\hat{\phi}_{u}^{\,}=&\,
\frac{1}{\mathrm{i}\hbar}\,
\left[
\hat{\phi}_{u}^{\,},
\widehat{H}_{\mathrm{C,J}}^{\,}
\right]
\nonumber\\
=&\,
+
\frac{1}{\hbar}\,
\sum_{v}
E_{u,v}^{\mathrm{(C)}}\,
\left(
\hat{n}_{v}^{\,}
-
n_{v}^{\mathrm{(bg)}}
\right)
+
\frac{U_{u}^{\mathrm{ext}}(t)}{\hbar},
\label{eq:mofidied coupled equations of motion b}
\end{align}
{obeyed by the wire $u$.}
These modified coupled equations of motion correspond
to choosing the coupling
\begin{equation}
\begin{split}
&
\widehat{H}_{\,}^{\prime}(t)=
\sum_{u}
\left[
-
\hbar\,
I_{u}^{\mathrm{ext}}(t)\,
\hat{\phi}_{u}^{\,}
+
U_{u}^{\mathrm{ext}}(t)\,
\left(
\hat{n}_{u}^{\,}
-
n_{u}^{\mathrm{(bg)}}
\right)
\right],
\end{split}
\label{eq:mofidied coupled equations of motion c}
\end{equation}
with units $\left[I_{u}^{\mathrm{ext}}(t)\,\right]=[\mathrm{time}]_{\,}^{-1}$ and $
\left[U_{u}^{\mathrm{ext}}(t)\,\right]=\mathrm{energy}$.
Additionally, the external current must obey,
\begin{equation}
\sum_{u}
I_{u}^{\mathrm{ext}}(t)=0
\label{eq:mofidied coupled equations of motion d}
\end{equation}
\end{subequations}
if coupling to the environment respects the $\mathrm{U}(1)$ global
symmetry.
Inserting electric current and probing voltage drop ($U_{w}^{\mathrm{ext}}(t)=0$) 
corresponds to linear response theory with
\begin{subequations}
\begin{equation}
I_{w}^{\mathrm{ext}}(t)=
+I\,\delta_{w,u'}^{\,}
-I\,\delta_{w,v'}^{\,},
\end{equation}
{for wire $w$.}
This tells us that the choices for $\widehat{A}$ and $\widehat{B}$
operators in the Kubo formula are 
\begin{equation}
\widehat{A}\to
\left(
\hat{n}_{u}^{\,}
-
n_{u}^{\mathrm{(bg)}}
\right)
-
\left(
\hat{n}_{v}^{\,}
-
n_{v}^{\mathrm{(bg)}}
\right)
\end{equation}
and
\begin{equation}
\widehat{B}\to
\hat{\phi}_{u'}^{\,}
-
\hat{\phi}_{v'}^{\,},
\end{equation}
respectively. {With this choice for the operators
$\widehat{A}$ and $\widehat{B}$,
it is the retarded response function
\begin{equation}
\chi_{\beta;uv,u'v'}^{\,}(\omega)\equiv
C_{
0\,\beta\,
\big(
\hat{n}_{u}^{\,}
-
\hat{n}_{v}^{\,}
\big),
\big(
\hat{\phi}_{u'}^{\,}
-
\hat{\phi}_{v'}^{\,}
\big)  
  }^{\mathrm{R}}(\omega)
\end{equation}  }
\end{subequations}
{that enters the generic Kubo formula from Appendix
\ref{appendix:Linear response theory for a quantum Josephson array}.}

\begin{widetext}
{Assuming that the charging-energy matrix is diagonal
and that all background charges vanish,
Eq.~(\ref{eq:fluctuation-dissipation theorem})
delivers the integral representation}
\begin{subequations}
\label{eq:xi for any beta and omega}
\begin{align}
\chi_{\beta;uv,u'v'}^{\,}(\omega)=&\,
-\mathrm{i}
\int\limits_{-\infty}^{+\infty}
\frac{\mathrm{d}\omega_{\,}^{\prime}}{2\pi}\,
\left(
1
-
e^{-\beta\,\hbar\,\omega_{\,}^{\prime}}
\right)
\frac{1}{\hbar}\,
\left[
{\frac{\mathrm{PV}}{\omega-\omega_{\,}^{\prime}}}
-
\mathrm{i}\pi
\delta(\omega-\omega_{\,}^{\prime})
\right]\,
\omega'
\nonumber\\
&\,\times
\int\limits_{-\infty}^{+\infty}
\mathrm{d}t'\,
e^{+\mathrm{i}\omega'\,t'}\,
\left\langle
\left(
\hat{\phi}_{u\,\mathrm{H}}^{\,}
-
\hat{\phi}_{v\,\mathrm{H}}^{\,}
\right)(t')
\left(
\hat{\phi}_{u'\,\mathrm{H}}^{\,}
-
\hat{\phi}_{v'\,\mathrm{H}}^{\,}
\right)(0)
\right\rangle^{\,}_{0\,\beta}
\label{eq:xi for any beta and omega a}
\end{align}
{for the retarded response function,
the four-terminal current-charge susceptibility,
that quantifies the change}
\begin{equation}
\left\langle
\Big[
\hat n_u(\omega)
-
\hat n_v(\omega)
\Big]
\right\rangle_{0\,\beta}=
\sum_{u',v'}
\chi_{\beta;uv,u'v'}^{\,}(\omega)
\left[
I^{\mathrm{ext}}_{u'}(\omega)
-
I^{\mathrm{ext}}_{v'}(\omega)
\right]
\label{eq:xi for any beta and omega b}
\end{equation}
\end{subequations}
in the thermal expectation value
of
$
\hat n_u(\omega)
-
\hat n_v(\omega)
$
of the unperturbed system to leading order in perturbation theory.
In the DC limit $\omega\to0$,
Eq.\ (\ref{eq:xi for any beta and omega a})
simplifies to
\begin{align}
&
\lim_{\omega\to0}\,
\chi_{\beta;uv,u'v'}(\omega)=
+
\frac{\mathrm{i}}{\hbar}\,
\int\limits_{-\infty}^{+\infty}
\frac{\mathrm{d}\omega_{\,}^{\prime}}{2\pi}\,
\left(
1
-
e^{-\beta\,\hbar\,\omega_{\,}^{\prime}}
\right)\,
\int\limits_{-\infty}^{+\infty}
\mathrm{d}t'\,
e^{+\mathrm{i}\omega'\,t'}\,
\left\langle
\left(
\hat{\phi}_{u\,\mathrm{H}}^{\,}
-
\hat{\phi}_{v\,\mathrm{H}}^{\,}
\right)(t')
\left(
\hat{\phi}_{u'\,\mathrm{H}}^{\,}
-
\hat{\phi}_{v'\,\mathrm{H}}^{\,}
\right)(0)
\right\rangle^{\,}_{0\,\beta}.
\label{eq:DC limit for the retarded Green fct entering Kubo}
\end{align}
If the zero temperature limit is taken after the DC limit,
we can safely replace the multiplicative factor
$e^{-\beta\,\hbar\,\omega_{\,}^{\prime}}$ in the parenthesis
on the right-hand side of Eq.\
(\ref{eq:DC limit for the retarded Green fct entering Kubo})
by zero owing to the fact that the spectral function
for the pair of operators
$
\left(
\hat{\phi}_{u\,\mathrm{H}}^{\,}
-
\hat{\phi}_{v\,\mathrm{H}}^{\,}
\right)
$
and
$
\left(
\hat{\phi}_{u'\,\mathrm{H}}^{\,}
-
\hat{\phi}_{v'\,\mathrm{H}}^{\,}
\right)
$
in the frequency domain
vanishes for negative frequencies
[see Eq.\
(\ref{eq:J0betaABomega vanishes for megative omege if beta is infinity})]. 
Hence,
the DC limit of the retarded correlation function in the frequency domain
at zero temperature is the trace over the ground states given by
(we can drop the reference to the Heisenberg time evolution at equal times)
\begin{subequations}
\label{eq:DC followed T=0 limits for the retarded Green fct entering Kubo}
\begin{equation}
\lim_{\beta\to\infty}\,
\lim_{\omega\to0}\,
\chi_{\beta;uv,u'v'}(\omega)=
+
\frac{\mathrm{i}}{\hbar}\,
\lim_{\beta\rightarrow\infty}
\left\langle
\left(
\hat{\phi}_{u}^{\,}
-
\hat{\phi}_{v}^{\,}
\right)(0)
\left(
\hat{\phi}_{u'}^{\,}
-
\hat{\phi}_{v'}^{\,}
\right)(0)
\right\rangle^{\,}_{0\,\beta}.
\label{eq:DC followed T=0 limits for the retarded Green fct entering Kubo a}
\end{equation}
This corresponds to a phase-difference correlation function, a direct
probe of geometrical properties of the underlying graph.  In a free
theory, where the cosine potential can be linearized, this quantity
corresponds to the inverse of adjacency matrix $A$.  {The} correction
induced by a nonvanishing temperature is
\begin{equation}
\lim_{\omega\to0}
\chi_{\beta;uv,u'v'}(\omega)
-
\lim_{\beta\to\infty}\,
\lim_{\omega\to0}\,
\chi_{\beta;uv,u'v'}(\omega)=
S_{uv,u'v'}(\beta\,\hbar),
\label{eq:DC followed T=0 limits for the retarded Green fct entering Kubo b}
\end{equation}
where
\begin{equation}
S_{uv,u'v'}(\beta\,\hbar):=
-
\frac{\mathrm{i}}{\hbar}\,
\int\limits_{-\infty}^{+\infty}
\frac{\mathrm{d}\omega'}{2\pi}
e^{-\beta\,\hbar\,\omega'}\,
\int\limits_{-\infty}^{+\infty}
\mathrm{d}t'\,
e^{+\mathrm{i}\omega'\,t'}\,
\left\langle
\left(
\hat{\phi}_{u\,\mathrm{H}}^{\,}
-
\hat{\phi}_{v\,\mathrm{H}}^{\,}
\right)(t')
\left(
\hat{\phi}_{u'\,\mathrm{H}}^{\,}
-
\hat{\phi}_{v'\,\mathrm{H}}^{\,}
\right)(0)
\right\rangle^{\,}_{0\,\beta},
\label{eq:DC followed T=0 limits for the retarded Green fct entering Kubo c}
\end{equation}
\end{subequations}
is the Laplace transform at the time $\hbar/(k_{\mathrm{B}}\,T)$ of
the spectral function for the pair of operators
$
\left(
\hat{\phi}_{u\,\mathrm{H}}^{\,}
-
\hat{\phi}_{v\,\mathrm{H}}^{\,}
\right)
$
and
$
\left(
\hat{\phi}_{u'\,\mathrm{H}}^{\,}
-
\hat{\phi}_{v'\,\mathrm{H}}^{\,}
\right)
$
in the frequency domain.
Equation
(\ref{eq:DC followed T=0 limits for the retarded Green fct entering Kubo})
is the main result of this section.
The DC limit followed by the zero-temperature limit
of the Kubo formula
(\ref{eq:xi for any beta and omega b})
that is relevant to the setup of Fig.\ \ref{Fig:shea-PRL-97},
yields the equal-time two-point correlation function of
the superconducting phase operator $\hat{\phi}_{u}^{\,}$.
\end{widetext}

The simplest incarnation of the quantum Josephson array defined
in Appendix \ref{subsec:Definition of the quantum Josephson array}
has two characteristic energy scales,
the charging energy $E^{(\mathrm{C})}$
and the Josephson coupling $E^{(\mathrm{J})}$.
We are interested in the following three regimes of temperature.
In the regime of temperature
\begin{equation}
k_{\mathrm{B}}\,T\ll
E^{(\mathrm{C})}\ll
E^{(\mathrm{J})},
\end{equation}
we may approximate
the DC limit of the retarded correlation function in the frequency domain
(\ref{eq:DC limit for the retarded Green fct entering Kubo})
by the equal-time quantum statistical average
(\ref{eq:DC followed T=0 limits for the retarded Green fct entering Kubo a}).
In the regime of temperature
\begin{equation}
E^{(\mathrm{C})}\ll
k_{\mathrm{B}}\,T\ll
E^{(\mathrm{J})},
\end{equation}
we may approximate
the DC limit of the retarded correlation function in the frequency domain
(\ref{eq:DC limit for the retarded Green fct entering Kubo})
by the classical average at temperature $T$ with the Hamiltonian
in which $E^{(\mathrm{C})}$ is substituted by 0 and $E^{(\mathrm{J})}$
has been renormalized by quantum fluctuations.
In this classical limit of vanishing charging energies,
the Hamiltonian reduces to that of the renormalized classical
$XY$ model on the bipartite simple graph $G=(V,E)$,
where $V$ is the set of vertices and $E$ encodes
the nonzero Josephson couplings.
In the regime of large temperature
\begin{equation}
E^{(\mathrm{J})}\ll
k_{\mathrm{B}}\,T,
\end{equation}
the correction
(\ref{eq:DC followed T=0 limits for the retarded Green fct entering Kubo b})
is sizable and cannot be neglected.

\section{Conclusion and Outlook}

We introduced a framework to emulate graphs, and through them curved
spaces in arbitrary dimensions using {arrays of}
superconducting wire{s coupled by Josephson junctions}. In
particular, we showed that any triangulation of a Riemannian manifold, viewed as
a simple graph, can be implemented by assigning a superconducting wire
with a rigid phase to each vertex and coupling pairs of wires via
Josephson junctions whenever the corresponding vertices share an edge
in the triangulation. This construction naturally supports scalar
field theories on the resulting curved geometries, with the
superconducting phase serving as the scalar field.

Furthermore, we established a conservative upper bound on the system size,
$2N_{\star}=6.50\times 10^4$, under which phase rigidity can be
safely assumed, yielding a sufficient condition to represent each vertex by a
superconducting wire. When applied to the realization of massive
scalar theories, the bound in fact diverges.

As an application, we proposed array{s} of superconducting wires
as an experimentally accessible platform
to probe AdS$_{d+1}$/CFT$_d$
holographic duality in a controlled setting for arbitrary dimension
$d$. We presented in detail the particular implementation of a massive
scalar field theory on hyperbolic space mimicking AdS$_{2}$ and
studied the effects of {quenched} disorder in the Josephson couplings,
which manifest {themselves} as distortions {of the metric}.
The impact of this {quenched} disorder on the
boundary scaling exponents was examined both analytically,
via replica {random-phase approximation},
and numerically, 
showing that holographic duality remains robust even in the presence
of strong disorder. Finally, we argued that probing correlation
functions in such exotic geometries is equivalent to measuring
linear-response current-voltage characteristics in the array.

We now discuss several open directions. The first is conceptually
straightforward, namely to exploit the connectivity of the wire networks
considered here in other types of circuits, most notably classical
circuits. In that setting, rather than realizing scalar field theories
through superconducting phases, one could use the electric potentials
(voltages) on the wires as the scalar fields. Indeed, by adapting
classical-circuit platforms such as those considered in
Refs.~\cite{Lenggenhager22,Chen2023hyperbolic,Dey2024simulating}, and replacing
literal tessellations of hyperbolic space on the Poincar\'e disk with
the wiring architecture proposed here, one could realize a much
broader class of geometries and topologies, extending beyond
hyperbolic space in two dimensions to higher dimensions,
and more generally to arbitrary graph-based manifolds.

The second concerns the role of quantum fluctuations in the
superconducting wire networks. Throughout this work, we focused on the
classical regime, where the Josephson energy $E^{(\mathrm{J})}_{u,v}$
dominates over the charging energy $E^{(C)}_{u,v}$. It would be of
considerable interest to investigate the full quantum regime, in which
these energy scales are comparable, and to understand its consequences
for the applications explored in this paper.

Third, because superconducting wire arrays can realize essentially
any simple graph structures,
they provide a versatile platform for
engineering unconventional geometries in the laboratory. This opens
the possibility of experimentally studying systems defined on highly
nontrivial graphs, including Cayley trees, hyperbolic lattices, and
fractal structures, all within a unified and experimentally accessible
framework. Beyond the examples considered in this work, such
architectures may enable the exploration of a broad range of
graph-based phenomena in condensed matter physics, quantum
information, and statistical mechanics.

Fourth, while this work focused on static geometries, it is natural to
ask what lies beyond them. One intriguing possibility is to use
electrically measured scalar-field fluctuations as feedback signals to
modulate the Josephson couplings, thereby dynamically reshaping the
geometry itself. Such a matter-metric coupling would amount to an
artificial realization of ``dynamical gravity,'' even if not one obeying
the Einstein-Hilbert equations. Still, the prospect of studying such
systems experimentally may provide valuable insight into how scalar
fields can affect the geometry of the spaces they inhabit. Going a
step further, if the measured fluctuations could be digitized and
processed before being fed back into the Josephson couplings, one
might envision engineering effective dynamics that approach genuine
gravitational equations. While speculative, these are precisely the
kinds of directions we hope this work will help open.

%%%%%%%%%%%%%%%%%%%%%%%%%%%%%%%%%%%%%%%%%%%%%%%%%%%%%%%%%%%%%%%%%%%%%%%%%%%%%%%
%%%%%%%%%%%%%%%%%%%%%%%%%%%%%%%%%%%%%%%%%%%%%%%%%%%%%%%%%%%%%%%%%%%%%%%%%%%%%%%
\acknowledgments

We are grateful to Richard C. Brower and Emanuel Katz for valuable discussions.
%The numerical computations reported here were performed on the Boston University Shared Computing Cluster (SCC).
The work of M.D., G.D., J.O., A.P.H., and C.C.,  is supported by the DOE Grant DE-SC0026189. 

%%%%%%%%%%%%%%%%%%%%%%%%%%%%%%%%%%%%%%%%%%%%%%%%%%%%%%%%%%%%%%%%%%%%%%%%%%%%%%%
%%%%%%%%%%%%%%%%%%%%%%%%%%%%%%%%%%%%%%%%%%%%%%%%%%%%%%%%%%%%%%%%%%%%%%%%%%%%%%%

\appendix

\section{Quantum Josephson arrays}
\label{subsec:Definition of the quantum Josephson array}

When two metallic regions are separated by a potential barrier,
quantum tunneling enables single-electron exchange across the
interface. Josephson predicted, within the Bardeen-Cooper-Schrieffer
(BCS) theory of superconductivity, that an analogous process occurs
when two superconductors are similarly separated.
Quantum tunneling permits the exchange of Cooper pairs across the
barrier~\cite{josephson1962possible}. Applying a potential difference
across such a junction, known as a Josephson junction, drives a
dissipationless Cooper-pair current, a phenomenon called the Josephson
effect~\cite{josephson1962possible,Anderson63,Ambegaokar63}.

Arrays of Josephson junctions have been realized experimentally in two
broad classes: artificially fabricated tunnel junctions
\cite{Voss82,Webb83,Vanwees87} and proximity-induced arrays
where a normal metal serves as the weak link
\cite{Resnik81,Abraham82,Tinkham83,Kimhi84,Leemann86,Brown86,Martinoli87}.
In both cases, Josephson couplings are restricted to nearest-neighbor
pairs of superconducting islands, and the arrays realize the
two-dimensional classical XY model, undergoing the
Berezinskii-Kosterlitz-Thouless (BKT) transition at temperature
$T_{BKT}\ll T_{BCS}$. Truly quantum arrays (in which charging energy
drives a superconductor-to-insulator quantum phase transition at zero
temperature) were achieved only later
\cite{Geerligs89,Chen92,Vanderzant92,Vanderzant96}.
A qualitatively different geometry was proposed by Vinokur et al.\
\cite{Vinokur87} and studied by Sohn et al.\
\cite{Sohn93a,Sohn93b,Shea97}. {T}wo orthogonal sets of $N$ parallel
superconducting wires, coupled by a Josephson junction at every
crossing, naturally realize long-range rather than nearest-neighbor
couplings.

Superconducting wires are labeled by the letter{s $u,u',v,v'$
and are assigned the pair of conjugate operators 
$(\hat{\phi}_{u}^{\,},\hat{n}_{u}^{\,})$
that obey the 
equal-time algebra }
\begin{subequations}
\label{eq:def Josephson array}
\begin{equation}
\left[
\hat{\phi}_{u}^{\,},\hat{n}_{u'}^{\,}
\right]=
\mathrm{i}\,\delta_{u,u'}^{\,},
\quad
\left[
\hat{\phi}_{u}^{\,},\hat{\phi}_{u'}^{\,}
\right]=
\left[
\hat{n}_{u}^{\,},\hat{n}_{u'}^{\,}
\right]=0.
\label{eq:def Josephson array a}
\end{equation}
The Hilbert space is the bosonic Fock space
\begin{equation}
{\mathfrak{F}_{\mathrm{b}}}:=
\mathrm{span}
\Big\{
\bigotimes\limits_{{u}}|n_{u}^{\,}\rangle
\Big|
n_{u}^{\,}=\mathbb{N},
\quad
\hat{n}_{u}^{\,}\,|n_{u}^{\,}\rangle=
n_{u}^{\,}\,|n_{u}^{\,}\rangle
\Big\}.
\label{eq:def Josephson array b}
\end{equation}
The phase operator
$\exp\left(+\mathrm{i}\hat{\phi}_{u}^{\,}\right)$
is not quite unitary, for it obeys the algebra
\begin{equation}
e^{-\mathrm{i}\hat{\phi}_{u}^{\,}}\,
\hat{n}_{u'}^{\,}\,
e^{+\mathrm{i}\hat{\phi}_{u}^{\,}}=
\left(
\hat{n}_{u'}^{\,}\,
+
\delta_{u,u'}^{\,}
\right)
\label{eq:def Josephson array c}
\end{equation}
on
\begin{equation}
\mathfrak{H}\setminus\{|0,\cdots,0\rangle\},
\label{eq:def Josephson array d}
\end{equation}
while
\begin{equation}
e^{+\mathrm{i}\hat{\phi}_{u}^{\,}}\,|n_{u}^{\,}=0\rangle=
|n_{u}^{\,}=1\rangle,
\qquad
e^{-\mathrm{i}\hat{\phi}_{u}^{\,}}\,|n_{u}^{\,}=0\rangle=
0.
\label{eq:def Josephson array e}
\end{equation}
The many-body quantum Hamiltonian for a Josephson array
acts on the {bosonic} Fock space ${\mathfrak{F}_{\mathrm{b}}}$
and is defined by two contributions
\begin{equation}
\widehat{H}_{\mathrm{C,J}}^{\,}:=
\widehat{H}_{\mathrm{C}}^{\,}
+
\widehat{H}_{\mathrm{J}}^{\,}.
\label{eq:def Josephson array f}
\end{equation}
The first governs the electrostatic interaction between charges on the wires
\begin{equation}
\begin{split}
&
\widehat{H}_{\mathrm{C}}^{\,}:=
\frac{1}{2}
\sum_{{u,u'}}
E_{u,u'}^{\mathrm{(C)}}
\left(
\hat{n}_{u}^{\,}
-
n_{u}^{\mathrm{(bg)}}
\right)\,
\left(
\hat{n}_{u'}^{\,}
-
n_{u'}^{\mathrm{(bg)}}
\right),
\end{split}
\label{eq:def Josephson array g}
\end{equation}
with symmetric charging energy
$E_{u,u'}^{\mathrm{(C)}}=E_{u',u}^{\mathrm{(C)}}\geq0$
and background charge $n_{u}^{\mathrm{(bg)}}\geq0 $.
here,
$E_{u,u'}^{\mathrm{(C)}}$ denotes the electrical energy stored between wires
{$u\neq v$}
and
$E_{u,u}^{\mathrm{(C)}}$ 
that stored within wire
{$u$}. 
The second {contribution}
encodes {the} Josephson interactions
\begin{equation}
\begin{split}
&
\widehat{H}_{\mathrm{J}}^{\,}:=
\frac{1}{2}\,
\sum_{u,u'}
E_{u,u'}^{\mathrm{(J)}}
\left[
1
-
\cos
\left(
\hat{\phi}_{u}^{\,}
-
\hat{\phi}_{u'}^{\,}
-
A_{u,u'}^{\mathrm{(bg)}}
\right)
\right],
\end{split}
\label{eq:def Josephson array h}
\end{equation}
\end{subequations}
with Josephson couplings $E_{u,u'}^{\mathrm{(J)}}=
E_{u',u}^{\mathrm{(J)}}{\geq0}$
nonzero only when wires
$u$ and $u'$ cross.
As a single wire is not allowed to intersect itself
$
E_{u,u}^{\mathrm{(J)}}=0$.
{Finally,}
$A_{u,u'}^{\mathrm{(bg)}}=
-
A_{u',u}^{\mathrm{(bg)}}\in[0,2\pi)$
correspond to antisymmetric background gauge fields. 

{There are two classical limits. }
The first {limit} is set by fixing the Josephson couplings
$E_{u,u'}^{\mathrm{(J)}}=0$ {for any pair $u,u'$ of wires,}
{in which case the ground state is insulating
when all the background charges are set to zero}
due to {unfrustrated two-body} strong repulsions. The second
{limit is set by fixing}
all charging energies to zero
$E_{u,u'}^{\mathrm{(C)}}=0$ {for any pair $u,u'$ of wires,}
in which case we recover a classical XY model {
whose ground state is the ferromagnetic state when all background gauge
fields are set to zero.}

The only symmetry of the many-body quantum Hamiltonian
for all values of  couplings
is the consequence of the invariance of the algebra
(\ref{eq:def Josephson array a})
and Hamiltonian
(\ref{eq:def Josephson array f})
under the global $\mathrm{U}(1)$ transformation
\begin{equation}
\hat{\phi}_{u}^{\,}=
\hat{\phi}_{u}^{\prime}
+
\alpha,
\qquad
\hat{n}_{u}^{\,}=
\hat{n}_{u}^{\prime},
\label{eq:U(1) symmetry}
\end{equation}
for any choice of the number $\alpha\in[0,2\pi)$
{and for any wire $u$}.
This global $\mathrm{U}(1)$ symmetry implies
the {local} continuity equation
\begin{subequations}
\label{eq:continuity equation}
\begin{align}
&
\frac{\partial}{\partial t}
\hat{n}_{u}^{\,}=
\sum_{{v}}
\widehat{J}_{v,u}^{\,},
\label{eq:continuity equation a}
\\
&
\widehat{J}_{u,v}^{\,}:=
-
\frac{E_{u,v}^{\mathrm{(J)}}}{\hbar}\,
\sin
\left(
\hat{\phi}_{u}^{\,}
-
\hat{\phi}_{v}^{\,}
-
A_{u,v}^{\mathrm{(bg)}}
\right)=
-
\widehat{J}_{v,u}^{\,}.
\label{eq:continuity equation b}
\end{align}
\end{subequations}
Observe that the fact that
the particle current $\widehat{J}_{u,v}^{\,}$ from $v$ to $u$
is minus the particle current $\widehat{J}_{v,u}^{\,}$
between $u$ to $v$
implies that the global number operator
\begin{equation}
\widehat{N}:=
\sum_{{v}}
\hat{n}_{v}^{\,}
\end{equation}
is time independent.

{For any wire $u$,} the non-linear coupled equations of motions are 
\begin{subequations}
\label{eq:exact coupled equations of motion}
\begin{align}
\frac{\partial}{\partial t}
\hat{n}_{u}^{\,}\equiv&\,
\frac{1}{\mathrm{i}\hbar}\,
\left[
\hat{n}_{u}^{\,},
\widehat{H}_{\mathrm{C,J}}^{\,}
\right]
\nonumber\\
=&\,
+
\frac{1}{\hbar}\,
\sum_{{v}}
E_{u,v}^{\mathrm{(J)}}
\sin
\left(
\hat{\phi}_{v}^{\,}
-
\hat{\phi}_{u}^{\,}
-
A_{v,u}^{\mathrm{(bg)}}
\right),
\label{eq:exact coupled equations of motion a}
\\
\frac{\partial}{\partial t}
\hat{\phi}_{u}^{\,}\equiv&\,
\frac{1}{\mathrm{i}\hbar}\,
\left[
\hat{\phi}_{u}^{\,},
\widehat{H}_{\mathrm{C,J}}^{\,}
\right]
\nonumber\\
=&\,
+
\frac{1}{\hbar}\,
\sum_{{v}}
E_{u,v}^{\mathrm{(C)}}\,
\left(
\hat{n}_{v}^{\,}
-
n_{v}^{\mathrm{(bg)}}
\right).
\label{eq:exact coupled equations of motion b}
\end{align}
\end{subequations}
On the one hand,
any state such that the expectation value in this state
of the right-hand side of
Eq.\ (\ref{eq:exact coupled equations of motion a})
[Eq.\ (\ref{eq:exact coupled equations of motion b})]
is nonvanishing and constant in time implies
a strongly fluctating expectation value of the local number operator
(the local phase operator).
On the other hand,
any state such that the expectation value in this state
of the local number operator (the local phase operator)
is constant in time implies no net particle current
flowing through any site 
(the expectation value in this state of the local number operator is the
local background charge).

\section{Proof of the upper bounds on the phase differences}

\subsection{Proof of the upper bound (\ref{eq:upper bound})}
\label{appssubsec:Proof of the upper bound A}

We consider the energy
(\ref{eq:def Josephson array on square lattice e})
in the spin-wave approximation, i.e.,
we expand the arguments of the cosine of the classical ferromagnetic
XY model to second order in its argument,
\begin{equation}
\begin{split}
E_{\Lambda,G}(J):=&\,
\frac{J}{2}
\sum_{i=1}^{N_1}
\sum_{j=1}^{N_2}
b_{i,j}^{\,}
\left(
\phi_{i,j}^{\,}
-
\varphi_{i,j}^{\,}
\right)^{2}
\\
&\,
+
\frac{J_{\mathrm{w}}^{\,}}{2}
\sum_{i=1}^{N_1}
\sum_{j=1}^{N_2}
\left(
\phi_{i,j}^{\,}
-
\phi_{i+1,j}^{\,}
\right)^{2}
\\
&\,
+
\frac{J_{\mathrm{w}}^{\,}}{2}
\sum_{i=1}^{N_1}
\sum_{j=1}^{N_2}
\left(
\varphi_{i,j}^{\,}
-
\varphi_{i,j+1}^{\,}
\right)^{2}
\\&\,
+
\frac{m^{2}}{2} \sum_{i=1}^{N_1}
\sum_{j=1}^{N_2}
\left(\phi^{2}_{i,j}+\varphi^{2}_{i,j}\right),
\end{split}
\label{eq: energy-general-matrix}
\end{equation}
where $b_{i{,}j}$ is a  
matrix {taking values 0 or 1}
and $J_{\mathrm{w}^{\,}}^{\,}$ is
fixed to an arbitrary value.
We also added mass regulators.

The energy (\ref{eq: energy-general-matrix}) is strictly positive
provided there exists at least one pair $i,j$ with
$i=1,\ldots,N_1$ and $j=1,\ldots,N_2$ such that
$\phi^{2}_{i,j}+\varphi^{2}_{i,j}>0$.
If we introduce the real-valued vector
$\bm{X}\in\mathbb{R}^{2N_1\,2N_2}$
whose entries are the superconductig phases $\phi$ and $\phi$
indexed by the pair of indices
$i=1,\ldots,N_1$ and $j=1,\ldots,N_2$,
we can write the energy (\ref{eq: energy-general-matrix})
as the positive definite quadratic form
\begin{equation}
E_{\Lambda,G}(J)=
\bm{X}^{\mathsf{T}}\,K(J)\,\bm{X}
\end{equation}
with $K(J)$ a
$(4N_1\,N_2)\times(4N_1\,N_2)$
real-valued symmetric matrix.
By inspection, we observe that
\begin{equation}
J'>J \ \Longrightarrow\ 
\bm{X}^{\mathsf T}\big[K(J')-K(J)\big]\bm{X}\geq 0,
\label{psd-ordering}
\end{equation}
from which follows \cite{horn2013matrix} that
\begin{equation}
J'>J \ \Longrightarrow\ 
\bm{X}^{\mathsf T}\big[K(J')^{-1} -K(J)^{-1}\big]\bm{X}\leq 0.
\end{equation}
The choice 
\begin{equation}
\bm{X}=(0,\cdots, 0,\phi_{i,j},0,\cdots,0,-\phi_{i',j},0,\cdots, 0)
\end{equation}
delivers the upper bound
\begin{align}
\left\langle
\left(
\phi_{i,j}^{\,}
-
\phi_{i',j}^{\,}
\right)^{2}
\right\rangle\leq
\lim\limits_{J\to0}
\left\langle
\left(
\phi_{i,j}^{\,}
-
\phi_{i',j}^{\,}
\right)^{2}
\right\rangle.
\end{align}

\subsection{Proof of the upper bound 
(\ref{eq:condition on N for small phase fluctuations})}
\label{appssubsec:Proof of the upper bound B}

To put an estimate for the right-hand side of the inequality
(\ref{eq:upper bound a}), we use the spin-wave approximation as in
Sec.~\ref{appssubsec:Proof of the upper bound A}.
Expanding the arguments of the cosine of the
classical ferromagnetic nearest-neighbor $XY$ model along a chain made
of $M$ sites yields the partition function
\begin{subequations}
\label{eq:def partition function}
\begin{align}
\begin{split}
Z_{M}^{\mathrm{har}}(K_{\mathrm{w}}^{\,}):=&\,
\left[
\prod\limits_{j=1}^{M}
\int\limits_{\mathbb{R}}
\mathrm{d}\theta_{j}^{\,}
\right]
e^{
-
\frac{K_{\mathrm{w}}^{\,}}{2}
\sum\limits_{j=1}^{M}
\left(\theta_{j}^{\,}-\theta_{j+1}^{\,}\right)^{2}
  }
\\
=&\,
\left[
\prod\limits_{k\in\mathrm{BZ}}^{k\geq0}
\int\limits_{\mathbb{C}}
\frac{
\mathrm{d}\tilde{\theta}_{k}^{*}
\mathrm{d}\tilde{\theta}_{k}^{\,}
     }
     {
2\mathrm{i}
     }
\right]
e^{
-
K_{\mathrm{w}}^{\,}
\sum\limits_{k\in\mathrm{BZ}}\!
(1-\cos k)\,
\tilde{\theta}_{k}^{*}\,
\tilde{\theta}_{k}^{\,}
  },
\end{split}
\label{eq:def partition function a}
\end{align}
where
\begin{equation}
\tilde{\theta}_{k}^{\,}:=
\frac{1}{M^{1/2}}
\sum_{j=1}^{M}
e^{-\mathrm{i}k\,j}
\theta_{j}^{\,}=
\tilde{\theta}_{-k}^{*}
\end{equation}
and
\begin{equation}
\frac{M}{2\pi}\,k=
-\left\lfloor\frac{M}{2}\right\rfloor,
-\left\lfloor\frac{M}{2}\right\rfloor+1,
\ldots,
+\left\lfloor\frac{M}{2}\right\rfloor-1
\end{equation}
\end{subequations}
defines the Brillouin Zone (BZ) of the chain.
In this harmonic approximation,
we may replace the right-hand side of the upper bound
(\ref{eq:upper bound a})
by
\begin{align}
(\Delta\theta)^{2}(m):=&\,
\left\langle
\left(
\theta_{j}^{\,}
-
\theta_{j+m}^{\,}
\right)^{2}
\right\rangle
\nonumber\\
=&\,
\left\langle
\left(
\theta_{j}^{\,}\,
\theta_{j}^{\,}
+
\theta_{j+m}^{\,}\,
\theta_{j+m}^{\,}
-
2
\theta_{j}^{\,}\,
\theta_{j+m}^{\,}
\right)
\right\rangle
\nonumber\\
=&\,
2
\left\langle
\left(
\theta_{j}^{\,}\,
\theta_{j}^{\,}\,
-
\theta_{j}^{\,}\,
\theta_{j+m}^{\,}
\right)
\right\rangle
\nonumber\\
=&\,
\frac{2}{M}
\sum_{k',k\in\mathrm{BZ}}
e^{\mathrm{i}(k'+k)j}
\left(
1
-
e^{\mathrm{i}km}
\right)
\left\langle
\tilde{\theta}_{k'}^{\,}\,
\tilde{\theta}_{k}^{\,}
\right\rangle
\nonumber\\
=&\,
\frac{2}{M}
\sum_{k\in\mathrm{BZ}}^{k\neq0}
\frac{
1
-
e^{\mathrm{i}km}
     }
     {
K_{\mathrm{w}}^{\,}
(1-\cos k)
     }
\nonumber\\
=&\,
\frac{2\,|m|}{K_{\mathrm{w}}^{\,}}
\end{align}
for any $m=1,\ldots,M-1$.
If we demand that the end-to-end variance of the superconducting phase
along the linear chain is at least one order of magnitude smaller than
$\pi^{2}$, i.e.,
\begin{equation}
(\Delta\theta)^{2}(M)\leq10^{-1}\times\pi^{2}\approx1,
\end{equation}
then Eq.\ (\ref{eq:condition on N for small phase fluctuations})
follows.

\subsection{Proof of the upper bound 
(\ref{eq:condition for rigid phases of crosses})}
\label{appssubsec:Proof of the upper bound C}

Equation (\ref{eq:condition for rigid phases of crosses})
follows from replacing 
in the partition function
(\ref{eq:def partition function a})
the linear chain by a square lattice with lattice spacing $\mathfrak{a}$
and cardinality $N_1 N_2$.

We can also understand
Eq.\ (\ref{eq:condition for rigid phases of crosses})
at the level of the approximation by which
we replace the lattice action
\begin{equation}
S_{\mathrm{lat}}=
\frac{K_{\mathrm{w}}}{2}
\sum_{\langle ij\rangle}
\left(\phi_{i}-\phi_{j}\right)^{2},
\end{equation}
with the sum over all directed pair of nearest neighbors on the square lattice,
by the continuum action
\begin{equation}
S_{\mathrm{cont}}=
\frac{K_{\mathrm{w}}}{2}
\int\mathrm{d}^{2}\bm{r}\,
\left[
\left(\partial_{r_x}\phi\right)^{2}
+
\left(\partial_{r_y}\phi\right)^{2}
\right].
\label{eq:classical-continuum-action}
\end{equation}
The two point-function
\begin{equation}
\left\langle
\Big(
\phi(\bm{r})
-
\phi(\bm{r}')
\Big)^{2}
\right\rangle_{S_{\mathrm{cont}}}=
\frac{1}{\pi\,K_{\mathrm{w}}}
\ln\left|\frac{\bm{r}-\bm{r'}}{\mathfrak{a}}\right|
\end{equation}
with $\mathfrak{a}$ is the ultraviolet cutoff follows.
Condition (\ref{eq:condition for rigid phases of crosses})
follows from demanding that
\begin{equation}
\frac{1}{\pi\,K_{\mathrm{w}}}
\ln\left|\frac{L}{\mathfrak{a}}\right|\sim
10^{-1}\,\pi^{2}
\end{equation}
with $L$ the infrared cutoff.

The classical continuum dimensionless action
(\ref{eq:classical-continuum-action})
is to be compared to the quantum continuum dimensionless action
in imaginary time
\begin{subequations}
\begin{equation}
S_{\mathrm{qu}}^{1d}:=
\frac{1}{2}
\int\frac{\mathrm{d}\tau}{\hbar}
\int\frac{\mathrm{d}x}{\mathfrak{a}}
\left[
\frac{\hbar^{2}}{E^{(\mathrm{C})}}
\left(\partial_{\tau}\phi\right)^{2}  
+
E^{(\mathrm{J})}\,\mathfrak{a}^{2}\,
\left(\partial_{x}\phi\right)^{2}  
\right]
\end{equation}
that approximates at zero temperature
a linear array of superconducting
islands with the charging energy $E^{(\mathrm{C})}$
that are coupled by the nearest-neighbor Josephson coupling
\begin{equation}
E^{(\mathrm{J})}\equiv J_{\mathrm{w}}.
\end{equation}
\end{subequations}
If we do the change of variable
\begin{subequations}
\begin{equation}
\tau:=
\sqrt{\frac{\hbar^{2}}{E^{(\mathrm{C})}\,\,E^{(\mathrm{J})}\,\mathfrak{a}^{2}}}\,
y=
\frac{\hbar}{\sqrt{E^{(\mathrm{C})}\,\,E^{(\mathrm{J})}}\,\mathfrak{a}}\,
y\equiv
\frac{1}{\omega_{\mathrm{jp}}\,\mathfrak{a}}\,
y
\end{equation}
where we have introduced the ``Josephson plasma energy''
\begin{equation}
\hbar\,\omega_{\mathrm{jp}}:=
\sqrt{E^{(\mathrm{C})}\,\,E^{(\mathrm{J})}},
\label{eq:Josephson-plasma-energy}
\end{equation}
we may write
\begin{equation}
S_{\mathrm{qu}}^{1d}:=
\frac{1}{2}\,
\sqrt{\frac{E^{(\mathrm{J})}}{E^{(\mathrm{C})}}}\,
\int\mathrm{d}y
\int\mathrm{d}x
\left[
\left(\partial_{y}\phi\right)^{2}  
+
\left(\partial_{x}\phi\right)^{2}  
\right].
\label{eq:quantum-1d-continuum-action}
\end{equation}
\end{subequations}

If the classical dimensionless action
(\ref{eq:classical-continuum-action})
is to be identified with the quantum dimensionless action
(\ref{eq:quantum-1d-continuum-action}),
we must do the identification
\begin{equation}
\frac{E^{(\mathrm{J})}}{k_{\mathrm{B}}\,T}\equiv
\sqrt{\frac{E^{(\mathrm{J})}}{E^{(\mathrm{C})}}}
\ \Longleftrightarrow\
k_{\mathrm{B}}\,T\equiv\hbar\,\omega_{\mathrm{jp}}.
\label{eq:kBT-sim-hbar-plasma-frequency}
\end{equation}

\section{Tessellations of homogeneous spaces}
\label{Appendix subsec: Tessellations of homogeneous spaces}

The goal of this section is to introduce by way of  examples
regular tessellations of spaces of constant curvature, i.e.,
spherical, Euclidean, and hyperbolic spaces.  The main results that we
we shall review are 
(i) the enumeration of all regular tessellations by tuples
of positive integers called Schl\"afli symbols
\cite{Schlaefli52,Coxeter73}
and
(ii) the construction of these tessellations by means of their Coxeter groups
\cite{Coxeter31,Coxeter32,Coxeter34a,Coxeter34b,Coxeter35,Todd36,Witt41,Coxeter94,Bourbaki2006groupes,Grove96,Muehlherr26}.

Tessellations are coverings of spaces using one or
more geometrical shapes, the tessellating cells, such that there are
neither gaps nor overlaps between the tessellation cells.
Tessellations of two- and three-dimensional spaces are called tilings
and honeycombs, respectively.

The most familiar examples of tessellations are those of
two-dimensional Euclidean space, i.e., the flat plane. 
One may use a covering made up of uniform quadratic tiles
as in Fig.\
\ref{Fig:Tilings 3,6 4,4 6,3}, say.

In the following, we {
are going to treat} regular tessellations
of spaces with constant curvature, i.e., spherical, Euclidean, and
hyperbolic spaces. We will motivate this choice from a physical
perspective and describe their characteristic properties. The
so-called Schl\"afli symbols will be discussed as they offer a
compact way to enumerate and classify regular tessellations in any
dimension. We shall then study their Coxeter groups, the tools
needed to program the Josephson arrays of superconducting wires
on the simple graph that tessellates
homogeneous space.

\begin{figure}[t!]
\centering
\begin{minipage}[t]{0.3\linewidth}
\raggedright
(a)\\
\vspace{2mm} 
\hspace{2mm}
\includegraphics[width=0.8\linewidth]{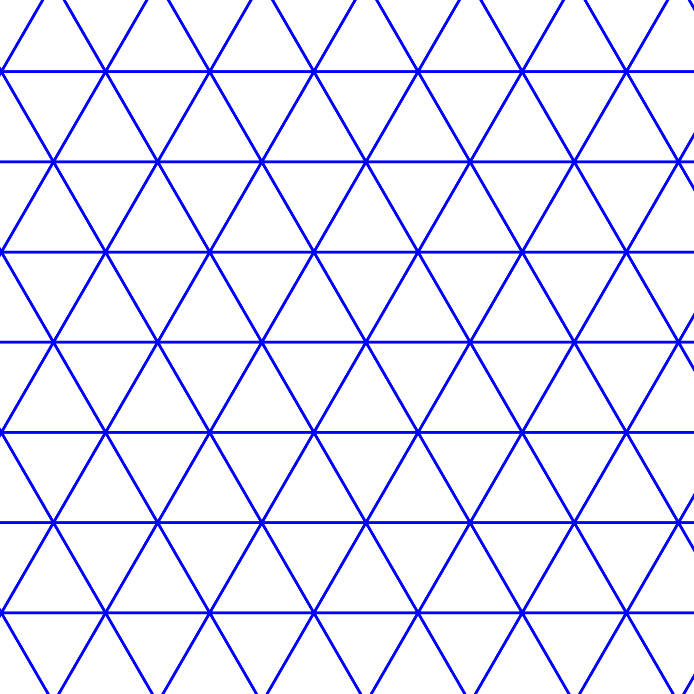}
\end{minipage}
\hfill
\begin{minipage}[t]{0.3\linewidth}
\raggedright
(b)\\
\vspace{2mm} 
\hspace{2mm}
\includegraphics[width=0.8\linewidth]{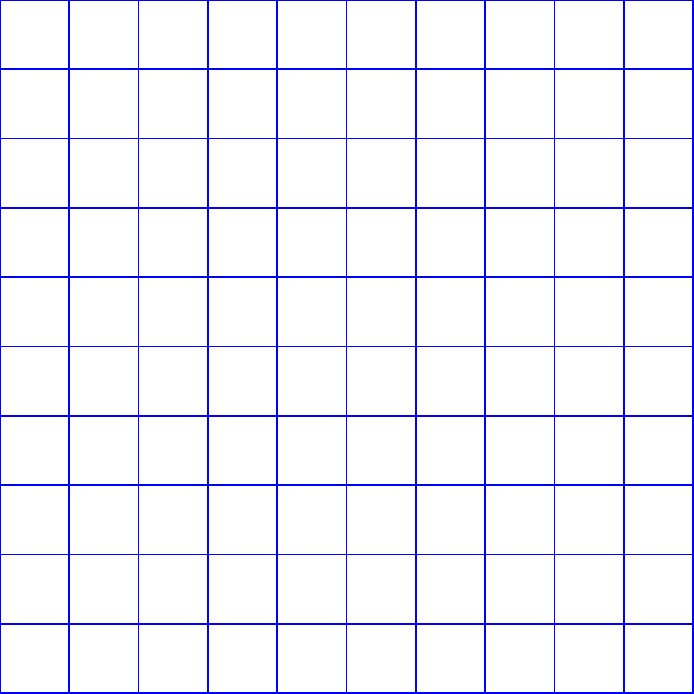}
\end{minipage}
\hfill
\begin{minipage}[t]{0.3\linewidth}
\raggedright
(c)\\
\vspace{2mm} 
\hspace{2mm}
\includegraphics[width=0.8\linewidth]{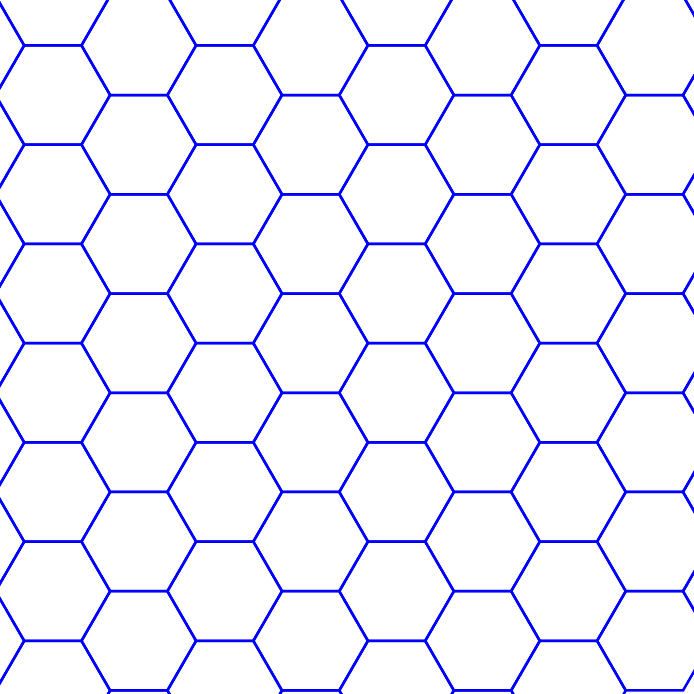}
\end{minipage}
\caption{\label{Fig:Tilings 3,6 4,4 6,3}
(a) Triangular tiling $\{3,6\}$,
(b) Quadratic tiling $\{4,4\}$,
(c) Hexagonal tiling $\{6,3\}$.
        }
\end{figure}

\subsubsection{Regular tessellations of spherical, Euclidean, and hyperbolic spaces}
\label{subsec:Regular tessellations of Euclidean Hyperbolic space}

Euclidean space is a set of points such that for any two points, there necessarily exists a translation (rotation) under which one point is the image by translation (rotation) of the other point.
Because of this property, the action of translations and rotations
on Euclidean space is transitive or, equivalently,
Euclidean space is homogeneous under the ``group action'' of
translations, rotations, and reflections.
These operations correspond exactly to the set of isometries
of Euclidean space, the so-called Euclidean group $\mathrm{E}(3)$.
By constraining Euclidean space, i.e.,
considering submanifolds thereof, we can create new spaces.
A new metric is gained by restricting the Euclidean metric
to the submanifold, i.e., by taking the pullback.

For example, one can take the subset $\sf{S}^{d}$ consisting of unit
vectors in $(d+1)$-dimensional Euclidean space.
This means, for
vectors $v\in\mathbb{R}^{d+1}$, the condition
\begin{equation}
v^{2}_{1}+v^{2}_{2}+\cdots+v^{2}_{d}+v^{2}_{d+1}=1,
\label{eq:def d-sphere}
\end{equation}
must hold. The pullback of the $(d+1)$-dimensional Euclidean metric
onto this set yields the so-called spherical metric.
The set of unit vectors from $\mathbb{R}^{d+1}$ define the
$d$-dimensional spherical space $\mathbb{S}^{d}$
with sectional curvature $+1$.
Now, the group of orthonormal transformations
$\mathrm{O}(d+1)$ that leaves condition
(\ref{eq:def d-sphere}) invariant and a fortiori
$\mathbb{S}^{d}$, act transitively on $\mathbb{S}^{d}$.

We may also consider submanifolds of non-Euclidean spaces.
Take $(d+1)$-dimensional Minkowski space
$\mathbb{M}^{d+1}\equiv\mathbb{R}^{d,1}$.
The hyperboloid is defined as the
upper sheet of vectors with Minkowski norm $-1$.
That means, for any
vector $v\in\mathbb{R}^{d,1}$,
the condition
\begin{equation}
-v^{2}_{1}+v^{2}_{2}+\cdots+v^{2}_{d}+v^{2}_{d+1}=-1
\end{equation}
must hold. The pullback of the Minkowski norm onto the hyperboloid
yields a metric of constant and negative sectional curvature. This is one way to define the $d$-dimensional hyperbolic space $\mathbb{H}^{d}$
with sectional curvature $-1$.
Its symmetry group is the pseudo-orthogonal group
$\mathrm{O}(1,d)$, which acts transitively on $\mathbb{H}^{d}$.

Regular tessellations are the ``most symmetric'' ones. Crucially, they
preserve the ambient space's property of homogeneity in the following
discrete sense. The tessellation's symmetry group (which is a discrete
subgroup of the ambient space's continuous symmetries) acts
transitively on all its geometric components (e.g., vertices, edges,
faces, etc.). As a consequence,
the field of view from any arbitrarily chosen vertex,
edge, face, etc.\ of the tessellation is the same, in that any
neighboring vertices, edges, faces, etc.\ can be obtained by the
action of the tessellation's symmetry group.

% \begin{figure}[t!]
% \begin{center}
% \includegraphics[width=0.5\linewidth]{FIGURES/FIG_RectangularTiling.png}
% \end{center}
% \caption{\label{Fig:RectangularTiling}
% Irregular tiling of the Euclidean plane using true rectangles. 
%          }
% \end{figure}

\subsubsection{Regular tilings}
In two dimensions, the definition of regular tilings reduces to the
following: A tiling is regular if
(1) there is only one type of tile,
(2) the tile is a regular polygon, and
(3) the tiles are edge-to-edge
(i.e., edges touch only edges, and vertices only vertices).

A tiling is a
tessellation of a two-dimensional space.
The tiling cell is itself a two-dimensional
geometrical object,
and in the regular case it must be a regular polygon.
A polygon is built out of edges,
which are one-dimensional geometrical objects. 
Edges meet at vertices, which are zero-dimensional geometrical objects.
Thus, a tiling is associated to a nested sequence of
geometrical objects of increasing dimensions: vertex, edge, tiling
cell.

In the case of honeycombs, space is three-dimensional. The
tessellating cell is three-dimensional, it has two-dimensional faces
meeting pairwise at one-dimensional edges, and zero-dimensional
vertices at which at least three faces are meeting.  The nested
sequence of geometrical objects associated to the honeycomb consists
of vertex, edge, face, and honeycomb cell.

Such nested sequences of geometric objects are called flags. As
dimensionality is increased, higher-dimensional objects are appended
to the sequence. The nomenclature for an $d$-dimensional polytope is:
vertex [$0$D], edge [$1$D], face [$2$D], cell [$3$D], ..., $j$-face
[$j$D], ..., peak [$(d-3)$D], ridge [$(d-2)$D], facet [$(d-1)$D], the
polytope itself [$d$D].

A $d$-dimensional tessellation is regular if its symmetry group acts
transitively on all of its flags. Since the symmetries of
tessellations are always isometries, this implies that (i) regular
tessellations are made up of only one type of cell, (ii) the
tessellation cell must be a regular $d$-dimensional polytope, and
(iii) polytopes must be arranged facet-to-facet.  We emphasize that
the tessellation polytope needs not be a compact subset of
$\mathbb{R}^{d}$.

\subsubsection{Regular polytopes as spherical tessellations}

Consider a polygon with $k$ corners. 
The circle that contains the polygon and touches each of its corners
is called its circumcircle.
Projecting the edges of the polygon outward onto the circumcircle
yields a tessellation of the $1$-sphere. If the polygon is regular,
then so is this spherical tessellation. One can also do the
converse. Consider a one-dimensional spherical space. It is always
isometric to the $1$-sphere $\mathbb{S}^{1}$
and can thus be represented as the unit circle in the
plane. Subdividing this circle into $k\geq3$ segments of equal length
realizes a regular tessellation.  Fix vertices at every meeting point
of different tessellation segments.  ``Straightening'' the circle arcs
between two consecutive vertices leads to a polygon with $k$
edges/corners.

The same exercise can be repeated in three dimensions. In this case,
we are either tessellating the $2$-sphere $\mathbb{S}^{2}$ or fixing a
circumsphere of some convex polyhedron. Hereto, we may identify
regular convex polyhedra with regular tessellations of
two-dimensional spherical space by projecting the circumscribing
$2$-sphere onto the inscribed convex polyhedron and vice et versa.
These regular polyhedra are known as the Platonic solids.  There
exist only five Platonic solids in three dimensions.

In any dimension $d$, every regular convex $d$-polytope (e.g.,
polygons and polyhedra) can be uniquely identified with some regular
tessellation of the $(d-1)$-dimensional spherical space, i.e., the
$(d-1)$-sphere $\mathbb{S}^{d-1}$.

\subsubsection{Schl\"afli symbols}
As we are going to review briefly, all regular tessellations of some
given $d$-dimensional space can be concisely denoted by some
ordered $d$-tuple
\begin{equation}
\{r^{\,}_{1}, r^{\,}_{2},
\cdots,r^{\,}_{i},\cdots,r^{\,}_{d}\}
\end{equation}
of integers $r^{\,}_{i}\geq3$.
These are called the Schl\"afli symbols
of a regular tessellation of some $d$-dimensional space.

Schl\"afli symbols are defined recursively, which is why we shall
introduce them successively in $1$, $2$, and $3$ dimensional spaces
before giving the general definition. This approach will also allow us
to derive constructively important bounds on the total number of
regular tessellations of Euclidean and spherical space.

%\subsubsection{Regular tessellations of one-dimensional spaces}

For one-dimensional spaces, the set of Schl\"afli symbols that we consider is labeled by the integers $p=3,4,5,\ldots$ and denoted by
\begin{equation}
\{p\}.
\end{equation}
The geometrical object associated to the Schl\"afli symbol
$\{p\}$ is the regular convex polygon with $p$ edges. 
As any regular polygon corresponds to a regular tessellation of the $1$-sphere 
$\mathbb{S}^{1}$, any Schl\"afli symbol $\{p\}$
corresponds to some tessellation of one-dimensional spherical space.
For example, the Sch\"afli symbol $\{3\}$
is recognized as an equilateral triangle,
while the  Sch\"afli symbol $\{6\}$ is recognized as a regular hexagon.

\begin{figure}[t!]
\centering

\begin{minipage}[t]{0.45\linewidth}
\raggedright
(a)\\
\includegraphics[width=\linewidth]{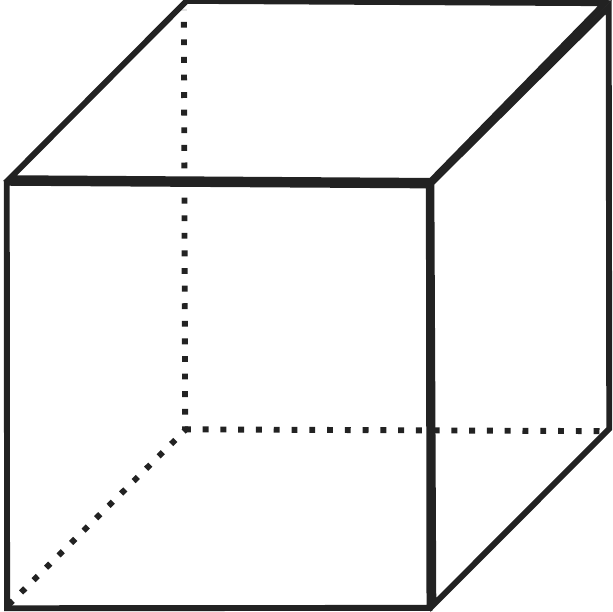}
\end{minipage}
\hfill
\begin{minipage}[t]{0.45\linewidth}
\raggedright
(b)\\
\includegraphics[width=\linewidth]{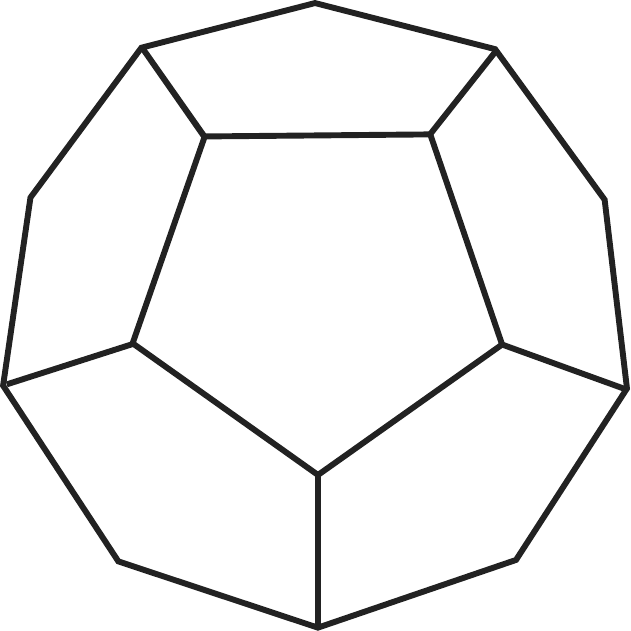}
\end{minipage}

\caption{\label{Fig: Tilings 4,3 and 5,3}
(a) The cube is a three-dimensional regular polyhedron made of
$V=8$, $E=12$, and $F=6$ vertices, edges, and faces, respectively.
The corresponding Schl\"afli symbol is $\{4,3\}$.
(b) The dodecahedron is a three-dimensional regular polyhedron made of
$V=20$, $E=30$, and $F=12$ vertices, edges, and faces, respectively.
The corresponding Schl\"afli symbol is $\{5,3\}$.
}
\end{figure}

%\subsubsection{Regular tessellations of two-dimensional spaces}

For two-dimensional spaces of constant sectional curvature, the set of
Schl\"afli symbols that we consider are labeled by two integers
$p,q=3,4,5,\ldots$ and denoted by
\begin{equation}
\{p,q\}.
\end{equation}
The geometrical object associated to the Schl\"afli symbol $\{p,q\}$
is constructed as follows. Start with $q$ copies of the polygon $\{p\}$.  Choose one vertex from
one of the $q$ polygons $\{p\}$. Attach to the vertex the $q-1$ remaining polygons so that (i) they all
share the vertex as a corner and (ii) all the $q$ edges meeting
at the  vertex are shared by exactly two polygons. This process
is iterated for each vertex that is not shared by the initial $q$
polygons until one receives a tiling that has at every vertex $q$
regular $p$-gons (regular polygons with $p$ edges) sharing a corner
and each edge is shared by exactly two polygons.  Observe here that
the Schl\"afli symbol $\{p,q\}$ living in two dimensions relies on the
object $\{p\}$ that lives in one dimension (as a tessellation). This is a first hint as to
the recursive nature of the Schl\"afli symbols.

The Schl\"afli symbols $\{3,6\}, \{4,4\}, \{6,3\}$ are all
tessellations of the flat plane, as can be verified by explicit
construction. They are shown in Figs.\
\ref{Fig:Tilings 3,6 4,4 6,3}(a),
\ref{Fig:Tilings 3,6 4,4 6,3}(b),
and \ref{Fig:Tilings 3,6 4,4 6,3}(c).
The geometrical object thus obtained is made of an
infinite but countable number of vertices, edges, and polygons. For
each of these three tilings, there exists a group made of
translations and point group transformations (i.e., generated by
rotations, inversions, and reflections) that act transitively on the
vertices, edge, and polygons of the tiling.

The Schl\"afli symbols $\{4,3\}$ and $\{5,3\}$ have two
interpretations.  On the one hand, they are 
regular tilings of a two-dimensional space with finite
numbers of vertices and edges.
The two-dimensional space in question is the two-dimensional
sphere that circumscribes the cube and dodecahedron, respectively,
very much in the same way as a regular polygon $\{p\}$ is a regular
tessellation of the circle. On the other hand, they can be thought of
as describing the cube and dodecahedron, respectively, which are
three-dimensional geometrical objects very much in the same way as a
regular polygon $\{p\}$ is also a two-dimensional geometrical
object. Indeed, any of the faces of the cube (dodecahedron) is a
regular polygon with $p=4$ ($p=5$) 
corners at which $q=3$
($q=3$) faces meet.  The cube (dodecahedron) is a three-dimensional
regular polyhedron made of $V=8\,(20)$, $E=12\,(30)$, and $F=6\,(12)$
vertices, edges, and faces, respectively, as shown in Fig.\ \ref{Fig:
  Tilings 4,3 and 5,3}. For each of these two tilings, there exists a
group made of rotations, inversions, and reflections that act
transitively on the vertices, edges, and polygons of the tiling.

Suppose that we start with the regular tiling $\{p,q\}=\{4,3\}$ of the
two-dimensional sphere and we increase $q$ holding $p=4$ fixed. We
encounter first the regular tiling $\{4,4\}$. This is a regular tiling
of $\mathbb{E}^{2}$ that can be interpreted as the square
lattice. What about the regular tiling $\{4,q\}$ with
$q=5,6,7,\ldots$? Increasing $q$ corresponds to ``squeezing'' additional
regular $4$-gons around each vertex. This squeezing can only be
accommodated if the two-dimensional homogeneous space being tiled
regularly has intrinsic negative curvature. Since this is done
at every vertex of the tessellation, the curvature is constant
everywhere. Thus, the tessellated space must be the hyperbolic
plane.  The value $q=4$ when $p=4$ can thus be thought of as a
critical value that separates spherical from hyperbolic geometry.
To visualize tilings of curved two-dimensional spaces on paper, we shall
use the Poincar\'e disk representation of the hyperbolic plane.  The
tilings $\{3,7\}$, $\{4,5\}$, and $\{6,4\}$ in
Figs.\
\ref{Fig:hyperbolic tilings 3,7 4,5 6.5}(a),
\ref{Fig:hyperbolic tilings 3,7 4,5 6.5}(b),
and \ref{Fig:hyperbolic tilings 3,7 4,5 6.5}(c)
are the hyperbolic extensions of the triangular, square, and hexagonal
tilings of the Euclidean plane in Figs.\
\ref{Fig:Tilings 3,6 4,4 6,3}(a),
\ref{Fig:Tilings 3,6 4,4 6,3}(b), and
\ref{Fig:Tilings 3,6 4,4 6,3}(c), respectively.

\begin{figure}[t!]
\centering

\begin{minipage}[t]{0.3\linewidth}
\raggedright
(a)\\[-1mm]
\hspace{2mm}
\includegraphics[width=\linewidth]{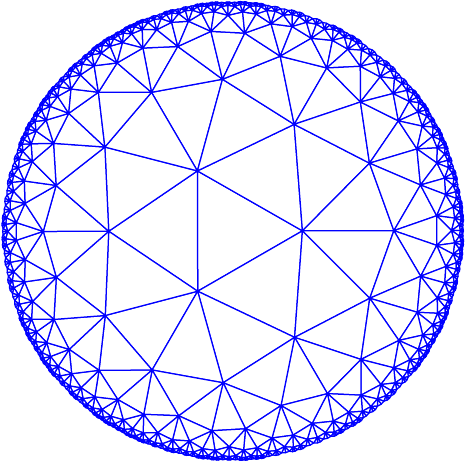}
\end{minipage}
\hfill
\begin{minipage}[t]{0.3\linewidth}
\raggedright
(b)\\[-1mm]
\hspace{2mm}
\includegraphics[width=\linewidth]{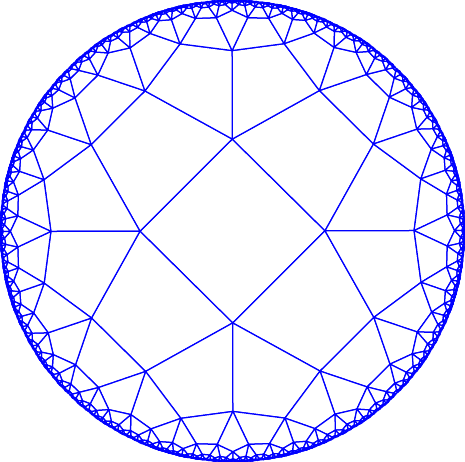}
\end{minipage}
\hfill
\begin{minipage}[t]{0.3\linewidth}
\raggedright
(c)\\[-1mm]
\hspace{2mm}
\includegraphics[width=\linewidth]{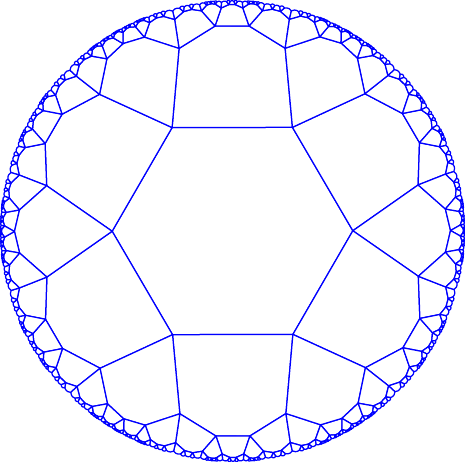}
\end{minipage}

\caption{\label{Fig:hyperbolic tilings 3,7 4,5 6.5}
(a) Hyperbolic tiling $\{3,7\}$,
(b) Hyperbolic tiling $\{4,5\}$,
(c) Hyperbolic tiling $\{6,4\}$.
}
\end{figure}

%\label{Proposition for d=2}
Let $\{p,q\}$ with $p,q=3,4,5,\ldots$
be an arbitrary Schl\"afli symbol. The corresponding tiling covers 
two dimensional Euclidean space ($\mathbb{E}^{2}$)
if and only if
\begin{subequations}
\label{eq:condition on p,q for tiling of plane, sphere, hyper}
\begin{equation}
\frac{1}{p}+\frac{1}{q}=\frac{1}{2}.
\label{eq:condition on p,q for tiling of plane, sphere, hyper a}
\end{equation}
If the rational number on the left-hand side
of Eq.\ (\ref{eq:condition on p,q for tiling of plane, sphere, hyper a})
is greater than $1/2$,
\begin{equation}
\frac{1}{p}+\frac{1}{q}>\frac{1}{2},
\label{eq:condition on p,q for tiling of plane, sphere, hyper b}
\end{equation}
then the tiling $\{p,q\}$ covers two-dimensional spherical space
($\mathbb{S}^{2}$).
If the rational number on the left-hand side
of Eq.\ (\ref{eq:condition on p,q for tiling of plane, sphere, hyper a})
is smaller than $1/2$,
\begin{equation}
\frac{1}{p}+\frac{1}{q}<\frac{1}{2},
\label{eq:condition on p,q for tiling of plane, sphere, hyper c}
\end{equation}
\end{subequations}
then the tiling $\{p,q\}$ covers
two-dimensional hyperbolic space ($\mathbb{H}^{2}$).

\begin{figure}[t!]
\begin{center},
\includegraphics[width=0.9\linewidth]{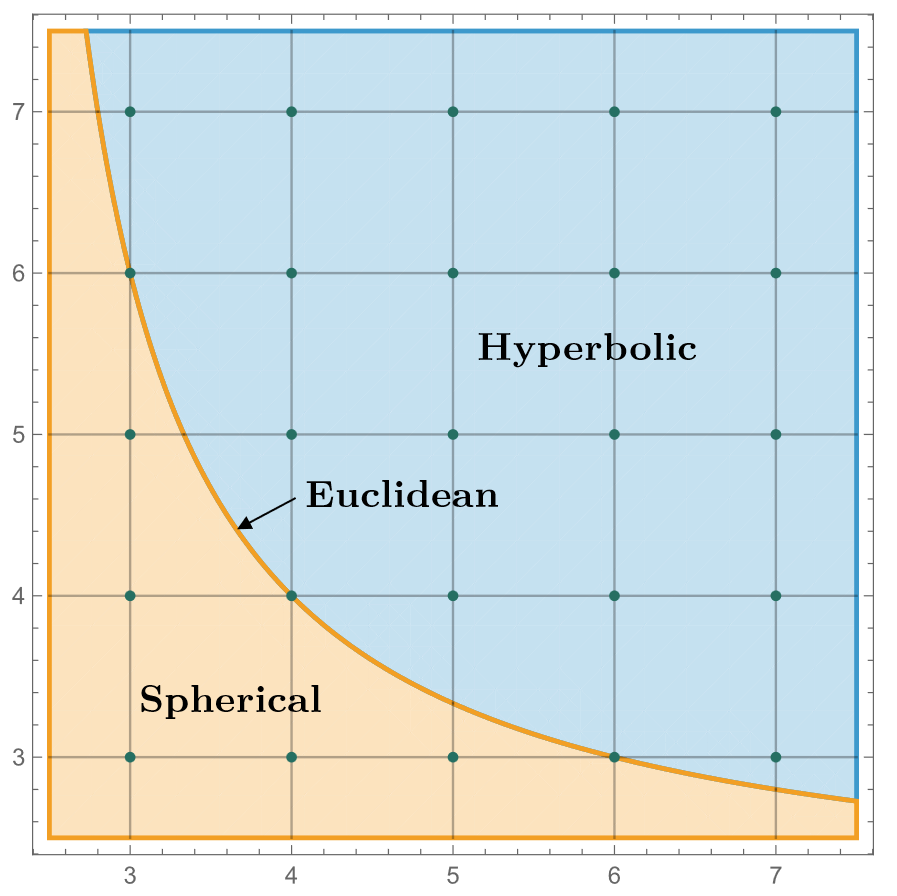}
\end{center}
\caption{\label{Fig:2dCurvatureCriterionPlot}
Region plot for the function, $\frac{1}{p}+\frac{1}{q}$.
Integer coordinates realize regular tilings of two-dimensional space
enumerated by the Schl\"afli symbols.
Region labels correspond to the curvature of the tiled space.
The interface of the orange and blue region corresponds to the Euclidean plane.
        }
\end{figure}

Application of criteria
(\ref{eq:condition on p,q for tiling of plane, sphere, hyper})
on the set of Schl\"afli symbols
$\{\, \{p,q\} \,|\, p,q=3,4,5,\ldots\}$
yields Fig.\ \ref{Fig:2dCurvatureCriterionPlot}, 
from which one infers the following.
\begin{enumerate}
\item
There are exactly 5 regular tilings of the two-dimensional sphere
$\mathbb{S}^{2}$. These are the 5 regular (convex) polyhedra, also
known as the Platonic solids. They are: $\{3,3\}$ (tetrahedron),
$\{4,3\}$ (cube), $\{3,4\}$ (octahedron), $\{5,3\}$ (dodecahedron),
and $\{3,5\}$ (icosahedron).
\item
There are exactly 3 regular tilings of the two-dimensional Euclidean
plane $\mathbb{E}^{2}$. They are: $\{6,3\}$ (triangular tiling),
$\{4,4\}$ (square tiling), and $\{3,6\}$ (hexagonal tiling).
\item
There are infinitely many regular tilings of the two-dimensional
hyperbolic plane $\mathbb{H}^{2}$.
The Python package \textit{Hypertiling}~\cite{schrauth_2025_16777106}
provides efficient tools for constructing and manipulating two-dimensional
\{p,q\} tessellations.
\end{enumerate}

%\subsubsection{Regular tessellations of three-dimensional spaces}

For three-dimensional homogeneous spaces,
the set of Schl\"afli symbols that we consider are
labeled by three integers $p,q,r=3,4,5,\ldots$ and denoted by
\begin{equation}
\{p,q,r\}.
\end{equation}
The geometrical object associated to the Schl\"afli symbol $\{p,q,r\}$
is constructed recursively as follows.
For some three-dimensional Schl\"afli symbol $\{p,q,r\}$, we take $r$
copies of the object $\{p,q\}$ and fix them around one edge
face-to-face. To make sense of this as a regular tessellation, we ask
that $\{p,q\}$ correspond to some compact polyhedron. This is repeated to
yield a honeycomb such that at every edge there are $r$ identical
regular polyhedra with exactly two polyhedra sharing a face.  (For
comparison, in the two-dimensional case, polygons were fixed at
vertices instead of edges, while edges instead of faces were shared by
exactly two polygons.)

As our aim is to tessellate three-dimensional spaces using a
tessellating cell that is made of a finite number of vertices, we will
only consider Schl\"afli symbols $\{p,q,r\}$ such that the
tessellating cell $\{p,q\}$ can be associated to a regular
polyhedron. This imposes a constraint on the possible values of
$(p,q,r)\in(\mathbb{N}^{\,}_{\geq3})^{3}$. This constraint is the
criterion
(\ref{eq:condition on p,q for tiling of plane, sphere, hyper})  
for the coanstant curvature of two-dimensional tilings,
i.e., $\{p,q\}$  must 
represent a spherical tiling.  Noting that reversing a Schl\"afli
symbol yields a tessellation's dual, one may deduce that $\{r,q\}$
must also represent a spherical tiling.

As in the two-dimensional case, there is a useful formula to determine
whether a three-dimensional Schl\"afli symbol represents a curved
tessellation.

%\label{Proposition for d=3}
Let $\{p,q,r\}$ with $p,q,r=3,4,5,\ldots$ be an arbitrary
Schl\"afli symbol. The corresponding tessellation fills Euclidean
space if and only if
\begin{subequations}
\label{Statement proposition 2}
\begin{equation}
\cos\left(\frac{\pi}{q}\right)= 
\sin\left(\frac{\pi}{p}\right)\,
\sin\left(\frac{\pi}{r}\right).
\label{Statement proposition 2 a}
\end{equation}
If the real number on the left-hand side
of Eq.\ (\ref{Statement proposition 2 a})
is smaller than the right-hand side,
\begin{equation}
\cos\left(\frac{\pi}{q}\right)<
\sin\left(\frac{\pi}{p}\right)\,
\sin\left(\frac{\pi}{r}\right),
\label{Statement proposition 2 b}
\end{equation}
then the tessellation $\{p,q,r\}$ fills three-dimensional spherical
space ($\mathbb{S}^{3}$).
If the real number on the left-hand side
of Eq.\ (\ref{Statement proposition 2 a})
is larger than the right-hand side,
\begin{equation}
\cos\left(\frac{\pi}{q}\right)>
\sin\left(\frac{\pi}{p}\right)\,
\sin\left(\frac{\pi}{r}\right),
\label{Statement proposition 2 c}
\end{equation}
\end{subequations}
then the tessellation $\{p,q,r\}$ fills three-dimensional
hyperbolic space ($\mathbb{H}^{3}$).

Application of criteria 
(\ref{Statement proposition 2})
on the set of Schl\"afli symbols
\begin{equation}
\big\{\, \{p,q,r\} \ \big|\ p,q,r=3,4,5,\ldots\big\}
\label{eq:def set all d=3 Schlaefli symbols}
\end{equation}
subject to the constraint that $\{p,q\}$ and $\{r,q\}$
correspond to polyhedra delivers the following.
\begin{enumerate}
\item 
There are exactly six regular tessellations of three-dimensional
spherical space, which correspond to the six regular
convex 4-polytopes.
\item 
There is exactly one regular tessellation of three dimensional space:
$\{4,3,4\}$ (cubic honeycomb).
\item 
There are exactly four compact regular tessellations of hyperbolic space.
\end{enumerate}

%\subsubsection{Regular tessellations of arbitrary-dimensional spaces}

Schl\"afli symbols in arbitrary dimensions are defined recursively. Let
$d\in \mathbb{N}^{\,}_{\geq2}$ be the dimensionality of space. The
Schl\"afli symbols we consider are of the form
\begin{equation}
\{r^{\,}_{1},r^{\,}_{2},\cdots, r^{\,}_{d-1},r^{\,}_{d}\}, 
\qquad 
r^{\,}_{1},\ldots,r^{\,}_{d} \geq 3. 
\end{equation}
The associated tessellation of $d$-dimensional
spaces with constant sectional curvature
is constructed as follows.  The tessellation cell is characterized by the
$(d-1)$-dimensional Schl\"afli symbol
$\{r^{\,}_{1},\cdots,r^{\,}_{d-1}\}$. We demand that at every ridge
[$(d-2)$-dimensional object] of the tessellation, one finds
$r^{\,}_{d}$ copies of $\{r^{\,}_{1},\cdots,r^{\,}_{d-1}\}$ arranged facet-to-facet.

For a two-dimensional
space with constant sectional curvature,
the tiling cell is a polygon (a tessellation of one-dimensional
homogeneous space) and the ridges are vertices (zero-dimensional
geometrical objects).  For a three-dimensional
space with constant sectional curvature,
the honeycomb cell is a polyhedron (a tessellation of
two-dimensional space with constant sectional curvature,)
and the ridges are edges (one-dimensional geometrical objects).

For a four‑dimensional space with constant sectional curvature,
the tessellating cell is a 4-polytope (i.e., a tessellation of
three-dimensional space with constant sectional curvature), and the
ridges are its two-dimensional faces.

As we saw by way of examples in the two- and three-dimensional cases,
each additional dimension introduces a new criterion to decide if the
tessellated space has positive, vanishing, or negative constant
curvature. Moreover, the criteria found in lower dimensions carry over
as constraints on the admissible Schl\"afli symbols. 
The resulting higher-dimensional tessellations of spherical (i.e.,
$d+1$-polytopes), Euclidean, and hyperbolic homogeneous spaces have
been tabulated, see Refs.\ \cite{mcmullen2002abstract,conway2008symmetries,grunbaum1987tilings,grunbaum1981spherical}.

For any $d\geq2$, the Schl\"afli symbol
\begin{equation}
\{4,3,\cdots,3,4\} \hbox{ with  $d-2$ occurences of 3} 
\end{equation}
represents a regular tessellation of $d$-dimensional Euclidean space. It
is the $d$-hypercubic tessellation generalizing the square tiling and
cubic honeycomb to arbitrary dimensions.

Similarly, the Schl\"afli symbol
\begin{equation}
\{4,3,\cdots,3,3\} \hbox{ with  $d-1$ occurences of 3} 
\end{equation}
is a regular tessellation of $d$-dimensional spherical space.  As
such, it corresponds to a $(d+1)$-polytope, the $(d+1)$-hypercube to
be precise.

\subsection{Construction of regular tessellations with Coxeter groups}

We now turn to the explicit construction of regular tessellations.
Our goal is to construct a simple graph that delivers a
regular tessellation of spherical, Euclidean, or hyperbolic space
by way of an algorithm that can be implemented efficiently on a computer.
Equipped with the adjacency matrix of the simple graph,
we can then program an array of Josephson-coupled superconducting
wires as a hardware platform to realize the $XY$ model on a spherical,
Euclidean, or hyperbolic space.
The key observation here
is that regular tessellations can be constructed by
reflecting points in space with respect to particular arrangements of mirror
hyperplanes, provided these reflections realize certain Coxeter groups.

\begin{figure}[t!]
\centering

\begin{minipage}[t]{0.45\linewidth}
\raggedright
(a)\\[-1mm]
\hspace{2mm}
\includegraphics[width=0.8\linewidth]{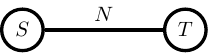}
\end{minipage}
\hfill
\begin{minipage}[t]{0.45\linewidth}
\raggedright
(b)\\[-1mm]
\hspace{2mm}
\includegraphics[width=0.8\linewidth]{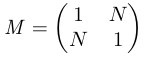}
\end{minipage}

\caption{\label{Fig:DN as Coxeter group}
Representations of the dihedral group $\mathrm{D}_{N}$
as a Coxeter group, with $r_{1}=S$ and $r_{2}=T$.
(a) Coxeter diagram.
(b) Coxeter matrix.
}
\end{figure}

\begin{figure}[t!]
\centering

\begin{minipage}[t]{0.45\linewidth}
\raggedright
(a)\\[-1mm]
\hspace{2mm}
\includegraphics[width=0.9\linewidth]{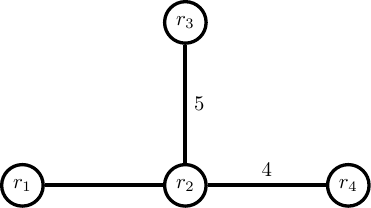}
\end{minipage}
\hfill
\begin{minipage}[t]{0.45\linewidth}
\raggedright
(b)\\[-1mm]
\hspace{2mm}
\includegraphics[width=0.8\linewidth]{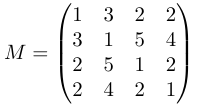}
\end{minipage}

\caption{\label{Fig:complicated example Coxeter group representation}
Representations of a Coxeter group $G$ generated by four reflections.
(a) Coxeter diagram.
(b) Coxeter matrix.
}
\end{figure}

\subsubsection{Coxeter groups, matrices, and diagrams}

Before defining Coxeter groups, we introduce group presentations
by way of two examples. Consider a finitely generated group. 
Its presentation is an enumeration of its generators, as well as a list of
algebraic relations large enough to characterize the entire group structure.
Take for example the cyclic group
\begin{subequations}
\label{eq:definition DN}
\begin{equation}
\mathbb{Z}_{N}^{\,}:=
\left\{e,g,g_{\,}^{2},g_{\,}^{3},\cdots,g_{\,}^{N-1}\right\}.
\label{eq:definition DN a}
\end{equation}
It may be more concisely described through its presentation
\begin{equation}
\mathbb{Z}_{N}^{\,}\equiv
\left\langle g \; \left| \; g_{\,}^{N} = e\right.\right\rangle.
\label{eq:definition DN b}
\end{equation}
\end{subequations}

Consider next the $N$-th order dihedral group $\mathrm{D}_{N}^{\,}$,
which describes the symmetries of a regular $N$-gon.
It contains $2N$ elements. Here, each symmetry is either a rotation,
a reflection, or their product.
In fact, each element of $\mathrm{D}_{N}^{\,}$ is
generated by a rotation $R$ by $2\pi/N$
around the origin and a reflection $S$
with respect to some axis of symmetry.
The group presentation is then
\begin{equation}
\mathrm{D}_{N}^{\,}=
\left\langle
R, S \ \big|\ R^{N} = S^{2} = e, \, S\,R\,S^{-1} = R^{-1}
\right\rangle.
\label{eq:presentation DN bis}
\end{equation}
The first pair of identities denotes the fact that rotation and
reflection are group elements of order $N$ and $2$, respectively.
The last identity states that conjugating the rotation by reflection
yields the counter-rotation.  One may also define the group element
\begin{subequations}
\begin{equation}
T\coloneqq S\,R
\label{eq:def element T=SR}
\end{equation}
and recast the presentation (\ref{eq:presentation DN bis}) as
\begin{equation}
\mathrm{D}_{N}^{\,}=
\left\langle
S, T \ \big|\ S^{2} = T^{2} = e, \, (S\,T)^{N} = e
\right\rangle.
\end{equation}
\end{subequations}
Here, $T$ represents the reflection with respect to the mirror axis of
$S$ shifted by the angle $2\pi/N$. As $S\,T$ realizes
a rotation by the angle $2\pi/N$, it is an element of order $N$.

Coxeter groups are algebraic generalizations of the dihedral groups.
Coxeter groups of finite rank $n=1,2,\ldots$
are finitely generated groups characterized by presentations
of the form
\begin{subequations}
\label{eq:def Coxeter group generated by n objects}
\begin{equation}
\left\langle
r_{1}^{\,}, r_{2}^{\,}, \cdots, r_{n}^{\,}
\ \left| \
(r_{i}^{\,}\,r_{j}^{\,})^{m_{ij}^{\,}} = e
\right.\right\rangle,
\label{eq:def Coxeter group generated by n objects a}
\end{equation}
where any element $m_{ij}^{\,}$
from the set
\begin{equation}
\left\{m_{ij}^{\,}\ \big|\ i,j=1,\ldots,n\right\}
\label{eq:def Coxeter group generated by n objects b}  
\end{equation}
of $n^{2}$ integers must obey the rules
\begin{align}
&
m_{ij}^{\,}=
m_{ji}^{\,},
\qquad
i,j=1,\ldots,n,
\label{eq:def Coxeter group generated by n objects c}
\\
&
m_{ii}^{\,}=1,
\qquad
i=1,\ldots,n,
\\
&
m_{ij}^{\,}\geq2,
\qquad
1\leq i<j\leq n.
\label{eq:def Coxeter group generated by n objects d}
\end{align}
\end{subequations}
The number $n$ is the number of generators of the Coxeter group.
The integers $m_{ij} =m_{ji}$ define a $n\times n$ symmetric matrix $M$,
the Coxeter matrix,
that concisely encodes the composition rules of the Coxeter group.
The condition $m_{ii}=1$
captures the fact that reflections are involutions, i.e., they square
to identity.  The relation $(r_{i}^{\,}\,r_{j}^{\,})^{m_{ij}} = e$ for $i\neq j$
corresponds to the geometric fact that two consecutive reflections whose
mirror axes meet at the angle $\pi/m_{ij}^{\,}$ is
equivalent to a rotation by the angle $2\pi/m_{ij}^{\,}$.
Note that if
$m_{ij}^{\,}=2$, then the reflections $r_{i}^{\,}$ and $r_{j}^{\,}$
commute with one another.

A real reflection group is
a subgroup of the orthogonal group of a
real finite-dimensional inner-product space
that is generated by reflections,
whereby a reflection is a linear transformation
that fixes a hyperplane pointwise (the mirror) and
sends a vector (a root vector) to its negative.
Thus, any real reflection group
whose mirrors meet at Coxeter angles, i.e., angles of the form
$\pi/m$ with $m\geq2$ an integer,
can be expressed as a Coxeter group.

Coxeter groups can also be represented visually as Coxeter diagrams
as follows.
Given an $n\times n$ Coxeter matrix, identify the $n$ generators
of the Coxeter group with the $n$ vertices of a weighted complete graph,
i.e., a graph such that (i) any pair of distinct vertices is
connected by one and only one undirected edge and (ii)
each undirected edge is labeled by the corresponding
element of the Coxeter matrix.
A Coxeter diagram is obtained from the weighted complete graph
by erasing all edges of weight 2 and omitting labels on edges of weight 3.
For the dihedral group $\mathrm{D}_{N}^{\,}$, $n=2$ and $m_{12}^{\,}=N$
as is shown in Fig.\ \ref{Fig:DN as Coxeter group}. 
Figure \ref{Fig:complicated example Coxeter group representation}
shows a Coxeter group generated by four reflections.

\begin{figure}[t!]
\centering

\begin{minipage}[t]{0.45\linewidth}
\raggedright
(a)\\[14mm]
\hspace{2mm}
\includegraphics[width=\linewidth]{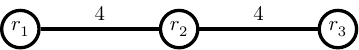}
\end{minipage}
\hfill
\begin{minipage}[t]{0.45\linewidth}
\raggedright
(b)\\[-1mm]
\hspace{2mm}
\includegraphics[width=\linewidth]{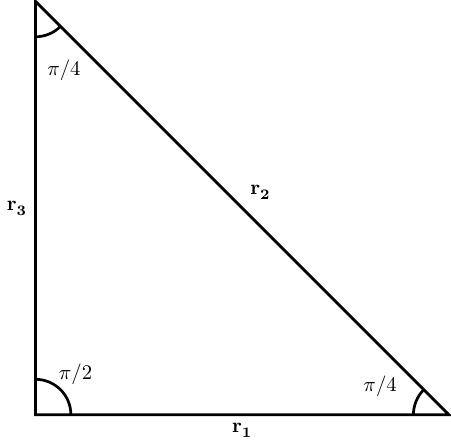}
\end{minipage}

\caption{\label{Fig:G44 construction of isocele Coxeter triangle}
(a) Coxeter diagram of $G_{\{4,4\}}$.
(b) Fundamental domain and mirror axes of the action of $G_{\{4,4\}}$
on $\mathbb{E}^{2}$.
}
\end{figure}

\subsubsection{Tits representation of Coxeter groups}
\label{subsubsec:Tits representation of Coxeter groups}

We begin by making use
of the following abstract definition of a Coxeter system.
We then give the conditions for which a Coxeter system is
naturally associated to a space of constant curvature
via the so-called Tits representation.

A Coxeter system of rank $n=1,2,\ldots$
consists of a group $G$ together with a
distinguished subset of generators $S=\{r_{1}^{\,},\cdots,r_{n}^{\,}\}$
whose elements obey the relations
(\ref{eq:def Coxeter group generated by n objects}).
A fundamental result,
known as the Tits representation
\cite{Humphreys1992,Grove96,Bjorner2005},
states that every
Coxeter system $(G,S)$ of rank $n$
can be realized faithfully by linear reflections in
a real vector space of dimension $n$ as follows.
There exists $n$ linearly independent root vectors
\begin{subequations}
\label{eq:Tits representation}
\begin{equation}
\alpha_{i}^{\,}\in\mathbb{R}_{\,}^{n},
\qquad
i=1,\ldots,n,
\label{eq:Tits representation a}
\end{equation}
and a symmetric bilinear form
\begin{equation}
\langle\cdot,\cdot\rangle:\mathbb{R}^{n}\times\mathbb{R}^{n}\to\mathbb{R}
\label{eq:Tits representation b}
\end{equation}
such that each generator $r_{i}^{\,}$ from the Coxeter system
$r_{i}^{\,}$ is faithfully represented by
\begin{equation}
\begin{split}
R_{i}^{\,}:\mathbb{R}_{\,}^{n}\to&\,\mathbb{R}_{\,}^{n},
\\
v\mapsto&\,R_{i}^{\,}(v):=
v
-
2\,
\frac{\langle v            ,\alpha_{i}^{\,}\rangle}
     {\langle\alpha_{i}^{\,},\alpha_{i}^{\,}\rangle}\,
\alpha_{i}^{\,}.
\end{split}
\label{eq:Tits representation c}
\end{equation}
Moreover, a representation of the bilinear form
(\ref{eq:Tits representation b})
is fixed by demanding that the $n\times n$
Gram matrix of the $n$ roots
(\ref{eq:Tits representation a})
has the matrix elements
\begin{equation}
\langle\alpha_{i}^{\,},\alpha_{j}^{\,}\rangle:=
-\cos(\pi/m_{ij}^{\,})
\qquad
i,j=1,\ldots,n,
\label{eq:Tits representation d}
\end{equation}
\end{subequations}
i.e.,
the inner-product structure
(\ref{eq:Tits representation b})
encodes the entire Coxeter matrix.

A key property of the Tits representation is that every generator of the
Coxeter system leaves invariant the bilinear form, i.e.,
\begin{equation}
\langle R_{i}^{\,}(u),R_{i}^{\,}(v)\rangle=
\langle u,v\rangle,
\qquad
u,v\in\mathbb{R}^{n}.
\label{eq:reflection invariance Tits bilinear form}  
\end{equation}

The $n$-dimensional root space arising in the Tits representation is an
algebraic construction. Its purpose is to encode the relations of the
Coxeter matrix through a system of reflections with respect to the
bilinear form. Each generator $r_{i}^{\,}$ with $i=1,\ldots,n$
of the Coxeter system
is assigned a root vector $\alpha_{i}^{\,}\in\mathbb{R}^{n}$
and a linear mirror plane
\begin{equation}
\alpha_{i}^{\perp}:=
\left\{
v\in\mathbb{R}^{n}\ |\
\langle v,\alpha_{i}^{\,}\rangle=0
\right\}\subset\mathbb{R}^{n},
\label{eq:def alphaiperp}
\end{equation}
i.e., the $(n-1)$-dimensional linear subspace that is point-wise
invariant under the action of $R_{i}^{\,}$. By construction,
two mirror planes $\alpha_{i}^{\perp}$ and $\alpha_{j}^{\perp}$ 
intersect with the dihedral angles $\pi/m_{ij}^{\,}$
prescribed by the Coxeter matrix.

A Coxeter system acquires a geometric interpretation
when the signature of the Gram matrix (\ref{eq:Tits representation d})
fixes the type of geometry;
spherical when it is positive definite,
Euclidean when it is semi-positive definite with one null direction,
and hyperbolic when it has Lorentzian signature.
We may then relate the Tits representation to spaces of constant curvature 
by embedding the latter into the $n$-dimensional Tits space
$\mathbb{R}^{n}$ as level sets of the bilinear form
(\ref{eq:Tits representation b}). Accordingly,
the $(n-1)$-dimensional sphere,
Euclidean space,
or hyperbolic space are realized respectively as
\begin{subequations}
\begin{align}
&  
\mathbb{S}^{\,n-1}
:=\{\,v\in\mathbb{R}^{n}\;|\;\langle v,v\rangle=+1\,\},
\\
&
\mathbb{E}^{\,n-1}
:=\{\,v\in\mathbb{R}^{n}\;|\;\langle v,w\rangle=0,\;
                          w\in\mathrm{ker}\,B\setminus\{0\}\},
\\
&
\mathbb{H}^{\,n-1}
:=\{\,v\in\mathbb{R}^{n}\;|\;\langle v,v\rangle=-1,\;v_{1}^{\,}>0\,\},
\end{align}
\end{subequations}
where $B:\mathbb{R}^{n}\to\mathbb{R}^{n}$
is the linear map represented by the Gram matrix
(\ref{eq:Tits representation d})
and $v_{1}^{\,}$ denotes one Cartesian coordinate of the vector
$v\in\mathbb{R}^{n}$.
The linear mirror planes
(\ref{eq:def alphaiperp})
are
$(n-1)$-dimensional hyperplanes in the Tits space.  Their geometric
counterparts in $\mathbb{S}^{\,n-1}$, $\mathbb{E}^{\,n-1}$, or
$\mathbb{H}^{\,n-1}$ are obtained by taking the intersections
\begin{equation}
H_{i}^{\,}:=\alpha_{i}^{\perp}\cap\mathbb{X}_{\,}^{n-1},
\qquad
\mathbb{X}_{\,}^{n-1}\in
\{\mathbb{S}_{\,}^{n-1},\mathbb{E}_{\,}^{n-1},\mathbb{H}_{\,}^{n-1}\},
\end{equation}
with $i=1,\ldots,n$.
Each $H_{i}^{\,}$ is an $(n-2)$-dimensional hypersurface
in the corresponding constant-curvature space and serves as the geometric
mirror for the reflection generated by $r_{i}^{\,}$.
In this way the linear
reflections $R_{i}^{\,}$ restricts naturally to a geometric reflection across
$H_{i}^{\,}$ for any $i=1,\ldots,n$.
Moreover, the $n$ mirrors $H_{i}^{\,}$
bound an $(n-1)$-simplex in
$\mathbb{S}^{\,n-1}$,
$\mathbb{E}^{\,n-1}$,
or $\mathbb{H}^{\,n-1}$,
with dihedral angles $\pi/m_{ij}^{\,}$ prescribed by
the Coxeter matrix.
This $(n-1)$-simplex serves as a fundamental domain
for the action of the Coxeter group
on the corresponding constant-curvature space.
It is this geometric realization that underlies the
construction of regular tessellations from Coxeter diagrams.

\subsubsection{Regular tessellations via Coxeter diagrams}
\label{Sec. regular tessellations via coxeter}

In order to construct the graph associated with the tessellation
of $\mathbb{X}^{d}\in\{\mathbb{S}^{d},\mathbb{E}^{d},\mathbb{H}^{d}\}$,
it is often convenient to represent the fundamental $d$-simplex
not in the linear space $\mathbb{R}^{n+1}$ as in the Tits representation,
but instead in the constant-curvature space
$\mathbb{X}^{d}\in\{\mathbb{S}^{d},\mathbb{E}^{d},\mathbb{H}^{d}\}$.
We refer to this embedding of the
fundamental domain into $\mathbb{X}^{d}$ as the kaleidoscopic map.
The resulting $d$-simplex in $\mathbb{X}^{d}$,
inherits the mirror structure encoded by the Coxeter diagram, so that
reflections in its faces correspond precisely to the generators of the
Coxeter group. A point chosen in the interior of this simplex serves
as a generating vertex. By repeatedly reflecting it across the mirrors
corresponding to the nodes of the Coxeter diagram, one obtains its
full orbit under the group action. This orbit determines the
vertices, and, similarly, the edges, of the resulting regular
tessellation. 

We illustrate this procedure with an example in $d=2$-dimensional space.
Let $G_{\{4,4\}}^{\,}$ be the Coxeter group associated to the Coxeter diagram
from Fig.\ \ref{Fig:G44 construction of isocele Coxeter triangle}(a).
It has $d+1=3$ generators, which we represent by
three one-dimensional mirror axes lying in $\mathbb{E}^{2}$. 
We arrange these mirrors in the shape of an isosceles right-angled triangle,
whereby each edge is in one-to-one correspondence with the
generators of $G_{\{4,4\}}^{\,}$, as is done in
Fig.\ \ref{Fig:G44 construction of isocele Coxeter triangle}(b).
Furthermore, each corner of this right-angled triangle,
i.e., the intersection of two different edges labeled by
$r_{i}^{\,}$ and $r_{j}^{\,}$ with $1\leq i<j\leq3$, respectively,
is assigned the numerical value $\pi/m_{ij}^{\,}$, see Fig.\
\ref{Fig:G44 construction of isocele Coxeter triangle}(b).
This $d=2$-simplex is nothing but the fundamental domain in
$\mathbb{E}_{\,}^{2}$
on which the reflections are acting. 
What makes the case of $G_{\{4,4\}}^{\,}$ special is that the values
$\pi/m_{12}^{\,}=\pi/m_{23}^{\,}=\pi/4$
and
$\pi/m_{13}^{\,}=\pi/2$
labeling the three internal angles of the formal triangle match
those of an isosceles right-angled triangle in two-dimensional Euclidean
space $\mathbb{E}^{2}$.

\begin{figure}[t!]
\centering

\begin{minipage}[t]{0.5\linewidth}
\raggedright
(a)\\[14mm]
\hspace{2mm}
\includegraphics[width=0.9\linewidth]{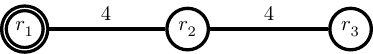}
\end{minipage}
\hfill
\begin{minipage}[t]{0.48\linewidth}
\raggedright
(b)\\[-1mm]
\hspace{2mm}
\includegraphics[width=0.9\linewidth]{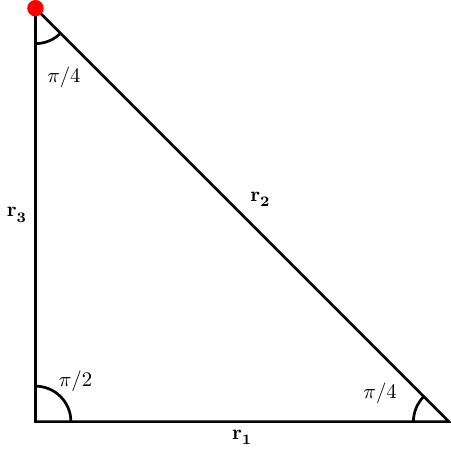}
\end{minipage}
\caption{
\label{Fig:G44 Coxeter diagram and fundamental domain with 1 active mirror}
(a) Augmented Coxeter diagram of $G_{\{4,4\}}$
with a single active mirror $r_{1}$.
(b) Fundamental domain and mirror axes of the action of $G_{\{4,4\}}$
on $\mathbb{E}^{2}$. Since the generators $r_{2}, r_{3}$
are inactive and $r_{1}$ is active, the generating vertex (red node)
must lie on the intersection of the mirrors associated to
$r_{2}$ and $r_{3}$. This fixes its position.
}
\end{figure}

Let $p\in\mathbb{E}^{2}$ be some point in the flat plane.
The orbit of $p$ under the (reflection) action of $G_{\{4,4\}}^{\,}$ is the set
\begin{equation}
G_{\{4,4\}}^{\,}\cdot p\coloneqq
\left\{
g\cdot p\ \ \left| \, g\in G_{\{4,4\}}^{\,}
\right.
\right\}.
\end{equation}
Any mirror $r_{i}^{\,}$ whose action on $p$ is trivial, i.e.,
$r_{i}^{\,}\cdot p=p$,
is called inactive. Conversely,
any mirror $r_{j}^{\,}$ whose action on $p$ is non trivial, i.e.,
$r_{j}^{\,}\cdot p\neq p$,
is called active. Given the pair
$G_{\{4,4\}}^{\,}$ and $p\in\mathbb{E}^{2}$, we modify the
Coxeter diagram of $G_{\{4,4\}}^{\,}$ by drawing two circles around those
generators that are active. If the point $p$ is chosen as the intersection of
the mirror lines labeled 
in Fig.\
\ref{Fig:G44 Coxeter diagram and fundamental domain with 1 active mirror}(b),
we get the augmented Coxeter diagram in Fig.\
\ref{Fig:G44 Coxeter diagram and fundamental domain with 1 active mirror}(a).
We call this point the generating vertex for one active mirror.
For each combination of more than one active mirrors, we call the unique point
belonging to the triangle in Fig.\
\ref{Fig:G44 construction of isocele Coxeter triangle}
that is equidistant from all the active mirrors the generating vertex.

The point fixed by the augmented Coxeter diagram is called the
generating vertex. We can interpret its orbit under
the action of the Coxeter group as the vertices of a graph.
Analogously, we can think of the generating vertex and its reflection
with respect to the only active mirror as the tuple corresponding to the
generating edge of the graph. All other edges are found by taking
the component-wise group action on the original tuple.

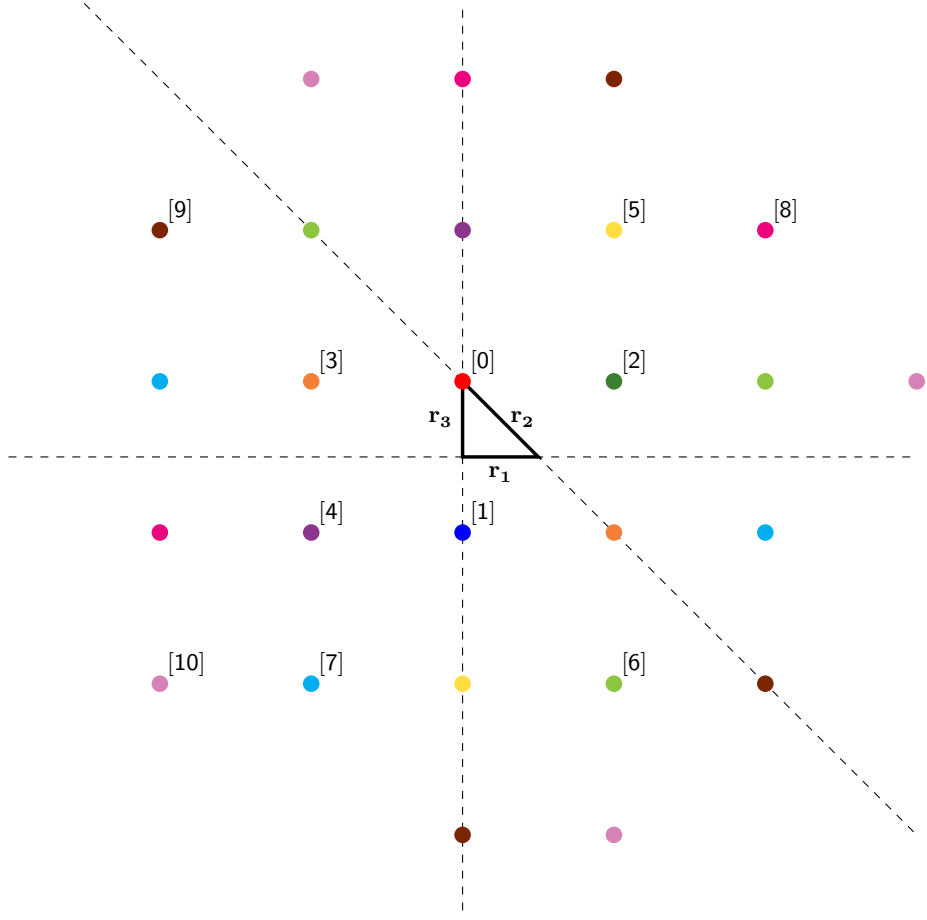
\begin{figure*}[t!]
\centering
\begin{tikzpicture}
% Coordinates for triangle vertices
\coordinate (A) at (0,0);
\coordinate (B) at (1,0);
\coordinate (C) at (0,1);

% Draw the triangle
\draw[line width=1.3pt] (A) -- node[below] {$\mathbf{r_{1}^{\,}}$} (B) -- node[above, right] {$\mathbf{r_{2}^{\,}}$} (C) -- node[left] {$\mathbf{r_{3}^{\,}}$} cycle;

% Draw mirrors
\draw[dashed] (-6,0) -- (6,0);
\draw[dashed] (0,-6) -- (0,6);
\draw[dashed] (-5,6) -- (6,-5);

% Zero Generation
\node[circle, fill=red, inner sep=2.2pt, label={[label distance=-4pt]above right:$\mathsf{[0]}$}] at (0,1) {};

% First Generation
\node[circle, fill=blue, inner sep=2.2pt, label={[label distance=-4pt]above right:$\mathsf{[1]}$}] at (0,-1) {};

% Second generation
\node[circle, fill=OliveGreen, inner sep=2.2pt, label={[label distance=-4pt]above right:$\mathsf{[2]}$}] at (2,1) {};

% Third generation
\node[circle, fill=Orange, inner sep=2.2pt, label={[label distance=-4pt]above right:$\mathsf{[3]}$}] at (-2,1) {};
\node[circle, fill=Orange, inner sep=2.2pt] at (2,-1) {};

% Fourth generation
\node[circle, fill=Fuchsia, inner sep=2.2pt, label={[label distance=-4pt]above right:$\mathsf{[4]}$}] at (-2,-1) {};
\node[circle, fill=Fuchsia, inner sep=2.2pt] at (0,3) {};

% Fifth generation
\node[circle, fill=Goldenrod, inner sep=2.2pt, label={[label distance=-4pt]above right:$\mathsf{[5]}$}] at (2,3) {};
\node[circle, fill=Goldenrod, inner sep=2.2pt] at (0,-3) {};

% Sixth generation
\node[circle, fill=LimeGreen, inner sep=2.2pt, label={[label distance=-4pt]above right:$\mathsf{[6]}$}] at (2,-3) {};
\node[circle, fill=LimeGreen, inner sep=2.2pt] at (-2,3) {};
\node[circle, fill=LimeGreen, inner sep=2.2pt] at (4,1) {};

% Seventh generation
\node[circle, fill=Cyan, inner sep=2.2pt, label={[label distance=-4pt]above right:$\mathsf{[7]}$}] at (-2,-3) {};
\node[circle, fill=Cyan, inner sep=2.2pt] at (4,-1) {};
\node[circle, fill=Cyan, inner sep=2.2pt] at (-4,1) {};

% Eigth generation
\node[circle, fill=RubineRed, inner sep=2.2pt, label={[label distance=-4pt]above right:$\mathsf{[8]}$}] at (4,3) {};
\node[circle, fill=RubineRed, inner sep=2.2pt] at (-4,-1) {};
\node[circle, fill=RubineRed, inner sep=2.2pt] at (0,5) {};

% Ninth generation
\node[circle, fill=Brown, inner sep=2.2pt, label={[label distance=-4pt]above right:$\mathsf{[9]}$}] at (-4,3) {};
\node[circle, fill=Brown, inner sep=2.2pt] at (4,-3) {};
\node[circle, fill=Brown, inner sep=2.2pt] at (2,5) {};
\node[circle, fill=Brown, inner sep=2.2pt] at (0,-5) {};

% Tenth generation
\node[circle, fill=Thistle, inner sep=2.2pt, label={[label distance=-4pt]above right:$\mathsf{[10]}$}] at (-4,-3) {};
\node[circle, fill=Thistle, inner sep=2.2pt] at (-2,5) {};
\node[circle, fill=Thistle, inner sep=2.2pt] at (2,-5) {};
\node[circle, fill=Thistle, inner sep=2.2pt] at (6,1) {};

%% Mark and label the angles
%\pic [draw, "$\pi/2$", angle eccentricity=1.6, angle radius=0.6cm] {angle = B--A--C};
%\pic [draw, "$\pi/4$", angle eccentricity=2.1, angle radius=0.6cm] {angle = C--B--A};
%\pic [draw, "$\pi/4$", angle eccentricity=2.1, angle radius=0.6cm] {angle = A--C--B};
\end{tikzpicture}
\caption{
\label{Fig:Iterative generation of square lattice points via coxeter}
Iterative construction of the orbit of the generating vertex under the
action of the Coxeter group $G_{\{4,4\}}^{\,}$.  The generating vertex
is marked in red and labeled by $[0]$. Vertices of the $m$-th
generation are constructed by applying $m$ generators of the Coxeter
group to the generating vertex. A representative of
each generation is labeled by $[m]$ and the vertices are colored as
follows: red - [0], dark blue - [1], dark green - [2], orange - [3],
dark purple - [4], yellow - [5], light green - [6], light blue -[7],
pink - [8], brown - [9], light purple - [10].
        }
\end{figure*}

\begin{figure*}
\centering
\begin{tikzpicture}
% Coordinates for triangle vertices
\coordinate (A) at (0,0);
\coordinate (B) at (1,0);
\coordinate (C) at (0,1);

% Draw the triangle
\draw[line width=1.3pt] (A) -- node[below] {$\mathbf{r_{1}^{\,}}$} (B) -- node[above, right] {$\mathbf{r_{2}^{\,}}$} (C) -- node[left] {$\mathbf{r_{3}^{\,}}$} cycle;

% Draw mirrors
\draw[dashed] (-6,0) -- (6,0);
\draw[dashed] (0,-6) -- (0,6);
\draw[dashed] (-5,6) -- (6,-5);

\node[circle, fill=red, inner sep=2.2pt, label={[label distance=-4pt]above right:$\mathsf{}$}] (A) at (0,1) {};

% First Generation
\node[circle, fill=red, inner sep=2.2pt,label={[label distance=-4pt]above right:$\mathsf{}$}] (B) at (0,-1) {};

% Second generation
\node[circle, fill=black, inner sep=2.2pt,label={[label distance=-4pt]above right:$\mathsf{}$}] (C) at (2,1) {};

% Third generatiblackon
\node[circle, fill=black, inner sep=2.2pt, label={[label distance=-4pt]above right:$\mathsf{}$}] (D) at (-2,1) {};
\node[circle, fill=black, inner sep=2.2pt, label={[label distance=-4pt]above right:$\mathsf{}$}] (E) at (2,-1) {};

% Fourth geneblackrablacktionblack
\node[circle, fill=black, inner sep=2.2pt, label={[label distance=-4pt]above right:$\mathsf{}$}] (F) at (-2,-1) {};
\node[circle, fill=black, inner sep=2.2pt, label={[label distance=-4pt]above right:$\mathsf{}$}] (G) at (0,3) {};

% Fifth generatiblackon
\node[circle, fill=black, inner sep=2.2pt, label={[label distance=-4pt]above right:$\mathsf{}$}] (H) at (2,3) {};
\node[circle, fill=black, inner sep=2.2pt, label={[label distance=-4pt]above right:$\mathsf{}$}] (I) at (0,-3) {};

% Sixth generblackatblackion
\node[circle, fill=black, inner sep=2.2pt, label={[label distance=-4pt]above right:$\mathsf{}$}] (J) at (2,-3) {};
\node[circle, fill=black, inner sep=2.2pt, label={[label distance=-4pt]above right:$\mathsf{}$}] (K) at (-2,3) {};
\node[circle, fill=black, inner sep=2.2pt, label={[label distance=-4pt]above right:$\mathsf{}$}] (L) at (4,1) {};

% Seventh generablacktioblack
\node[circle, fill=black, inner sep=2.2pt, label={[label distance=-4pt]above right:$\mathsf{}$}] (M) at (-2,-3) {};
\node[circle, fill=black, inner sep=2.2pt, label={[label distance=-4pt]above right:$\mathsf{}$}] (N) at (4,-1) {};
\node[circle, fill=black, inner sep=2.2pt, label={[label distance=-4pt]above right:$\mathsf{}$}] (O) at (-4,1) {};

% Eigth generatiblackon
\node[circle, fill=black, inner sep=2.2pt, label={[label distance=-4pt]above right:$\mathsf{}$}] (Q) at (4,3) {};
\node[circle, fill=black, inner sep=2.2pt, label={[label distance=-4pt]above right:$\mathsf{}$}] (R) at (-4,-1) {};
\node[circle, fill=black, inner sep=2.2pt, label={[label distance=-4pt]above right:$\mathsf{}$}] (S) at (2,5) {};

% Ninth generatiblackon
\node[circle, fill=black, inner sep=2.2pt, label={[label distance=-4pt]above right:$\mathsf{}$}] (T) at (-4,3) {};
\node[circle, fill=black, inner sep=2.2pt, label={[label distance=-4pt]above right:$\mathsf{}$}] (U) at (4,-3) {};
\node[circle, fill=black, inner sep=2.2pt, label={[label distance=-4pt]above right:$\mathsf{}$}] (P) at (0,-5) {};

% Tenth generatiblackon
\node[circle, fill=black, inner sep=2.2pt, label={[label distance=-4pt]above right:$\mathsf{}$}] (V) at (-4,-3) {};
\node[circle, fill=black, inner sep=2.2pt, label={[label distance=-4pt]above right:$\mathsf{}$}] (X) at (-2,5) {};
\node[circle, fill=black, inner sep=2.2pt, label={[label distance=-4pt]above right:$\mathsf{}$}] (Y) at (0,5) {};
\node[circle, fill=black, inner sep=2.2pt, label={[label distance=-4pt]above right:$\mathsf{}$}] (Z) at (2,-5) {};
\node[circle, fill=black, inner sep=2.2pt, label={[label distance=-4pt]above right:$\mathsf{}$}] (ZZ) at (6,1) {};

%%% Edges %%%

% Zeroth generation

\draw[ultra thick, red] (A) -- (B) node[near end, left] {$\mathsf{[0]}$};

% First generation

\draw[ultra thick, blue] (A) -- (C) node[midway, above] {$\mathsf{[1]}$};

% Second generation

\draw[ultra thick, OliveGreen] (A) -- (D) node[midway, above] {$\mathsf{[2]}$};
\draw[ultra thick, OliveGreen] (B) -- (E);

% Third generation

\draw[ultra thick, Orange] (A) -- (G);
\draw[ultra thick, Orange] (B) -- (F) node[midway, below] {$\mathsf{[3]}$};
\draw[ultra thick, Orange] (E) -- (C);

% Fourth generation

\draw[ultra thick, Fuchsia] (D) -- (F);
\draw[ultra thick, Fuchsia] (B) -- (I);
\draw[ultra thick, Fuchsia] (C) -- (H) node[midway, right] {$\mathsf{[4]}$};

% Fifth generation

\draw[ultra thick, Goldenrod] (C) -- (L);
\draw[ultra thick, Goldenrod] (G) -- (H);
\draw[ultra thick, Goldenrod] (D) -- (K) node[midway, left] {$\mathsf{[5]}$};
\draw[ultra thick, Goldenrod] (E) -- (J); 

% Sixth generation

\draw[ultra thick, LimeGreen] (K) -- (G);
\draw[ultra thick, LimeGreen] (E) -- (N);
\draw[ultra thick, LimeGreen] (O) -- (D);
\draw[ultra thick, LimeGreen] (I) -- (J);
\draw[ultra thick, LimeGreen] (F) -- (M) node[midway, left] {$\mathsf{[6]}$};

% Seventh generation

\draw[ultra thick, Cyan] (R) -- (F);
\draw[ultra thick, Cyan] (M) -- (I);
\draw[ultra thick, Cyan] (G) -- (Y);
\draw[ultra thick, Cyan] (L) -- (N);
\draw[ultra thick, Cyan] (Q) -- (H) node[midway, above] {$\mathsf{[7]}$};

% Eigth generation

\draw[ultra thick, RubineRed] (S) -- (H);
\draw[ultra thick, RubineRed] (Q) -- (L);
\draw[ultra thick, RubineRed] (O) -- (R);
\draw[ultra thick, RubineRed] (T) -- (K);
\draw[ultra thick, RubineRed] (J) -- (U) node[midway, above] {$\mathsf{[8]}$};
\draw[ultra thick, RubineRed] (I) -- (P);

% Ninth generation

\draw[ultra thick, Mahogany] (K) -- (X);
\draw[ultra thick, Mahogany] (Y) -- (S);
\draw[ultra thick, Mahogany] (ZZ) -- (L);
\draw[ultra thick, Mahogany] (T) -- (O);
\draw[ultra thick, Mahogany] (V) -- (M);
\draw[ultra thick, Mahogany] (J) -- (Z);
\draw[ultra thick, Mahogany] (N) -- (U) node[midway, right] {$\mathsf{[9]}$};
\end{tikzpicture}
\caption{\label{Fig:Iterative generation of square lattice edges via coxeter}
Iterative construction of the orbit of the generating edge under the
action of the Coxeter group $G_{\{4,4\}}^{\,}$.  The generating edge
is marked in red and labeled by $[0]$. Edges of the $m$-th generation
are constructed by applying $m$ generators of the Coxeter group to the
generating edge. A representative of each generation is labeled by $[m]$
and the edges are colored as follows:
red - [0], dark blue - [1], dark green - [2], orange - [3], dark
purple - [4], yellow - [5], light green - [6], light blue -[7], pink -
[8] brown - [9].
        }
\end{figure*}
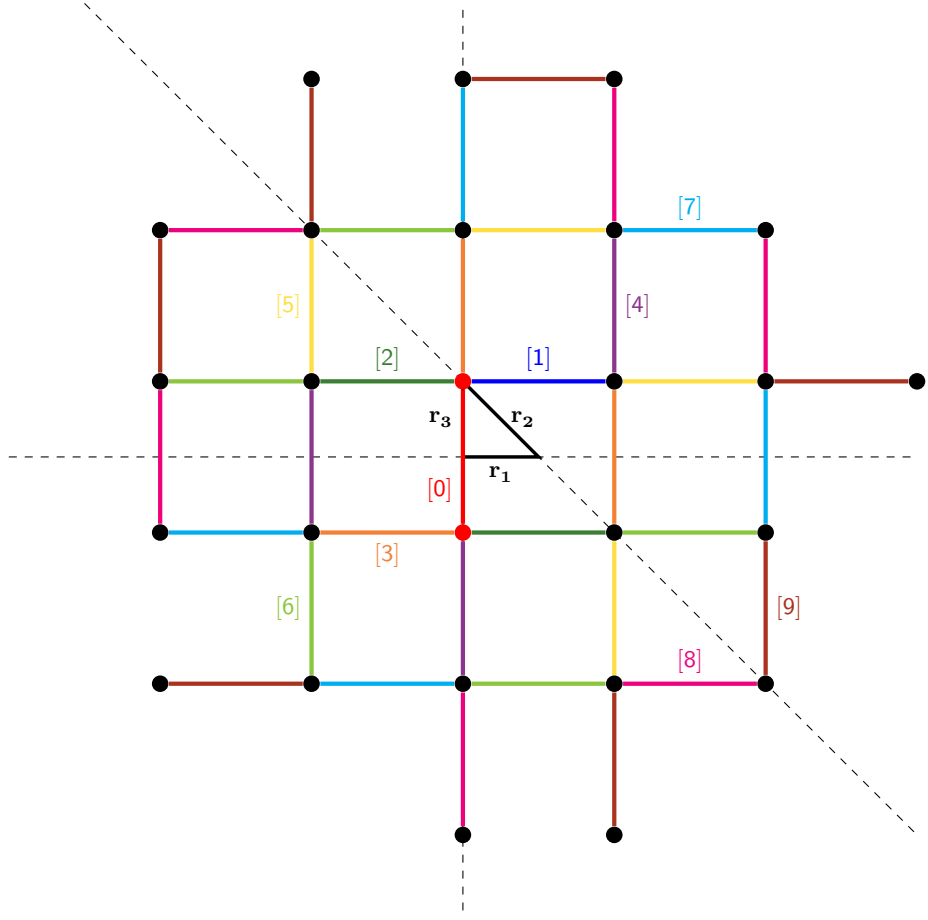

In the case of $G_{\{4,4\}}^{\,}$, we fix $r_{1}^{\,}$ to be the
only active mirror, i.e., the generating vertex $p\in\mathbb{E}^{2}$
lies on the intersection of the mirror axes associated to
$r_{2}^{\,}$ and $r_{3}^{\,}$. We then proceed by iteratively
applying the group generators $r_{1}^{\,},r_{2}^{\,},r_{3}^{\,}$ to
the generating vertex $p$ and its subsequent images, see
Fig.\
\ref{Fig:Iterative generation of square lattice points via coxeter}.
Since there is only one generating mirror, the generating
edge $(p,\, r_{1}^{\,}\!\cdot \!p)$ is unique.  Subsequent edges are
generated analogously to the construction of the vertices, see
Fig.\ \ref{Fig:Iterative generation of square lattice edges via coxeter}.
From these two figures, we infer that we have locally covered $\mathbb{E}^{2}$
with squares.  Continuously iterating the process yields a quadratic
tiling of the Euclidean plane. The Coxeter group $G_{\{4,4\}}^{\,}$
generates the regular tessellation $\{4,4\}$.

\begin{figure}[t!]
\centering

\begin{minipage}[t]{0.5\linewidth}
\raggedright
(a)\\[14mm]
\hspace{2mm}
\includegraphics[width=0.9\linewidth]{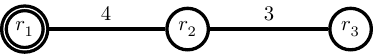}
\end{minipage}
\hfill
\begin{minipage}[t]{0.48\linewidth}
\raggedright
(b)\\[-1mm]
\hspace{2mm}
\includegraphics[width=0.9\linewidth]{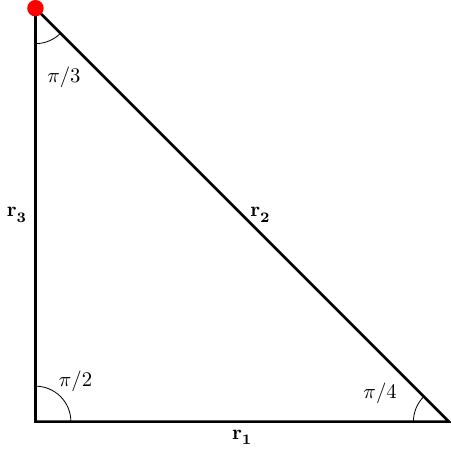}
\end{minipage}

\caption{\label{Fig:G43 Coxeter diagram and fundamental domain without active mirrors}
(a) Augmented Coxeter diagram of $G_{\{4,3\}}$ with a single active mirror
$r_{1}$.
(b) Kaleidoscopic fundamental domain and mirror axes of the action
of $G_{\{4,3\}}$ on $\mathbb{S}^{2}$.
The generating vertex is marked in red. Note that the sum of the interior
angles of the triangle is larger than $\pi$.
}
\end{figure}

\begin{figure}[t!]
\centering

\begin{minipage}[t]{0.5\linewidth}
\raggedright
(a)\\[14mm]
\hspace{2mm}
\includegraphics[width=0.9\linewidth]{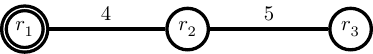}
\end{minipage}
\hfill
\begin{minipage}[t]{0.48\linewidth}
\raggedright
(b)\\[-1mm]
\hspace{2mm}
\includegraphics[width=0.9\linewidth]{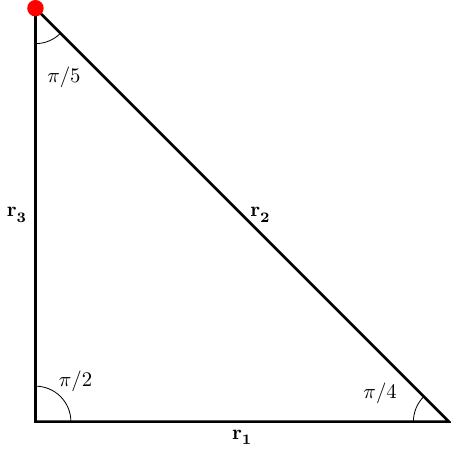}
\end{minipage}

\caption{\label{Fig:G45 Coxeter diagram and fundamental domain without active mirrors}
(a) Augmented Coxeter diagram of $G_{\{4,5\}}$ with a single active mirror
$r_{1}$.
(b) Kaleidoscopic fundamental domain and mirror axes of the action
of $G_{\{4,5\}}$ on $\mathbb{H}^{2}$. The generating vertex is marked
in red. Note that the sum of the interior angles of the triangle is
smaller than $\pi$.
}
\end{figure}

Consider now the related Schl\"afli symbols $\{4,3\}$ and
$\{4,5\}$ which tile the spherical and hyperbolic planes, respectively.
Let $G_{\{4,3\}}$ and $G_{\{4,5\}}$ be Coxeter groups defined
analogously as $G_{\{4,4\}}^{\,}$.  The respective Coxeter diagrams
and kaleidoscopic fundamental domains can be found in Figs.\
\ref{Fig:G43 Coxeter diagram and fundamental domain without active mirrors}
and
\ref{Fig:G45 Coxeter diagram and fundamental domain without active mirrors}.
One notes that the sum of the interior angles of the
triangular kaleidoscopic fundamental domain associated to $G_{\{4,3\}}^{\,}$
($G_{\{4,5\}}^{\,}$) is larger (smaller) than $\pi$; such a triangle can
only exist in spherical (hyperbolic) space.  As expected, the Coxeter
groups associated to the tilings $\{4,3\}$ and $\{4,5\}$ act on the
spherical and hyperbolic plane respectively.  As before, the
tessellations can be constructed by considering the orbits of the
generating vertex and edge.  In particular, one finds that the orbit
under the action of $G_{\{4,3\}}$ is finite, in accordance to the
compactness of the 2-sphere.

It turns out that this correspondence of tessellations to Coxeter
groups holds for any Schl\"afli symbol.  In higher dimensions, the
triangular kaleidoscopic fundamental domain is replaced by the
kaleidoscopic image of the fundamental domain
associated to the $d+1$ generating mirror hyperplanes.
We summarize this result as follows.

\begin{widetext}
Every regular tessellation defined by the Schl\"afli symbol
$\{q_{1}^{\,},q_{2}^{\,},\cdots, q_{d}^{\,}\}$
of a $d$-dimensional Riemannian manifold
$\mathbb{X}^{d}\in\{\mathbb{S}^{d},\mathbb{E}^{d},\mathbb{H}^{d}\}$
of constant curvature
can be constructed as the orbit of the generating vertex
associated to the augmented Coxeter diagram
\begin{equation}
\includegraphics[width=1\textwidth]{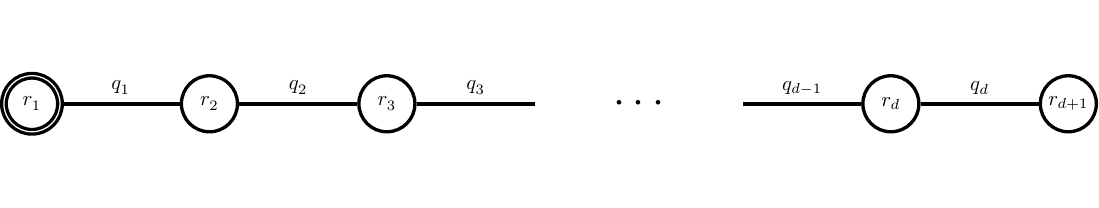}
\label{eq:thm1}
\end{equation}
where $\{r_{1}^{\,},\cdots,r_{d+1}\}$ are the generators of the Coxeter group.
\end{widetext}

\subsubsection{Constructing regular tessellations as simple graphs}

Regular tessellations of $d$-dimensional spaces are fully described by
tuples of $d$ positive integers, the Schl\"afli symbols.
Section \ref{Sec. regular tessellations via coxeter}
gave an algebraic representation of regular tessellations 
through augmented Coxeter diagrams.
The key insight is that the Schl\"afli symbols identify tessellations
descriptively, while the augmented Coxeter diagrams do so constructively.

Regular tessellations of $d$-dimensional spaces
can be easily stored on a computer.
As all edges are of the same length,
we do not need to record the distance between tessellation vertices.
As the incoming edges at a vertex are all evenly spaced,
their arrangement is implicitly fixed.
Thus, we need only keep track of which vertices are connected to one another.
In other words, a regular tessellation is encoded by the
adjacency matrix of a simple graph,
which we shall call the regular tessellation graph.

As regular tessellations can be realized algebraically through
a Coxeter group, we expect that group theory alone is sufficient to
construct a regular tessellation graph.  As the vertices of a regular
tessellation correspond to the orbit under the action of a
Coxeter group of a generating vertex, we can address all the vertices
of a regular tessellation graph by an element of the Coxeter group,
with the caveat that this address need not be unique.

Consider for example the group $G_{\{4,4\}}^{\,}$ with Coxeter diagram as in
Fig.\
\ref{Fig:G44 Coxeter diagram and fundamental domain with 1 active mirror}.
From Fig.\
\ref{Fig:Iterative generation of square lattice points via coxeter}
we infer that the fourth-generation purple vertex labeled by $\mathsf{[4]}$
can either be addressed by
\begin{subequations}
\begin{equation}
(r_{1}^{\,} \, r_{3}^{\,} \, r_{2}^{\,} \, r_{1}^{\,}) \cdot v
\end{equation}
or by
\begin{equation}
(r_{3}^{\,} \, r_{1}^{\,} \, r_{2}^{\,} \, r_{1}^{\,}) \cdot v,
\end{equation}
\end{subequations}
where $v$ denotes the generating vertex.

We resolve this issue by quotienting out this ambiguity
with the help of the orbit-stabilizer theorem.
%\begin{prop}[Orbit-Stabilizer]
Let $\mathrm{H}$ be a group acting on a set $X$.
We denote the orbit of $x\in X$ under the action of $\mathrm{H}$ by
\begin{subequations}
\label{eq:orbit-stabilizer theorem}
\begin{equation}
\mathrm{H}\mathbf{\cdot}x:=
\{h\cdot x \, | \, h \in \mathrm{H}\}
\end{equation}
and the stabilizer of $x$ in $\mathrm{H}$ by
\begin{equation}
\mathrm{H}_{x}^{\,}:=
\{h\in \mathrm{H} \, | \, h\cdot x =x\}.
\end{equation}
Then, there exists a bijection between the orbit $\mathrm{H}\mathbf{\cdot} x$
and the group $\mathrm{H}$ quotiented by the stabilizer $\mathrm{H}_{x}^{\,}$,
\begin{equation}
\mathrm{H}\mathbf{\cdot}x \, \overset{1:1}{\longleftrightarrow} \,
\mathrm{H}/\mathrm{H}_{x}^{\,}.
\end{equation}
\end{subequations}
In other words, the elements
$\{h \! \cdot  \! x\}_{h\in\mathrm{H}}$
are in  one-to-one correspondence with the left cosets
$\{h\,\mathrm{H}_{x}^{\,}\}_{h\in\mathrm{H}}$.
%\end{prop}

As an application of the orbit-stabilizer theorem
(\ref{eq:orbit-stabilizer theorem}),
consider the regular tessellation with the Schl\"afli symbol
$\{q_{1}^{\,},q_{2}^{\,},\cdots, q_{d}^{\,}\}$
that specifies the augmented Coxeter diagram in Eq.\
(\ref{eq:thm1}).
If we identify $\mathrm{H}$ with the Coxeter group $G$
and $x$ with the
generating vertex $v$ associated to the
augmented Coxeter diagram (\ref{eq:thm1}),
respectively,
we may uniquely address all the
vertices of the regular tessellation
$\{q_{1}^{\,},q_{2}^{\,},\cdots, q_{d}^{\,}\}$
with the elements of
$G/G_{v}^{\,}$. Now, we claim that
\begin{equation}
G_{v}^{\,}= \langle r_{2}^{\,}, \cdots, r_{d+1}^{\,} \rangle.
\end{equation}
The proof of this fact relies on viewing the Coxeter group as a root system
and can be found in Satz 2 of Ref.\ \cite{Witt41}, or
Theorem 5.4.1 of Ref.\ \cite{Grove2010-vf} for a more modern formulation.

\begin{widetext}
The following theorem holds.
%\begin{thm}[Regular tessellations as graphs via Coxeter diagrams]
Let $\{q_{1}^{\,},q_{2}^{\,},\cdots,q_{d}\}$ be the Schl\"afli symbol
of some regular tessellation of
$d$-dimensional space
$\mathbb{X}^{d}\in\{\mathbb{S}^{d},\mathbb{E}^{d},\mathbb{H}^{d}\}$
of constant curvature.
Then, there exists a Coxeter group $G$ with augmented Coxeter diagram
\begin{subequations}
\begin{equation}
\includegraphics[width=1\textwidth]{FIGURES/coxeter-augmented-thm1.pdf}
\label{eq:thm2}
\end{equation}
that generates this regular tessellation in the following sense:
\begin{enumerate}
\item 
Let
\begin{equation}
v_{0}^{\,}\in\mathbb{X}^{d}
\end{equation}
be the generating vertex and
\begin{equation}
E_{0}^{\,}=
[v_{0}^{\,}, \, r_{1}^{\,} \!\cdot \! v_{0}^{\,}]
\end{equation}
the generating edge designated in the augmented Coxeter diagram.
The orbit of $v_{0}^{\,}$ and $E_{0}^{\,}$ under the action of $G$ results
in a regular tessellation of $\mathbb{X}^{d}$
with curvature determined
by the Schl\"afli symbols.
\item\label{thm item 2}
The vertices of the tessellation
$\{q_{1}^{\,},\cdots,q_{d}^{\,}\}$ are in a one-to-one
correspondence with the cosets
\begin{equation}
t\,G_{v_{0}^{\,}}^{\,},
\qquad
t\in G,
\end{equation}
of the quotient $G/G_{v_{0}^{\,}}$, where
\begin{equation}
G_{v_{0}^{\,}}^{\,}=
\langle r_{2}^{\,},\cdots,r_{d+1}^{\,}\rangle< G
\end{equation}
is the stabilizer of $v_{0}^{\,}$.
\item
The coset
\begin{equation}
e\,G_{v_{0}^{\,}}^{\,} = G_{v_{0}^{\,}}^{\,}
\end{equation}
corresponds to the generating vertex, and
the unordered pair of cosets
\begin{equation}
\left(G_{v_{0}^{\,}}^{\,}, \, r_{1}^{\,} G_{v_{0}^{\,}}^{\,}\right)
\end{equation}
corresponds to the generating edge.
All edges of the tessellation are in one-to-one
correspondence with the unordered pairs of cosets
\begin{equation}
\left(t\,G_{v_{0}^{\,}}^{\,},\, (t\,r_{1}^{\,})\,G_{v_{0}^{\,}}^{\,}\right),
\qquad
t\in G.
\end{equation}
\item 
The simple graph resulting from the above correspondence characterizes
fully the original regular tessellation.
\end{enumerate}
\end{subequations}
%\end{thm}
As a corollary, in order to construct the simple graph
(and its adjacency matrix)
that is described by the regular tessellation
with the Schl\"afli symbol
$\{q_{1}^{\,},q_{2}^{\,},\cdots,q_{d}\}$,
we only need to identify
the cosets of the quotient $G/G_{v_{0}^{\,}}^{\,}$.
This can be done using the
Todd-Coxeter coset enumeration algorithm \cite{Todd36},
which has been implemented on computational platforms such as GAP
\cite{GAP4} and Python. %\cite{SymPy}.
\end{widetext}

\section{Kubo formula and fluctuation-dissipation theorem}
\label{appendix:Linear response theory for a quantum Josephson array}

{A review of the
Kubo formula and the fluctuation-dissipation theorem
is given.}

Consider the time-independent observable $\widehat{A}$ with
the unperturbed expectation value
\begin{subequations}
\label{eq:Kubo formula}
\begin{equation}
\langle\widehat{A}\rangle_{0\,\beta}^{\,}=
\frac{
\mathrm{Tr}_{\mathfrak{H}_{0}^{\,}}\,
e^{-\beta\,\widehat{H}_{0}^{\,}}\,
\widehat{A}
     }
     {
\mathrm{Tr}_{\mathfrak{H}_{0}^{\,}}
e^{-\beta\,\widehat{H}_{0}^{\,}}\,
\hphantom{\widehat{A}}
     }
\label{eq:Kubo formula a}
\end{equation}
for a closed system in thermodynamic equilibrium with the
conserved Hamiltonian $\widehat{H}_{0}^{\,}$.
Imagine that this closed system is coupled to the environment
by the perturbation
\begin{equation}
\Theta\left(t-t_{0}^{\,}\right)\,\widehat{H}'(t),
\qquad
\widehat{H}'(t)=f(t)\,\widehat{B},
\label{eq:Kubo formula b}
\end{equation}
with the operator $\widehat{B}$ time independent.
The Kubo formula quantifies the change of
the expectation value (\ref{eq:Kubo formula a}) of observable $\widehat A$
to leading order in perturbation theory. This change is given by

\begin{equation}
\left\langle\widehat{A}(t)\right\rangle_{0\,\beta}^{\,}\:=
\left\langle\widehat{A}\right\rangle_{0\,\beta}^{\,}
+
\int\limits_{\mathbb{R}} \mathrm{d}t'\,
C^{\mathrm{R}}_{0\,\beta\,\widehat{A},\widehat{B}}(t-t')\,
f(t')
\label{eq:Kubo formula c}
\end{equation}
in the time domain and
\begin{equation}
\left\langle\widehat{A}(\omega)\right\rangle_{0\,\beta}^{\,}=
2\pi
\left\langle\widehat{A}\right\rangle_{0\,\beta}^{\,}
\delta(\omega)
+
C^{\mathrm{R}}_{0\,\beta\,\widehat{A},\widehat{B}}(\omega)\,
f(\omega)
\label{eq:Kubo formula d}
\end{equation}
in the frequency domain.
Here, the retarded correlation function in the time domain is given by
(the Heisenberg time evolution is done with $\widehat{H}_{0}^{\,}$)
\begin{multline}
C^{\mathrm{R}}_{0\,\beta\,\widehat{A},\widehat{B}}(t-t')=
-\frac{\mathrm{i}}{\hbar}\,\Theta(t-t')
\\
\times \lim_{t_{0}^{\,}\to-\infty}\,
\left\langle\left[
\widehat{A}^{\,}_{\mathrm{H}}(t,t_{0}^{\,}),
\widehat{B}^{\,}_{\mathrm{H}}(t',t_{0}^{\,})
\right]\right\rangle_{0\,\beta}^{\,}.
\label{eq:Kubo formula e}
\end{multline}
\end{subequations}
From now on, $t'=t_{0}^{\,}=0$.
Our goal is to find a useful representation of {
\begin{equation}
C^{\mathrm{R}}_{0\,\beta\,\widehat{A},\widehat{B}}(\omega):=
\int\mathrm{d}t\,
e^{+\mathrm{i}\omega\,t}\,
C^{\mathrm{R}}_{0\,\beta\,\widehat{A},\widehat{B}}(t). 
\end{equation}
To this end}, we make use of the fluctuation-dissipation theorem.

In the time domain, the fluctuation-dissipation theorem states that
\begin{subequations}
\label{eq:fluctuation-dissipation theorem}
\begin{align}
C_{0\,\beta\,\widehat{A},\widehat{B}}^{\mathrm{R}}(t)=&\,
-\frac{\mathrm{i}}{\hbar}\,\Theta(t)\,
\int\limits_{-\infty}^{+\infty}
\frac{\mathrm{d}\omega_{\,}^{\prime}}{2\pi}\,
e^{-\mathrm{i}\omega_{\,}^{\prime}\,t}
\nonumber\\
&\,
\times
\left(
1 - e^{-\beta\,\hbar\,\omega_{\,}^{\prime}}
\right)\,
J_{0\,\beta\,\widehat{A},\widehat{B}}^{\vphantom{A}}(+\omega_{\,}^{\prime}),
\label{eq:fluctuation-dissipation theorem a}
\end{align}
where
\begin{align}
J_{0\,\beta\,\widehat{A},\widehat{B}}^{\vphantom{A}}(\omega)=&\,
\int\limits_{-\infty}^{+\infty}
\mathrm{d}t\,
e^{+\mathrm{i}\omega\,t}
\left\langle
\widehat{A}_{\mathrm{H}}^{\,}(t)\,
\widehat{B}_{\mathrm{H}}^{\,}(0)
\right\rangle^{\,}_{0\,\beta}
\nonumber\\
\equiv&\,
\int\limits_{-\infty}^{+\infty}
\mathrm{d}t\,
e^{+\mathrm{i}\omega\,t}\,
J_{0\,\beta\,\widehat{A},\widehat{B}}^{\vphantom{A}}(t).
\label{eq:fluctuation-dissipation theorem b}
\end{align}
In the frequency domain,
the fluctuation-dissipation theorem states that
\begin{align}
C_{0\,\beta\,\widehat{A},\widehat{B}}^{\mathrm{R}}&(\omega)=
\frac{1}{\hbar}
\int\limits_{-\infty}^{+\infty}
\frac{\mathrm{d}\omega_{\,}^{\prime}}{2\pi}\,
\left(
1-e^{-\beta\,\hbar\,\omega_{\,}^{\prime}}
\right)
\nonumber\\
\times&
\left[
\frac{\mathrm{PV}}{\omega-\omega_{\,}^{\prime}}
-
\mathrm{i}\pi\,
\delta(\omega-\omega_{\,}^{\prime})
\right]\,
J_{0\,\beta\,\widehat{A},\widehat{B}}^{\vphantom{A}}(\omega_{\,}^{\prime}).
\label{eq:fluctuation-dissipation theorem c}
\end{align}
\end{subequations}
\begin{proof}
The retarded Green function of operators
$\hat{A}$ and 
$\hat{B}$ as a function of $t$ 
has the integral representation
\begin{align}
C^{\mathrm{R}}_{0\,\beta\,\hat{A},\hat{B}}(t)\!=&\!
-\frac{\mathrm{i}}{\hbar}\Theta(t)\!
\left\langle
\left[
\hat{A}^{\,}_{\mathrm{H}}(t)\hat{B}^{\,}_{\mathrm{H}}(0)
\!-\!
\hat{B}^{\,}_{\mathrm{H}}(0)\hat{A}^{\,}_{\mathrm{H}}(t)
\right]
\right\rangle^{\,}_{0\,\beta}
\nonumber\\
=&\!
-\frac{\mathrm{i}}{\hbar}
\Theta(t)
\left[
J^{\vphantom{A}}_{0\,\beta\,\hat{A},\hat{B}}(+t)
-
J^{\vphantom{A}}_{0\,\beta\,\hat{B},\hat{A}}(-t)
\right]
\nonumber\\
=&\!
-\frac{\mathrm{i}}{\hbar}\Theta(t)
\int\limits_{-\infty}^{+\infty}
\frac{\mathrm{d}\omega^\prime}{2\pi}
e^{-\mathrm{i}\omega^\prime t}
\nonumber\\
&\!
\times
\left[
J^{\vphantom{A}}_{0\,\beta\,\hat{A},\hat{B}}(+\omega^\prime)
-
J^{\vphantom{A}}_{0\,\beta\,\hat{B},\hat{A}}(-\omega^\prime)
\right]
\label{appeq: Green RET in terms spectral fcts}
\end{align}
in terms of the spectral density function of operators $\hat{A}$ and $\hat{B}$.
Now,
$J^{\vphantom{A}}_{0\,\beta\,\hat{A},\hat{B}}(+\omega^\prime)$
and
$J^{\vphantom{A}}_{0\,\beta\,\hat{B},\hat{A}}(-\omega^\prime)$
are related. To see this denote by $\{|\mu\rangle\}$ the exact basis
of eigenstates of $\hat{H}^{\,}_{0}$ with eigenvalues $\{{E}^{\,}_{0\,\mu}\}$.
From the definition
(\ref{eq:fluctuation-dissipation theorem b}),
\begin{widetext}
\begin{equation}
\begin{split}
J^{\vphantom{A}}_{0\,\beta\,\hat{A},\hat{B}}(+\omega)=&\,
\int\limits_{-\infty}^{+\infty}
\mathrm{d}t\,
e^{+\mathrm{i}\omega t}
\langle
\hat{A}^{\,}_{\mathrm{H}}(t)\hat{B}^{\,}_{\mathrm{H}}(0)
\rangle^{\,}_{0\,\beta}
\\
=&\,
Z^{-1}_{0\,\beta}
\sum_{\mu,\nu}
\int\limits_{-\infty}^{+\infty}
\mathrm{d}t\,
e^{+\mathrm{i}(\hbar\omega+{E}^{\,}_{0\,\mu}-{E}^{\,}_{0\,\nu})t/\hbar}
e^{-\beta{E}^{\,}_{0\,\mu}}
\left\langle\mu\left|\hat{A}^{\,}_{\mathrm{H}}(0)\right|\nu\right\rangle
\left\langle\nu\left|\hat{B}^{\,}_{\mathrm{H}}(0)\right|\mu\right\rangle
\\
=&\,
2\pi
Z^{-1}_{0\,\beta}
\sum_{\mu,\nu}
e^{-\beta{E}^{\,}_{0\,\mu}}
\left\langle\mu\left|\hat{A}^{\,}_{\mathrm{H}}(0)\right|\nu\right\rangle
\left\langle\nu\left|\hat{B}^{\,}_{\mathrm{H}}(0)\right|\mu\right\rangle
\delta\left(\omega+\frac{E^{\,}_{0\,\mu}-E^{\,}_{0\,\nu}}{\hbar}\right)
\end{split}
\label{appeq: Lehmann J(+omega) AB}
\end{equation}
and
\begin{equation}
\begin{split}
J^{\vphantom{A}}_{0\,\beta\,\hat{B},\hat{A}}(-\omega)=&\,
\int\limits_{-\infty}^{+\infty}
\mathrm{d}t\,
e^{+\mathrm{i}\omega t}
\langle
\hat{B}^{\,}_{\mathrm{H}}(0)
\hat{A}^{\,}_{\mathrm{H}}(t)
\rangle^{\,}_{0\,\beta} 
\\
=&\,
Z^{-1}_{0\,\beta}
\sum_{\mu,\nu}
\int\limits_{-\infty}^{+\infty}
\mathrm{d}t\,
e^{+\mathrm{i}(\hbar\omega+{E}^{\,}_{0\,\nu}-{E}^{\,}_{0\,\mu})t/\hbar}
e^{-\beta{E}^{\,}_{0\,\mu}}
\left\langle\mu\left|\hat{B}^{\,}_{\mathrm{H}}(0)\right|\nu\right\rangle
\left\langle\nu\left|\hat{A}^{\,}_{\mathrm{H}}(0)\right|\mu\right\rangle
\\
=&\,
2\pi
Z^{-1}_{0\,\beta}
\sum_{\mu,\nu}
e^{-\beta{E}^{\,}_{0\,\mu}}
\left\langle\mu\left|\hat{B}^{\,}_{\mathrm{H}}(0)\right|\nu\right\rangle
\left\langle\nu\left|\hat{A}^{\,}_{\mathrm{H}}(0)\right|\mu\right\rangle
\delta\left(\omega+\frac{E^{\,}_{0\,\nu}-E^{\,}_{0\,\mu}}{\hbar}\right)
\\
=&\,
2\pi
Z^{-1}_{0\,\beta}
\sum_{\mu,\nu}
e^{-\beta{E}^{\,}_{0\,\nu}}
\left\langle\mu\left|\hat{A}^{\,}_{\mathrm{H}}(0)\right|\nu\right\rangle
\left\langle\nu\left|\hat{B}^{\,}_{\mathrm{H}}(0)\right|\mu\right\rangle
\delta\left(\omega+\frac{E^{\,}_{0\,\mu}-E^{\,}_{0\,\nu}}{\hbar}\right),
\end{split}
\label{appeq: Lehmann J(-omega) BA}
\end{equation}
\end{widetext}
where the canonical partition function is given by
\begin{align}
Z^{\vphantom{-1}}_{0\,\beta}:=
\sum_{\mu}
e^{-\beta{E}^{\,}_{0\,\mu}}.
\end{align}
Making use of the constraint of energy conservation in the
so-called Lehmann expansions 
(\ref{appeq: Lehmann J(+omega) AB})
and
(\ref{appeq: Lehmann J(-omega) BA}),
we infer that
\index{Lehmann expansion}
\begin{align}
J^{\vphantom{A}}_{0\,\beta\,\hat{B},\hat{A}}(-\omega)=
e^{-\beta\,\omega}
J^{\vphantom{A}}_{0\,\beta\,\hat{A},\hat{B}}(+\omega),
\label{appeq: relationship between J AB and J BA as fct omega}
\end{align}
for which Eq.\ (\ref{eq:fluctuation-dissipation theorem a})
follows.
\end{proof}
We observe that the Lehmann expansion 
(\ref{appeq: Lehmann J(+omega) AB})
implies that, in the zero temperature limit, the summation over $\mu$
is restricted to the linearly-independent  ground states, in which case
\begin{equation}
E^{\,}_{0\,\mu}-E^{\,}_{0\,\nu}\leq0
\end{equation}
for any $\nu$. It follows that
\begin{equation}
\lim_{\beta\uparrow\infty}J^{\vphantom{A}}_{0\,\beta\,\hat{B},\hat{A}}(\omega)=0,
\qquad
\forall\omega<0.
\label{eq:J0betaABomega vanishes for megative omege if beta is infinity}
\end{equation}

%%%%%%%%%%%%%%%%%%%%%%%%%%%%%%%%%%%%%%%%%%%%%%%%%%%%%%%%%%%%%%%%%%%%%%%%%%%%%%%
%%%%%%%%%%%%%%%%%%%%%%%%%%%%%%%%%%%%%%%%%%%%%%%%%%%%%%%%%%%%%%%%%%%%%%%%%%%%%%%
\bibliography{references-hyper}

\onecolumngrid
\setcounter{secnumdepth}{2}

\renewcommand{\appendixname}{Supplementary Material}

\appendix

\end{document}